\documentclass[aip,reprint]{revtex4-1}

\usepackage{amsmath}
\usepackage[utf8]{inputenc}
\usepackage{xcolor}
\DeclareUnicodeCharacter{0087}{\color{red}{BAD!!}}
\usepackage{graphics}
\usepackage{graphicx}
\usepackage{hyperref} 
\usepackage[T1]{fontenc} 
\draft 

\begin{document}


\title{Femtosecond Diffraction and Dynamic High Pressure Science} 

\author{Justin S. Wark}
\email[]{justin.wark@physics.ox.ac.uk}
\affiliation{Department of Physics, Clarendon Laboratory, University of Oxford, Parks Road, Oxford, OX1 3PU, United Kingdom}

\author{Malcolm I. McMahon}
\affiliation{SUPA, School of Physics and Astronomy, and Centre for Science at Extreme Conditions,
The University of Edinburgh, Edinburgh EH9 3FD, United Kingdom}

\author{Jon H. Eggert}
\affiliation{Lawrence Livermore National Laboratory, 7000 East Avenue, Livermore, California 94501, USA}

\date{\today}

\begin{abstract}
Solid state material at high pressure is prevalent throughout the Universe, and an understanding of the structure of matter under such extreme conditions, gleaned from x-ray diffraction, has been pursued for the best part of a century.  The highest pressures that can be reached to date (2 TPa) in combination with x-ray diffraction diagnosis have been achieved by dynamic compression via laser ablation~\cite{Lazicki2021}.  The past decade has witnessed remarkable advances in x-ray technologies, with novel x-ray Free-Electron-Lasers (FELs) affording the capacity to produce high quality single-shot diffraction data on time scales below 100 fsec.  We provide a brief history of the field of dynamic compression, spanning from when the x-ray sources were almost always laser-plasma based, to the current state-of-the art diffraction capabilities provided by FELs. We give an overview of the physics of dynamic compression, diagnostic techniques, and the importance of understanding how the rate of compression influences the final temperatures reached.  We provide illustrative examples of experiments performed on FEL facilities that are starting to give insight into  how materials deform at ultra-high strain rates, their phase diagrams, and the types of states that can be reached. We emphasize that there often appear to be differences in the crystalline  phases observed between use of static and dynamic compression techniques.  We give our perspective on both the current state of this rapidly evolving field, and some glimpses of how we see it developing in the near-to-medium term.

\end{abstract}


\maketitle 

\section{Introduction and Background}
\label{S:Intro}

\subsection{Background}
\label{SS:Background}

Over the past decade we have witnessed impressive developments in both dynamic high pressure science, and in x-ray physics, that when combined provide future potential to create, study and diagnose the structure and properties of matter in the laboratory under unprecedented conditions of pressure and temperature.  Previously unexplored regimes of the phase diagram will become accessible, giving rise to the opportunity to probe materials in the laboratory that have hitherto only existed deep within planets other than our own, be they within our own solar system or beyond~\cite{Duffy2019}. As well as the final states achieved, the deformation pathways that  materials  take to reach these conditions are also of interest in their own right,  will also be amenable to interrogation, and indeed need to be understood if the results of dynamic compression experiments are ultimately going to be meaningfully related to planetary systems. 

The pertinent development in dynamic high pressure science has been the crafting of techniques, based on short-pulse (typically nanosecond) laser-ablation, that afford in a reasonably controlled way, the creation of high pressure states of solid-state matter beyond the TPa ($\sim$10 Mbar) regime.  We note at the outset of our discussions that it is important to recognise that the rate at which the sample  (usually a solid foil, of order microns to tens-of-microns thick) is dynamically loaded plays a pivotal role in determining its conditions at peak pressure: if the compression wave formed by the nanosecond laser ablation process is allowed to steepen into a shock, then the sample's state will lie on the Hugoniot, and will be considerably hotter than if the pressure is applied more slowly, and within metals will lead to shock melting at a relatively small fraction of a TPa. In contrast, if the sample is compressed in a temporally-ramped manner, yet still on the timescale of nanoseconds,  the path taken by the sample will be closer to an isentrope, keeping it cooler for the same compression ratio. Such dynamical ramp compression approaches can achieve conditions that 
exceed the static pressures achievable in diamond anvil cells (DACs) by a considerable margin - for example diamond itself has been compressed in such a manner to pressures of order 5 TPa \cite{Smith2014} -- although in the absence of diffraction data - and up to 2 TPa with diffraction information being obtained with a laser-plasma generated x-ray source~\cite{Lazicki2021}.  Such pressures should be compared with those thought to exist at the centers of such planets as Neptune ($\sim$ 0.8 TPa)~\cite{Helled2010} and Saturn ($\sim$4 TPa)~\cite{Smith2014}.

\begin{figure}
     \includegraphics[width=0.95\columnwidth]{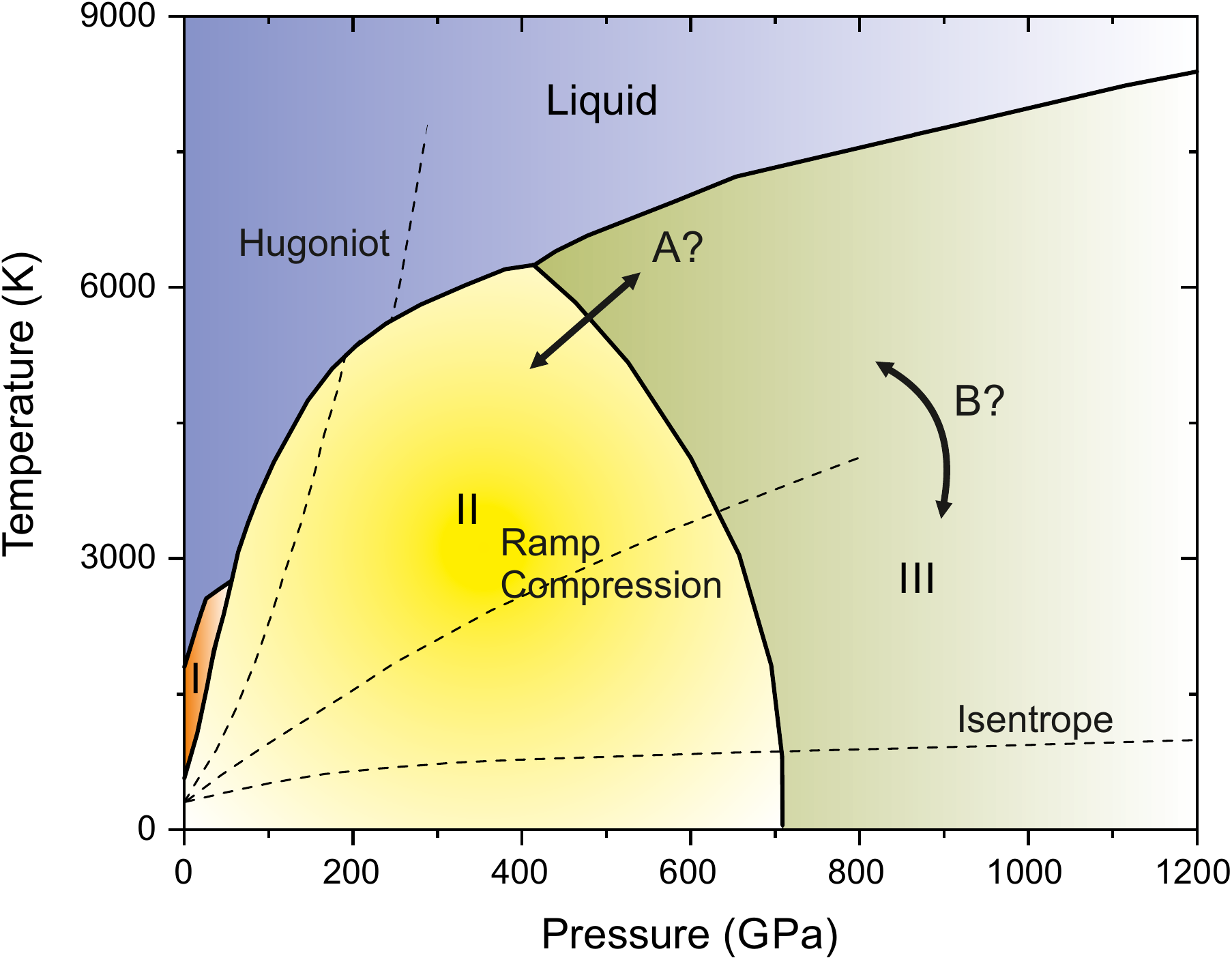}
 
    \caption {The phase diagram of a fictitious  material, showing stability fields of three different solid phases (I, II and III) and the liquid phase above the melt line. Also shown are the two P-T paths of the isentrope, accessible with ``ideal" ramp compression, and the Hugoniot, accessible with shock compression. The third path is an intermediate, and is more indicative of a ''real" ramp compression path. This exact trajectory of this pathway depends on properties such as material strength, and its endpoint could be anywhere along line B. Note that along the Hugoniot the sample melts at $\sim$200 GPa and phase III is therefore inaccessible to shock compression. Also note that phase-I is inaccessible to all dynamic compression techniques, and the II-III phase boundary at `A', can only be accessed using complex compression histories like a combination shock-ramp requiring precise control of the compression history.
    }
    \label{F:Schematic}
\end{figure}

Whilst it is proven that ramp compression on nanosecond timescales can keep material below the melt curve into the multi-TPa regime, the temperatures reached during such experiments have yet to be measured accurately.  As we shall discuss in section \ref{SS:Shock-Ramp-JSW}, a knowledge of the degree of heating above the isentrope, and a good understanding of the mechanisms behind, and magnitude of, the plastic work that causes it, remain outstanding issues within the field.  Indeed, our knowledge, or lack thereof, of the temperature rise of matter at high strain rates will be a {\it leitmotif} throughout this paper, and we illustrate this in Fig. \ref{F:Schematic}, where in schematic form we show a temperature-pressure plot of a fictitious, yet representative, material, that would shock melt around 200 GPa, but, via ramp compression, could be kept solid well beyond a TPa, traversing parts of the phase diagram where several polymorphic phase transitions may occur.  Note that on this diagram we have indicated that the path taken by ramp compression is unknown, owing to this large uncertainty in the temperatures reached, and what diffraction experiments can tell us to date is largely simply that the material remains in the solid state.  A further important uncertainty, also alluded to within this figure, is the position of the phase boundaries that one might ascribe to polymorphic phase transitions induced by dynamic compression.  As we shall discuss below, there is ample evidence that in many cases the pressures (or more accurately stresses) at which many transitions take place under dynamical compression are substantially different from those that are found to pertain on the far longer timescales of static experiments.  Whilst dynamic compression experiments can reach pressures unattainable by static techniques, and as such appear to enable us to reach, albeit transiently, conditions comparable to those within planetary cores, it is our view that the current lack of knowledge of both the temperatures actually reached, and the potential differences between dynamic and static phase diagrams, ought to engender a degree of caution when claiming direct relevance to planetary physics.

\begin{figure}
        \includegraphics[width=0.95\columnwidth]{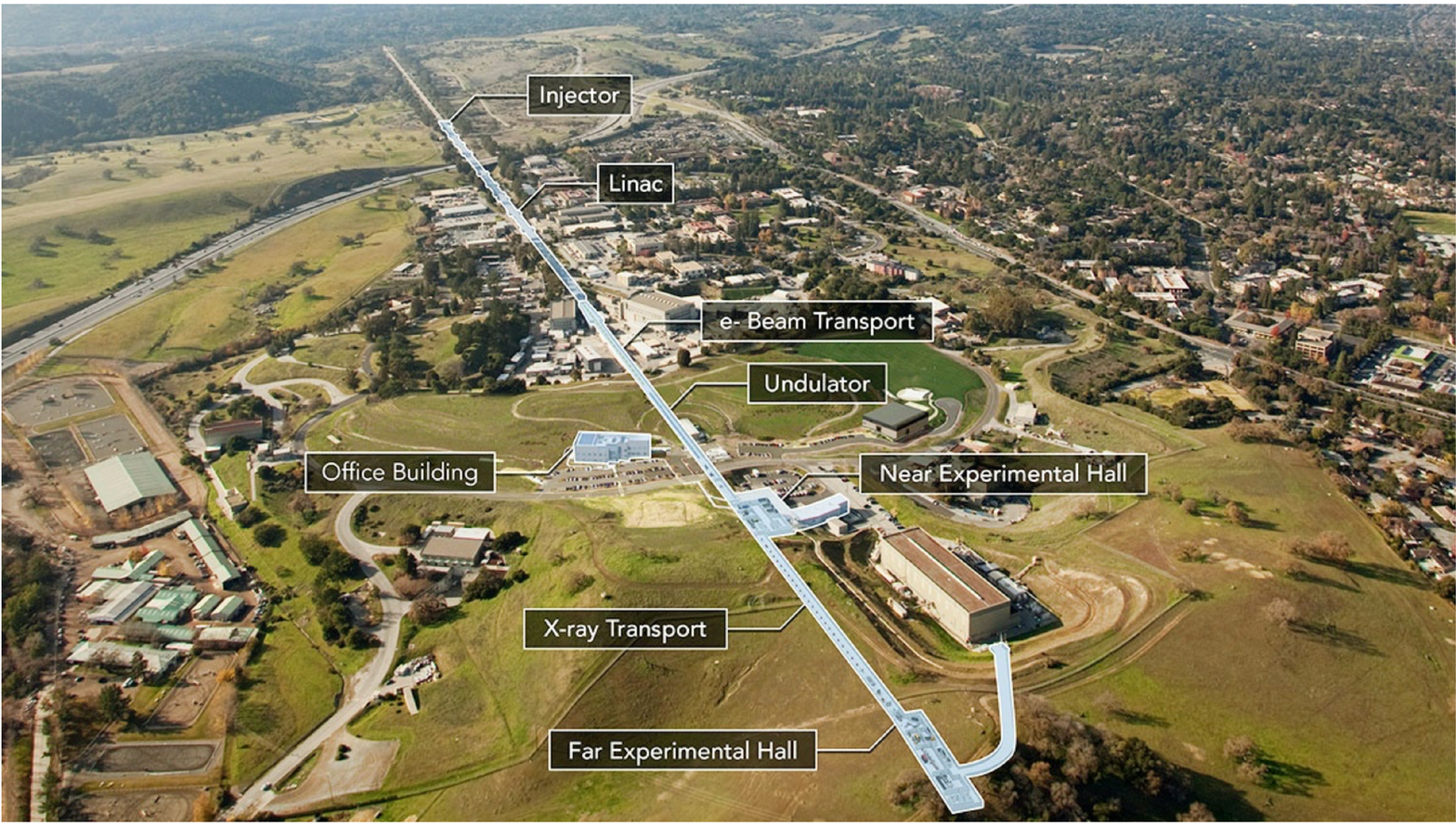}
    \caption {An aerial view of the Linac Coherent Light Source (LCLS) at SLAC National Accelerator Laboratory showing the location of the various parts of the machine such as the injector, linac, undulator, x-ray transport and the experimental hall.}
    \label{F:LCLS}
\end{figure}

In parallel with advances in methods to apply transient pressure in a controlled way, the field of time-resolved x-ray science, which itself has a long history associated with the development of several different types of sources, has been transformed by the successful demonstration of the world's first hard x-ray free electron laser (FEL).  In April 2009, in the course of a single morning, as a row of undulators 100 m long were slid into place with micron precision at the end of the linear accelerator at SLAC, x-ray pulses of a peak spectral brightness over a billion times greater than emitted by any previous man-made source were produced.  The Linac Coherent Light Source (LCLS), an aerial view of which is shown in Fig. \ref{F:LCLS}, produces x-ray pulses of sub 100-fsec duration, containing of order 1 mJ per pulse, at a rate of 120 Hz~\cite{Emma2010}.  These pulses are almost completely spatially coherent, and when exiting the undulators have divergences in the $\mu$rad regime.  Their bandwidth, if the pulse develops from noise within the electron bunches (so-called self amplified spontaneous emission -  SASE), is of order 0.5 \% (although we discuss techniques to improve this below).  An individual x-ray pulse contains of order 10$^{13}$ photons, more than sufficient to produce a clear x-ray diffraction pattern in a single shot, thus allowing detailed structural information to be obtained on a timescale shorter than even the fastest phonon period.  

Since LCLS first `lased', several other facilities around the world have come  on line, or are under development~\cite{Ishikawa2012,Allaria2012,Park2016,Milne2017,Decking2020,Zhao2017,Seddon2017}.  Pulsed optical lasers, capable of dynamically compressing samples as described above, have been placed alongside many of these FEL facilities,  and already several key diffraction experiments have provided fundamental insight into how matter behaves under dynamic compression.   Thus far these lasers have been of a relatively modest power compared with those utilising laser-plasma x-ray sources (see section \ref{SS:Some-History-JSW}), affording compression to only a fraction of a TPa, where at the lower end of the pressure range the differences between shock and ramp compression are less pronounced: indeed, to date the vast majority of dynamical compression experiments at FELs have utilised shock compression.  We discuss below developments of these facilities to attain the multi-TPa regime via ramp compression at repetition rates such that many thousands of individual compression-diffraction experiments can be performed in a reasonable time-frame.  The scientific opportunities that such facilities will offer motivate much of the thinking behind our discussions of future directions.

Within this paper we provide our perspective on how this area of research has developed over the past decade in the context of both the development of ramp-compression techniques, and in the rapid development of FELs,  what we perceive to be some of the major science drivers, and how we view prospects for future developments.  Given the huge breadth of the field of high energy density science, we will restrict our comments to studies of matter in the solid (and occasionally molten) state.  Whilst hot, dense plasmas can also be created by both high power optical or x-ray illumination~\cite{Vinko2012}, their study requires a description of matter from a different physical perspective, which is beyond the scope of this work.

The paper is laid out in the following manner.  Following this brief introduction,  in section \ref{SS:Some-History-JSW} we place the field of femtosecond diffraction using FELs as a tool for the study of matter under high pressure into a broader  historical perspective of dynamical compression of matter combined with diffraction in general, taking care to highlight those experiments where shock compression was employed, and those more recent ones where the material was kept closer to an isentrope.  In section \ref{S:Techniques} we provide an outline of the basic physics of how dynamic pressure is applied via laser ablation, measured via interferometric techniques, and briefly discuss the types of targets studied.  Following this in section  \ref{S:Work_to_date} we give a brief overview of work in this field to date, before in section \ref{S:Future} outlining what we perceive to be at least some of the major outstanding problems that need to be addressed in the future.  Much of our thinking is based, as noted above, on the need to understand the heating mechanisms that pertain, and temperatures that can be reached, via these dynamic compression techniques.  Given that the rate at which pressure is applied in these experiments plays a crucial role in the final state of the material, section \ref{SS:Shock-Ramp-JSW} concentrates on the basic pertinent physics, as presently understood, between dynamically compressing samples so rapidly that a shock wave is produced, and compressing them with a temporal pressure profile that, whilst still rapid (nanosecond timescales), is sufficiently slow that the compression path is cooler, lies closer to the isentrope, and potentially keeps the sample solid even at ultra-high pressures.  After presenting our perspectives, we note in section \ref{S:Tech_Dev} that future directions will rely to a large degree on further technical developments in FEL and driver technology.  We summarise and conclude in section \ref{S:Summary}.

\subsection{Historical Perspective}
\label{SS:Some-History-JSW}

The beginning of the study of solid state matter under high pressure in a controlled way under static conditions is usually ascribed to the pioneering and Nobel prize winning work of Percy Bridgman \cite{Bridgman1964}, whose invention of a novel type of seal, which ensured that the pressure in the gasket always exceeded that in the sample under pressure (thus leading to self-sealing), immediately enabled him to subject samples to pressures up to several GPa, and ultimately 40 GPa \cite{Bridgman1931,McMillan2005}.  So-called static high pressure research has advanced significantly since those early studies, with almost all work now making use of Diamond Anvil Cells \cite{Forman1972,Jayaraman1983,Piermarini2001}, conventional forms of which can reach pressures of order 400 GPa, whilst novel anvil shapes and two-stage forms can now achieve pressures of 500 - 600 GPa \cite{Dewaele2018,Jenei2018} and close to 1 TPa \cite{Dubrovinsky2012,Takeshi2018,Dubrovinskaia2016}, respectively.  On the other hand, the rise of dynamic compression of solids can in a large part be traced back to the work performed during the second world war at Los Alamos National Laboratory, which provided scientists with an unprecedented capability to shock-load materials explosively – but precisely – to several 10s of GPa, and accurately to diagnose their pressure and density. The decade following the war saw the publication of a number of seminal papers \cite{Minshall1955,Bancroft1956,Walsh1955} culminating in a review by Rice, McQueen, and Walsh \cite{Rice1958} in 1958 introducing the scientific community at large to this new branch of physics.

\begin{figure}
        \includegraphics[angle=270,width=0.95\columnwidth]{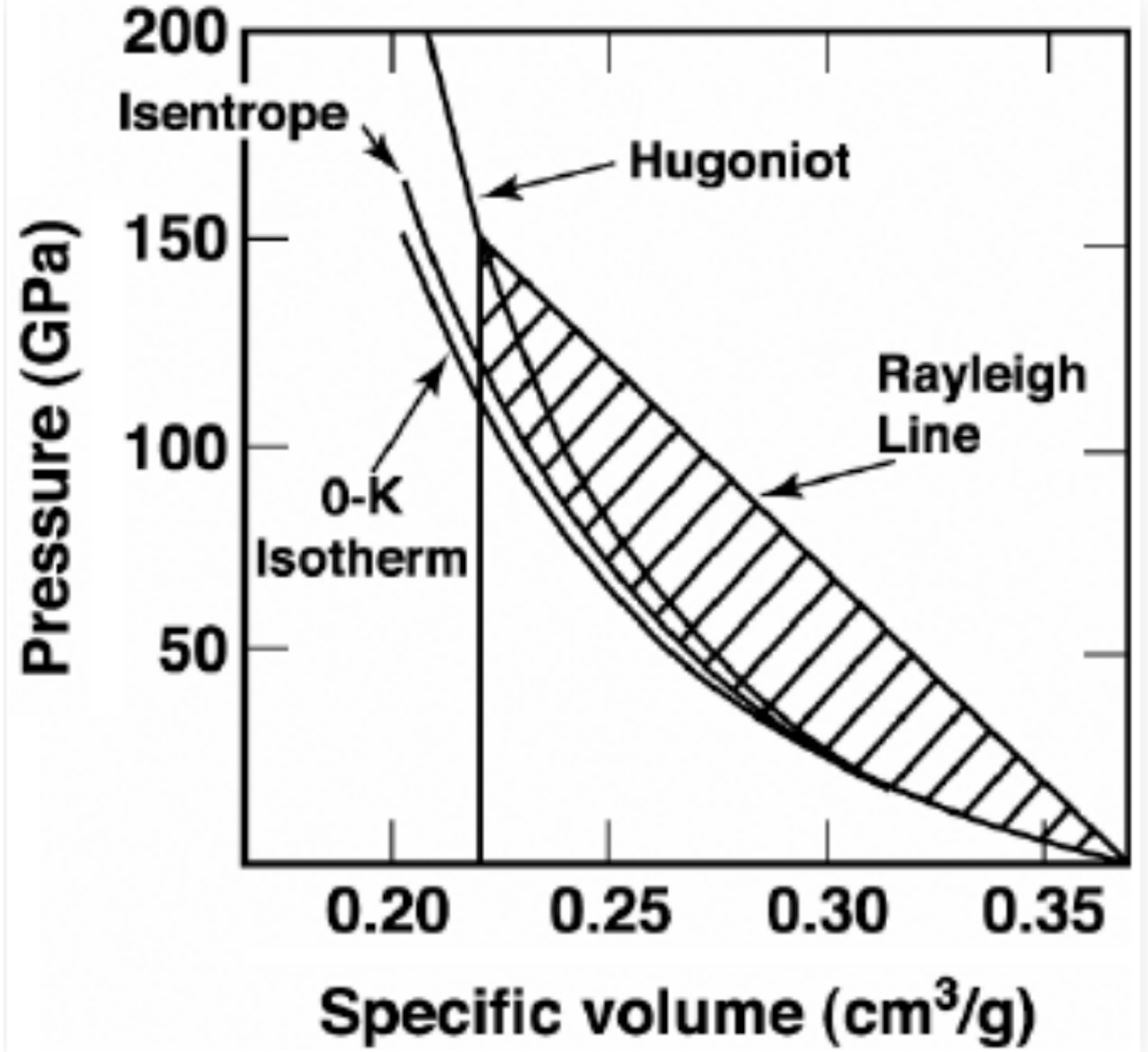}
    \caption {The Hugoniot, isentrope and 0 K isotherm of Al from the initial specific volume of crystal to specific volumes near 150 GPa pressure. Reprinted with permission from Nellis (1997)~\cite{Nellis2006}, Copyright 1997, VCH Publishers.}
    \label{F:Compression}
\end{figure}

Importantly the pressures reached under such dynamic (shock) compression could be gauged accurately because the material conditions in front of, and directly behind, a shock front must be related by the simple conservation equations derived in the 19$^{\rm th}$ century 
by Rankine and Hugoniot \cite{Rankine1870, Hugoniot1887}. These equations, that express conservation of mass, momentum, and energy link conditions upstream of the shock (with subscript 0), to those behind the shock front:
\begin{eqnarray}
    \rho_ 0 U_{S} &=& \rho (U_S - U_P ) \quad ,\\
    \sigma_{zz} - \sigma_0 &=& \rho_0 U_S U_P \quad , \\
    e - e_0 &=& \frac{1}{2} (\sigma_{zz} + \sigma_0)  \frac{\rho \rho_0}{\rho - \rho_0} \quad ,
\end{eqnarray}
where $\rho$ is the density, $U_S$ and $U_P$ the shock and particle velocity respectively, $\sigma_{zz}$ the longitudinal stress, and $e$ the energy per unit mass.  As one knows the initial density, stress, and energy, these three equations have five unknowns ($\rho$, $e$, $\sigma_{zz}$, $U_S$ and $U_P$), thus requiring a measurement of two parameters for closure.  Typically these were chosen for experimental reasons to be $U_S$ and $U_P$, obtained by the shock transit time, and measurements of surface or interface velocities upon shock breakout.  We note also that if it is assumed, as in the original work at Los Alamos~\cite{Rice1958}, that the relationship between $U_{S}$ and $U_{P}$ can be approximated as linear, i.e that $U_S = C_0 +s U_P$, where $C_0$ is roughly equal to the zero-pressure bulk sound speed, then the pressure on the Hugoniot is given by
\begin{equation}
    P_H ( \rho )= \rho_0 C_0^2 \kappa /(1 -s \kappa)^2 \quad,
\end{equation}
where $\kappa = 1-\rho_0 / \rho$.  Whilst such a linear relationship between shock and particle velocity is often posited, and indeed widely used to characterise the shock response of materials, care should be taken in its employment, as it is by no means a universal law ~\cite{Kerley2013}.  The locus of states that can be reached via shock compression is known as the Hugoniot.  For a given compression, owing to shock heating, the pressures reached lie above both the isotherm and the isentrope, as shown schematically in Fig. \ref{F:Compression}, where this is illustated in the case of aluminum.

The original static experiments of Bridgman studied the compressibility of matter by monitoring the motion of the compressing piston, as well as measuring electrical and thermal conductivities \cite{Bridgman1964}, whilst early measurements of shock-compressed materials relied almost exclusively on a study of the wave profiles within them, as evinced by measurements of the movement of the rear surface of a sample as the waves released from it.  As such, no direct structural information was ever obtained {\it in situ} from either technique, which could lead to some dispute as to how the sample had actually deformed or changed phase. For example, perhaps one of the most renowned early results in shock physics was the observation by Bancroft and co-workers in 1956 of a two-wave structure in iron when shocked above 13 GPa, which they attributed to a polymorphic phase transition \cite{Bancroft1956}.  In the absence of any direct measurement of structure, Bridgman disputed the claim, noting that he could find no evidence of any change in electrical resistivity in his studies of the same element up to 17.5 GPa \cite{Bridgman1956}  In the paper reporting those studies, Bridgman stated that ``{\it ..it seems to be a widely held opinion that transitions involving lattice type would be unlikely to occur in times as short as a few microseconds.}''  Fortunately for the field of dynamic compression, and indeed for the authors of this current paper, Bridgman would,  in this regard at least, be proven to be incorrect, although we shall return at a later juncture to what differences may remain between static and dynamic experiments.

The next few years heralded several key developments. Improvements in electrical measurements in pressure cells by Balchan and co-workers eventually led to confirmation that there indeed was a step change in the resistivity of iron \cite{Balchan1961} at precisely the pressures seen in the dynamic work of Bancroft, and the field of shock physics was vindicated. This confirmation of a change in electrical properties of iron at 13 GPa took place around the same time as the development of the DAC in 1958 \cite{Piermarini2001}, and its subsequent adaptation to allow it to be used in conjunction with x-ray diffraction in 1960, paving the way for a true study of structure under high pressure, and building on techniques developed at lower pressures~\cite{Jacobs1938a}.  X-ray diffraction could also be used in amorphous boron compression cells, and indeed these were used by Jamieson and Lawson to indicate that the transition seen by Balchan was indeed the bcc-hcp ($\alpha - \epsilon$) transition \cite{Jamieson1962} - the evidence being the appearance of a single diffraction line, the position of which was inconsistent with a bcc or fcc lattice (given the known volume), but could be reconciled as being the (10$\bar{1}$1) line of the hcp form.

The capability to study dynamically (at that time shock) compressed matter via x-ray diffraction was only developed close to a decade after its use by the static high pressure community.  Quentin Johnson at LLNL developed an x-ray diode that could produce flashes of x-rays tens of nanoseconds in duration, sufficiently bright for single-shot diffraction \cite{Johnson1967}, and he later obtained diffraction from explosively shocked LiF three years later \cite{Johnson1970}, proving that crystalline order was preserved behind the shock front, and subsequently used the same techniques to provide the first {\it in-situ} x-ray diffraction evidence for a shock-induced phase transition, observing new diffraction lines when pyrolytic boron nitride, which has a structure similar to that of graphite, was shocked to 2.45 GPa, with the new lines being consistent with the wurtzite phase~\cite{Johnson1972B}.  This was a landmark achievement in dynamic compression science, and the consensus had clearly moved on from the original views of Bridgman: as the authors of the work themselves stated {\it ``.....shock-wave Hugoniot measurements have established as a virtual certainty that shock pressures can induce crystal structure transformations within 10$^{-7}$sec. Until now, however, it has not been possible to inspect the actual crystal structure existing under dynamic conditions."}.

The next significant developments in the field came with the realisation,  very soon after the invention of the laser, that they could be used to generate shock waves in materials \cite{askaryan1962, Trainor1979}.  Subsequent development of the use of high-powered nanosecond optical lasers led to the capability both to shock crystals via laser ablation, but also, via a separate arm of the laser, to provide a synchronous quasi-monochromatic x-ray source.  This laser-based diffraction technique was first demonstrated in the mid-80s at pressures of just a few tens of GPa, when Wark and co-workers used a nanosecond laser to shock single crystals of silicon~\cite{Wark1987,Wark1989}, and has subsequently been developed and adapted for diffraction from  polycrystalline materials~\cite{Hawreliak2007} far into the TPa regime~\cite{Rygg2012}. Indeed, the application of transient pressure via laser ablation - where a thin surface layer of the target is vaporised and turned into a plasma by the high-intensity laser, introducing a pressure wave via Newton's third law, is now a standard technique for dynamic compression combined with pulsed x-ray diffraction, and is the basis for the majority of the science discussed within this paper. We thus provide a short, basic description of the underlying physical process in section \ref{S:Ablation}.

As noted, the x-ray source used alongside the early laser-ablation work was itself produced by a separate arm of the same laser, the energy contained within it focussed to a very small spot, of a diameter of order tens to hundreds of microns, onto a foil mounted some mm to cm from the shocked sample.  Typical laser irradiances on this foil are of order a few times 10$^{14}$ to 10$^{15}$ Wcm$^{-2}$. Irradiation of the foils at such intensities creates a laser-produced plasma that emits x-ray radiation isotropically.  The temperatures within these plasmas are sufficiently high that for mid-Z elements there is copious emission of the resonance line radiation of helium-like ions of the element of concern, which can be used to diffract from the compressed material. Whilst the highest pressures at which x-ray diffraction has been successfully employed to date have utilised these laser-plasma x-ray sources, these experiments suffer from several disadvantages.  Most notably, the x-ray source itself is far from spectrally pure, and is employed in a highly inefficient manner.  The resonance lines of the helium-like ions that are used as the x-ray source are accompanied by a spectrally broad pedestal of bremsstrahlung emission.  Furthermore, the x-rays are emitted isotropically, resulting in only a small fraction of them being used for diffraction after appropriate collimation (in the case of powder diffraction), or associated divergence (in the case of diffraction from single crystals).  Furthermore, the natural bandwidth of the resonance line is of order 0.5\%, as it actually comprises two main  lines - the main resonance line itself (associated with the 1s2p ($^1$P) $\rightarrow$ 1s$^2$ ($^1$S) transition), as well as a strong intercombination line (associated with the 1s2p ($^3$P) $\rightarrow$ 1s$^2$ ($^1$S) transition).  Stark broadening and spectrally close transitions from doubly-excited lithium-like ions (dielectronic satellites) also contribute to the linewidth, broadening it to 1.0-1.5\%\cite{coppari2019}.  In addition, as well as thermal bremsstrahlung radiation, the optical laser intensities required to produce the mid-Z helium-like ions are sufficiently great that often copious quantities of `hot' electrons are produced, with energies significantly greater than the thermal average.  The bremsstrahlung radiation from these hot electrons can produce an additional source of x-rays that is hard to filter out.  As a result, considerable effort needs to be made to remove background effects from any recorded diffraction pattern \cite{Rygg2020}.

Despite the difficulties associated with laser-plasma x-ray sources, the fact that they can be produced by some of the largest and most powerful lasers on the planet, such as the National Ignition Facility (NIF)~\cite{Hurricane2017,Rygg2020} at Lawrence Livermore National Laboratory, and the Omega Laser Facility at the University of Rochester~\cite{Boehly1997}, has resulted in them being used to provide diffraction information on solid-state matter at extremely high pressure, well into the TPa regime, and far beyond what is achievable with DACs.  As noted previously, keeping a material solid whilst compressing to such high pressures cannot be achieved via shock compression, but only if the material is loaded more slowly.  Much of the capability  that has led to this being possible is a result of the development of exquisite temporal pulse shaping techniques of the output of these large laser systems~\cite{Wisoff2004, Feigenbaum2013}, with such sculpting itself being necessary to control the amount of entropy produced when compressing laser-fusion targets~\cite{Robey2013}, that being the primary goal of the NIF and Omega systems.

Such developments have led to a host of studies, both in the shock and ramp compression regime, primarily on metals and materials of planetary interest \cite{Coppari2013,Wang2015,Lazicki2015,Wang2016,Denoeud2016,Polsin2018,Wicks2018}, with the highest pressure at which diffraction signals have been obtained to date being 2 TPa in a study of diamond~\cite{Lazicki2021}, showing that it does not change phase even when ramp compressed to such conditions.
Furthermore, to give some degree of closure to the long-standing debate on the response of iron to shock compression, given the importance of this element in the history of dynamic compression, it is worth noting that it was experiments employing laser plasma x-ray sources that led to the first conclusive measurement of its structure under shock conditions above 13 GPa, confirming that it does indeed transform from the bcc $\alpha$ to the hcp $\epsilon$ phase ~\cite{Kalantar2005,Hawreliak2006,Hawreliak2011}.

As noted above, laser-plasma x-ray sources have distinct disadvantages in terms of bandwidth and efficiency (being emitted isotropically), but perhaps most notable is the considerable bremsstrahlung background. In contrast, the spectrally pure, highly collimated and high-energy x-ray beams available at 3rd generation synchrotrons have many advantages over plasma-generated x-ray sources.  As a result, recent years have seen the development of a number of laser-compression experiments at synchrotrons, such as Spring-8, the APS and the ESRF. Such experiments use a single bunch of electrons in the storage ring to generate a $\sim$100 ps pulse of x-rays, which is well matched to the ns-lifetimes of the compressed states. Single pulse diffraction data from laser compressed samples have been obtained at the ESRF \cite{dAlmeida2002,Liss2009}, Photon Factory\cite{Ichiyanagi2007,Ichiyanagi2016} and Photon Factory Advanced Ring \cite{Hu2013}. At the Advanced Photon Source, a dedicated laser-compression hutch has  recently  been  completed  at  the  Dynamic  Compression Sector (DCS) (which also includes high-explosive and gas-gun compression hutches) specifically for diffraction and imaging experiments \cite{Wang2019}.  To date the laser-compression hutch has been used to study  a number of compressed metals to pressures as high as 383 GPa. \cite{Briggs2019b,Sharma2019,Sharma2020a,Sharma2020b,Beason2020}.

Given the success of laser-plasma and synchrotron based sources to study dynamically compressed matter, when hard x-ray FELs were starting to be planned close to two decades ago, it was evident that the study of dynamically compressed matter would be a prime area of scientific interest: indeed the study of shock compressed materials was mentioned in some of the earliest design reports for LCLS~\cite{Cornacchia1998}.  X-ray FELs have characteristics that provide them with certain distinct advantages over synchrotron and laser-plasma sources.  Typical pulse lengths are below 100 fsec -- this is shorter than even the fastest phonon period, providing a true snap-shot of the state of the lattice at a given instant. The output from such systems is also close to being fully spatially coherent, providing exquisite control over the size of the x-ray spot that can be used to interrogate the sample (although explicit use of the coherence has yet to be exploited).  Furthermore, whilst the inherent bandwidth of the SASE source is close to 0.5\%, recent developments in seeding technology~\cite{Amann2012,Inoue2019} have reduced this to 
a few times $10^{-5}$ without a significant reduction in the total energy in the beam which is typically of order 1 mJ.  As a result FELs have peak spectral brightnesses (photons s$^{-1}$ mrad$^{-2}$ mm$^{-2}$ Hz$^{-1})$) about a billion times greater than any synchrotron.

Thus during the planning stages of FELs such as LCLS it was evident that siting a high-power optical laser alongside the FEL facility would provide the potential to provide novel insights into dynamically compressed matter.  This realisation ultimately led to the construction of the   Materials in Extreme Conditions (MEC) end-station at LCLS.  This beamline started operations with an optical laser that could produce several 10s of Joules of 0.53 $\mu$m light in a pulse length of order 10 nsec.  Each arm of this two-beam system can operate at a repetition rate of one shot every 6 minutes. Many of the results from the intervening period that we present in section \ref{S:Work_to_date} have been obtained on this facility, although other facilities, such as one at SACLA in Japan, have also yielded interesting results, and the High Energy Density Science end-station at the European XFEL has recently started operation, with the optical lasers along side it, which will be used for dynamic compression, currently being commissioned.

At the present time dynamic compression studies have reached an interesting juncture.  The highest pressures at which solids can be diagnosed via diffraction, presently around 2 TPa~\cite{Lazicki2021}, can at the moment only be reached with  the extremely high power nanosecond lasers such as NIF and Omega, which have energies per pulse in the 10s of kJ to MJ regime, and have sophisticated pulse-shaping capabilities that allow greater control over the rate at which pressure is applied.  Although these lasers can achieve such pressures, they are large, stand-alone facilities, not sited alongside FELs, and diagnosis of crystalline structure is reliant on the far-from-optimum laser-plasma sources discussed above. Furthermore, these facilities typically have very low shot rates - of order once or twice per hour for Omega, and perhaps a similar number shots per day in the case of NIF: exploring the phase diagram of even a single compound at such facilities is a time-consuming and hugely-expensive endeavor.  In contrast, the nanosecond optical lasers that have hitherto been sited along side FELs, for example at the MEC end-station at LCLS, have contained only a few tens of Joules in a pulse length of several nanoseconds.  Given that this optical laser energy is many orders of magnitude lower than those available at NIF and Omega, the pressures that can be produced in the diffraction experiments are somewhat lower, although it should be noted that the spot into which the optical laser is focused at the FEL facilities is typically of order 300 $\mu$m in radius, whereas at NIF and Omega this dimension is often several times greater: this offsets somewhat the differences in beam energy, as it is irradiance (to the 2/3 power - see section \ref{S:Ablation}) that determines pressure.  The ability to use smaller focal spots at FELs is largely related to the fact that the x-rays themselves can be positioned and focused within the optical focal spot, in contrast to the situation with the divergent laser-plasma generated x-rays.  Furthermore, to date, the optical lasers used at FELs have not employed the impressive pulse-shaping characteristics of the larger laser systems.  Clearly developments in the technical capabilities of the optical lasers sited alongside the FEL facilities will be an important driver for future scientific directions - a point to which we shall return in section \ref{S:Tech_Dev}.  

Whilst dynamic compression experiments at FELs have yet to attain the multi-TPa pressures in solids achieved at the other stand-alone facilities, the higher shot rate and  purity of the diffracting x-ray beam have led to a series of impressive results that have improved our understanding of several areas of this important field, and which in turn provide further signposts to how it might develop in the future.  In particular, there are promising indications that FEL-based experiments are starting to provide crucial information on the deformation pathways taken by materials under dynamic compression, and hence an understanding of the plasticity mechanisms that determine the temperature of the compressed sample.  Furthermore, several ideas have emerged that show some promise for making {\it in situ} temperature measurements.  We also discuss these developments in section \ref{S:Future}.  Before giving a short overview of the FEL experiments that have hitherto been performed, we first provide a brief description of the basic physics and current measurement techniques that underpin dynamic compression.

\section{Techniques for achieving dynamic compression, and pressure measurements}
\label{S:Techniques}

\subsection{Laser Ablation}
\label{S:Ablation}
Within this overview we restrict ourselves to the dynamic application of pressure via laser ablation, and we concentrate on techniques to compress matter on the timescale of several nanoseconds.  Note that typical sound velocities in solids are in the km s$^{-1}$ regime -  equivalent to $\mu$m ns$^{-1}$.  The absorption length of mid-$Z$ elements to x-rays, which will limit the useful thickness of a target for diffraction in transmission geometry, is of order a few-to-tens of microns, and thus we note how the nanosecond timescale for dynamic compression arises as a natural choice.  

High power lasers with nanosecond pulse-lengths have been developed over many years, and much of the motivation for the development of lasers with the largest energies (with total energies in the 10s of kJ to MJ regime), such as the NIF and Omega referred to earlier, has been the quest for the demonstration of inertial confinement fusion (ICF)~\cite{Lindl1995,Haines1997}. As noted above, to date the lasers placed alongside FELs deliver smaller energies in their nanosecond pulses, typically in the regime of several tens of Joules.  These solid-state lasers are usually based on Nd:Glass or Nd:YAG technology, and have traditionally been flash-lamp pumped.  For dynamic compression experiments, the laser light is focused in vacuum onto the target of interest, with a spot size that is typically several times the thickness of the target (to minimize the effects of rarefaction of the compression wave from the edges of the irradiated region).  We are thus dealing with intensities such that we have several tens to hundreds of Joules in a few nanoseconds in spot sizes of order a 100 $\mu$m - that is to say irradiances of interest are of order 10$^{12}$ - 10$^{14}$ W cm$^{-2}$.

The detailed physics of how such intense laser light is eventually converted to a compression wave is a highly complex process, however simple scaling arguments can be derived from a few basic assumptions. At these intensities a plasma is formed on the surface of the irradiated target, which expands into the vacuum, decreasing in density with increasing distance from the as-yet unablated material.  The light that propagates through this expanding plasma is absorbed via inverse bremsstrahlung, whereby the ordered oscillatory motion of the electrons in the electric field of the laser is randomised by multiple small angle collisions with the ions.  However, the real part of the refractive index, $\mu$ of this plasma is dominated by the electron response, such that for light of frequency $\omega$,

\begin{equation}
\mu = \sqrt{   1- \frac{\omega_{\rm p}^2}{\omega^2}    } \quad ,
\end{equation}
where $\omega_{\rm p}  = \sqrt{ne^2/ ( \varepsilon_0 m_{\rm e} )}$ is the plasma frequency, determined by the local electron density.  Thus light can only propagate up to a certain electron number density, $n_c$, which we call the critical density, given by $n_c = \varepsilon_0  m_{\rm e} \omega^2/ e^2$, at which point it will be reflected.  Thus the laser light deposits energy up to the critical surface, producing a hot plasma up to that point.  Energy must be thermally transported down from the critical surface to the target to keep ablating material. We assume a steady-state situation ensues, by assuming that the flow velocity of the material at the critical surface is equal to the sound speed - that is Mach 1.  That is to say $v_c \approx \sqrt{(P/\rho_c)}$.  All the pieces are now in place to construct our simple model.  We equate a fraction, $\alpha$, of the laser intensity $I$ to flow down the temperature gradient to sustain the ablative flow, and give rise to the outflow rate of energy at the critical surface.  That is to say,
\begin{equation}
\alpha I \approx \rho_c v_c^3 = \rho_c \left( \frac{P}{\rho_c}  \right)^{3/2} \quad , \label{E:I1}
\end{equation} 
where
\begin{equation}
\rho_c \approx 2 m_{\rm p}n_c \approx \frac{2 m_{\rm p}m_{\rm e}\varepsilon_0 \omega^2}{e^2}  \quad  ,  \label{E:I2}
\end{equation}
and where $v_c$, $\rho_c$ and $n_c$ are the flow velocity, mass density, and electron density at the critical surface respectively, and $m_{\rm p}$ and $m_{\rm e}$ are the masses of a nucleon and electron respectively (and the factor of 2 on the left hand side in eqn \ref{E:I2} is roughly approximating the number of nucleons in an ion as twice the atomic number).
From equations (\ref{E:I1}) and (\ref{E:I2}) we find
\begin{equation}
P =  \left( \frac{8 \pi^2 m_{\rm p}m_{\rm e}\varepsilon_0 c^2}{ e^2} \right)^{1/3} (\alpha I / \lambda)^{2/3}   \quad . \label{E:Pablate}
\end{equation}
For example, for light of wavelength 1 $\mu$m, an irradiance of 10$^{13}$ Wcm$^{-2}$, and assuming $\alpha = 0.5$, the above equation  gives a pressure estimate of 200 GPa, which is surprisingly close to the experimental result, given the crudeness of the model.  We also see from this rather hand-waving approach the motivation for, whenever possible, efficient conversion of the laser light (which for the Nd systems has a wavelength of order 1 $\mu$m) to its second or third harmonic.  The increase in pressure that is obtained via the $\lambda^{-2/3}$ scaling can usually more than offset the conversion efficiency loses.  We also note that we predict a $I^{2/3}$ scaling of the pressure with intensity, and this lower than linear scaling will impact on the design of facilities aimed at providing access to the TPa regime at FELs.

\subsection{Pressure Measurements using VISAR}
\label{S:Visar}
In dynamic compression experiments information about the density can be obtained directly by measurement of the lattice parameters via x-ray diffraction and, as will be discussed, x-ray spectroscopic methods are being developed to constrain temperature.  For measuring the pressure  conditions within the target (or more accurately, longitudinal stress) we resort to measurements of the velocity of a surface - normally the rear surface of the target.  We previously commented in section \ref{SS:Some-History-JSW} on the Hugoniot equations, noting that closure can be achieved by the measurement of two quantities,  which are usually the shock and particle velocities.  Via such measurements over many years there now exists a series of equation of state (EOS) tables, such as the the SESAME database\cite{SESAME},  linking these quantities for a particular material.

The most straight-forward measurement in the situation of a steady-shock is a measurement of the rear surface of the target, because when the shock reaches the rear surface, as the rarefaction wave travels back into the target, it can be shown that the velocity of the free surface is approximately equal to twice the particle velocity.  More accurately the free surface velocity, $U_{fs}$, is given by
\begin{equation}
    U_{fs}= U_{P} + \int_{V(P_{H)}}^{V(P_0)} \left ( \frac{d P_S}{dV} \right )^{1/2} dV \quad ,
\end{equation}
where $P_H$ is the shock pressure, $P_S$ the pressure along the release path.

\begin{figure} 
\includegraphics[width=9cm, angle=0]{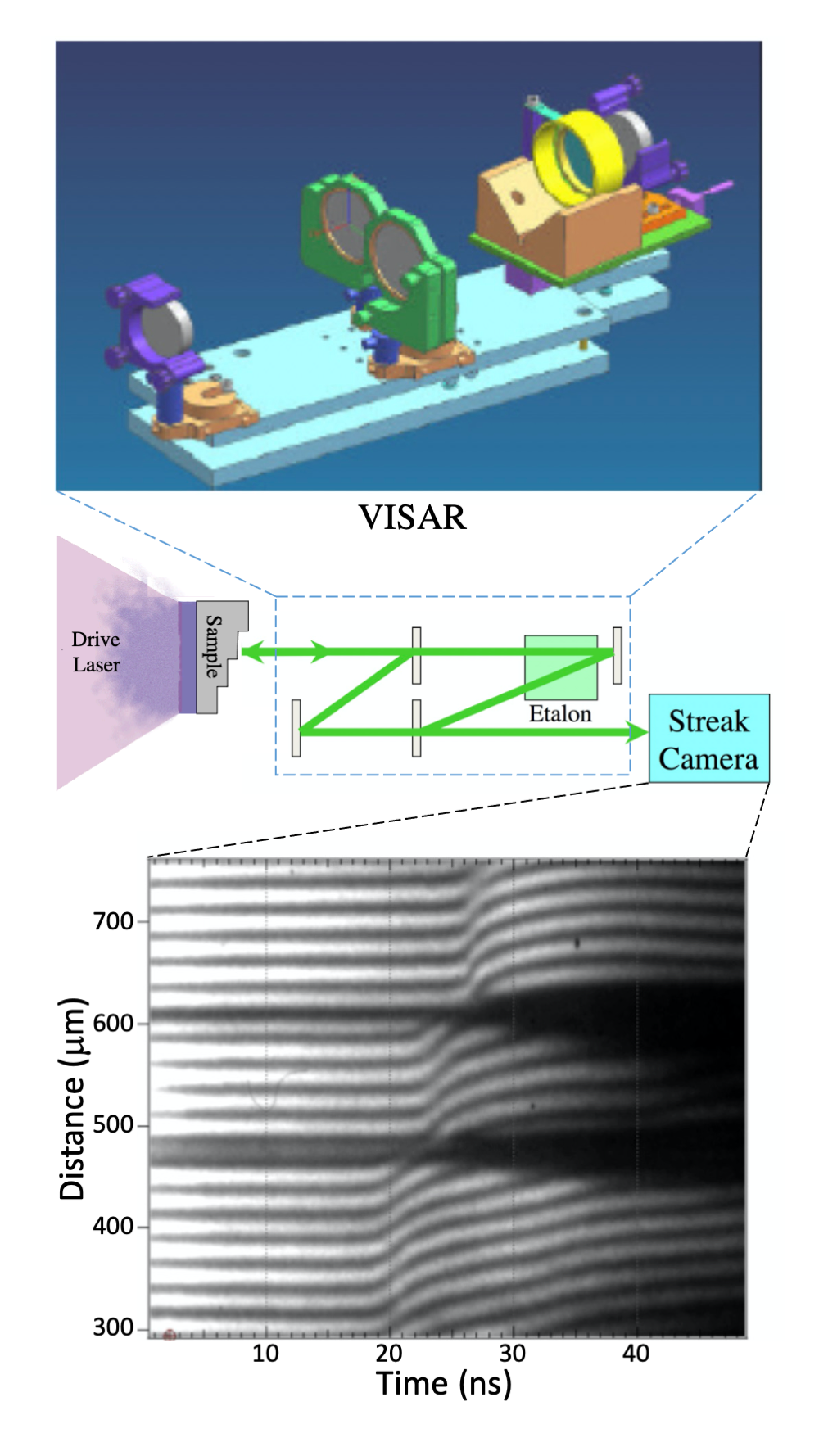}
\caption{\small  The line VISAR diagnostic takes reflected laser light and runs it through a Mach-Zehnder interferometer.  The spatial imaging properties of the line VISAR are essential to laser-driven compression experiments where shock planarity needs to be evaluated or where the velocity history varies with the sample position.  The figure shows a typical experimental set up, and raw streak camera data for a compression wave breaking out from a target comprising steps of three different thicknesses. }
\label{F:VISAR}
\end{figure}

Such interface velocity measures are generally performed by use of
the well-established  Velocity Interferometer System for Any Reflector (VISAR)~\cite{Barker1972,Celliers2004,Dolan2006} 
Various implementations of the technique exist: the line VISAR allows velocity profiles of a sample interface in one dimension (1-D) to be measured so that drive planarity or more spatially-complex samples (e.g. those with steps in their thickness) can be characterized.\cite{Celliers2004}  An implementation of the line VISAR commonly used is shown in Fig. \ref{F:VISAR}.  Other useful implementations of the VISAR are the 0-D point VISAR\cite{Hemsing1979,Dolan2006}, and the 2-D high-resolution VISAR\cite{Erskine2012,Erskine2014}.  

{
The line VISAR is a high-resolution optical system that projects a two-dimensional magnified image of the target onto a streak-camera detector as shown in  Fig. \ref{F:VISAR}.  The reflected laser-light signal passes through a pair of velocity interferometers before being recorded on the streak cameras. The interferometers superimpose a sinusoidal spatial modulation on the image: Doppler shifts in the reflected probe are manifested as shifts of these fringes at the interferometer outputs. The streak cameras record the central slit region of the field of view and sweep this signal in time across the output detector.  An optical delay is achieved by a combination of the refractive delay in the etalon and an additional translational offset of the etalon-mirror combination along a direction perpendicular to the mirror plane.   The translation distance, $d=h(1-\frac{1}{n})$, is calculated to place the image of the mirror plane as viewed through the etalon coincident with its initially-determined null position (i.e., zero path delay with the etalon removed).  The resulting optical time delay is $\tau=\frac{2h}{c}(1-\frac{1}{n})$ where $h$ is the etalon thickness, $n$  is the index of refraction of the etalon, and $c$  is the speed of light. The end mirror and the etalon in the delay arm of the interferometer are both mounted on a motorized translation stage.

When perfectly aligned (parallel output beams), the Mach-Zehnder interferometer has a uniform output intensity; a single phase across the output. For the line-imaging application, a fringe pattern is imposed on the output by slightly tilting the output beamsplitter, thus imposing a linear ramp in relative optical-path difference across the output field. Doppler shifts in the light passing through the interferometer result in changes in the optical phase at the output. These, in turn, appear as shifts in fringe position. The spatial frequency of the fringe pattern is arbitrary and is usually set to provide from 20 to 30 fringes across the output image (see Fig. \ref{F:VISAR}).

Such a velocity interferometer measures the two-point time-autocorrelation function of the input beam, and the apparent velocity can be related to the phase shift as, 
\begin{equation}
u_{app}(t) = \frac{\lambda}{d\tau (1+\delta)}\frac{\phi_v (t+\tau /2)}{2\pi} \label{E:u_app}
\end{equation}
the constant, $\frac{\lambda}{d\tau (1+\delta)}$, defines the velocity per fringe (VPF) for the VISAR, and $(1+\delta)$  accounts for an additional phase shift caused by dispersion in the delay etalon ($\delta=0.0315$ at 532 nm in a fused silica etalon).\cite{Celliers2004,Dolan2006}  

This apparent velocity  is accurate for Doppler shifts observed from free surfaces moving in vacuum, and we have already noted that for the break out of a steady shock, the free surface velocity measurement can readily provide a stress measurement based on previous tabulated measurements.  In practice many dynamic compression experiments are more complex than this simple scenario.  For example, in many cases, especially if one wishes to sustain the pressure in the material of interest for some time, a transparent window is placed at the rear surface of the target, such that the shock propagates into it, rather than breaking out into vacuum.  In this situation reflections will occur at the sample-window interface, and an impedance matching analysis is required to determine the pressure-particle velocity relationships, as shown in Fig. \ref{F:ImpMatch}.

Furthermore,  Hayes\cite{Hayes2001} showed that for a complex windowed target system, equation (\ref{E:u_app}) is equivalent to $u_{app}(t)=\frac{dZ}{dt}$ where $Z$ is the total optical thickness between the reflecting surface and the interferometer beam splitter (the only time-dependent path lengths are within the driven portion of the target).  If the reflecting surface probed is a single reflecting shock, then the actual velocity, $u_{act}$, and the apparent velocity, $u_{app}$, are related by $u_{act}=\frac{u_{app}}{n_0}$, where $n_0$ is the refractive index of the initial state.  Hayes\cite{Hayes2001} also showed that if the refractive index of a compressed, transparent window can be written as $n(\rho)=a+b\rho$ then $u_{act}=\frac{u_{app}}{a}$, independent of the time or position dependence of the compression wave.  Since the refractive index of many sample windows can be fit relatively well to $n(\rho)=a+b\rho$ it is often assumed that the apparent and actual velocities are proportional to each other.  In these special cases the “actual” VPF is, $VPF_{act}=\frac{\lambda}{2\tau(1+\delta)n^*}$ where $n^*$ is $n_0$ for a reflecting shock, or $a$ for a transparent window with an index defined by $n(\rho)=a+b\rho$.  

Thus far we have discussed the case of a steady shock, but we have already noted that ramp compression is an area of growing interest, as it keeps the material cooler.  In this situation, in the case of free surface breakout, or indeed a propagation of the compression wave from the target of interest into a transparent window, the velocity of the reflecting surface will  rise more gradually than in the case of the shock.  For such ramp compression situations often targets are used that have several steps in them, such that the ramp wave travels differing distances across the target (as seen in the raw data shown for a three-step target in Fig. \ref{F:VISAR}).  A ramp wave will generally steepen up as it propagates further, and it has been shown that, by considering the interaction of successive characteristics reaching the interfaces , the stress-density relation within the target can be deduced directly by recursive methods using the measurement of the interface velocities at the steps~\cite{Swift2019}.

\begin{figure} 
\includegraphics[width=9cm, angle=0]{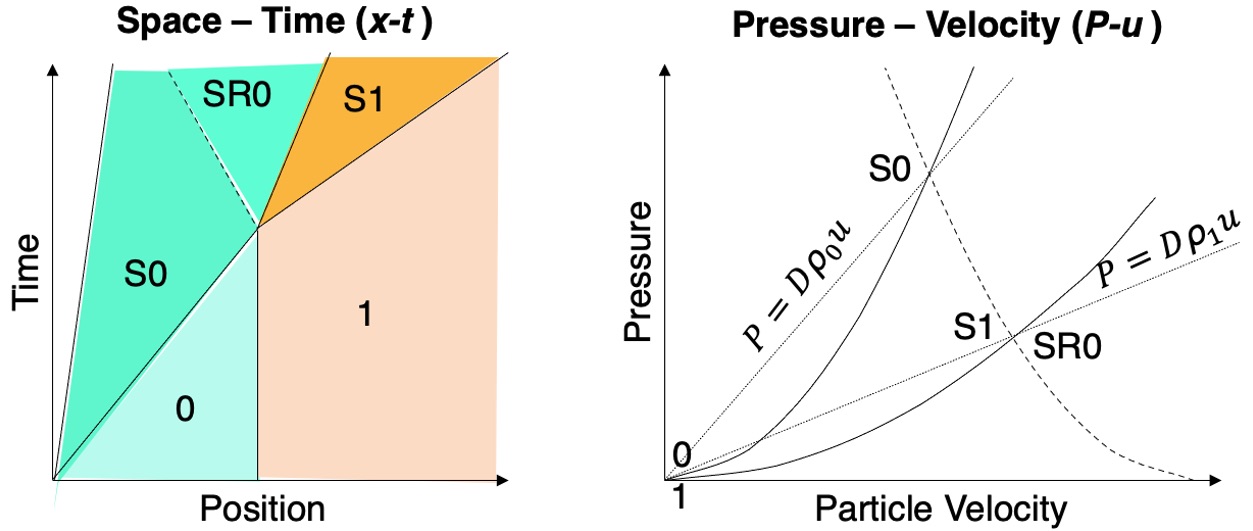}
\caption{\small  Impedance Matching:  Space-Time ($x-t$) and Pressure-Velocity ($P-u$) plots of a sample (0) and window (1) interface.  A shock takes the sample from state 0 to a shocked state S0.  When the shock S0 reaches the window another shock takes the window from state 1 to S1 and a release wave (SR0) is reflected back into the sample. The corresponding states are shown in the $P-u$ plot by the forward-moving solid Hugoniot curves S0 and S1, and by the backward-moving dashed release wave SR0. }
\label{F:ImpMatch}
\end{figure}

\subsection{Target Design}
\label{SS:Targets}

Traditional HED campaigns using large-laser drivers at facilities like NIF, Omega, Orion, or Laser Megajoule (LMJ) generally consist of 4-20 shots scheduled over several shot days. Once the shots are granted, a campaign places great significance on the success of every shot which requires ongoing modifications to the target and experimental design based on the results of previous shots.  Preparing targets and the requisite shot reviews at these facilities require months of planning, so that completion of a single campaign generally takes several years.\cite{Nikroo2016} By contrast, 3-5 day campaigns at a FEL consists of hundreds of shots taken every 5-10 minutes, and with the advent of high-repetition-rate lasers tens of thousands of shots taken at 0.1 to 10 Hz. Target fabrication methods have been, and are being, developed to meet these target needs.\cite{Prencipe2017} 

Laser-compression targets require an ablator layer designed to efficiently convert the intense laser energy into pressure and transmit this pressure to the sample; a sample layer to be compressed and probed; likely, a tamper (window) layer to equilibrate and maintain the sample pressure; and an appropriate interface for VISAR to measure the time-dependent velocity wave profile and accurately determine the sample pressure.  The conversion of laser energy into pressure, discussed in section \ref{S:Ablation}, is most efficient for low atomic number (Z) ablators so that plastic (CH), diamond, and beryllium ablators are commonly used. Another advantage of low-Z ablators is that the x-ray emission generated by the ablation plasma is at lower energy and can often be effectively absorbed by the ablator itself or by an additional pre-heat layer.  Samples are generally solid, but have included pre-compressed fluids or liquids on large laser facilities.\cite{Jeanloz2007,Celliers2018,Millot2019} Solid samples can be powdered,\cite{Gorman2018,Gorman2020} single crystal,\cite{Wark1987,Wark1989,Kalantar2005} highly-textured polycrystals,\cite{Wehrenberg2017} or even microscopic grains encased in an epoxy slurry.\cite{smith2022}  

Selection of appropriate ablators and tampers for specific samples and experimental goals generally require many considerations. The ablator material must be chosen to enable the compression type (shock, shock/ramp, pure ramp) and the maximum pressure required. The ablator thickness is critical to enable sufficient transverse smoothing of the pressure drive\cite{}, and to eliminate unwanted reverberations that can adversely affect the pressure history in the sample. At every interface within the sample package a portion of the compression wave will be reflected and a portion transmitted. Standard impedance-matching (Fig. \ref{F:ImpMatch}) considerations are used to simulate, calculate, or estimate the effect material choices, but invariably the choices must be checked using a hydrocode simulation to ensure that the target package is consistent with the experimental goals of the campaign.

The need for economic and efficient production of so many targets has spurred creative fabrication techniques.  The ablator, sample, and tamper layers are generally bonded together using a low viscosity epoxy such as Stycast 1266 (650 cP) or ÅngströmBond 9110LV (500 cP). Occasionally it is economical to directly deposit a textured sample onto the ablator or to coat a CH film onto a sample film. If an epoxy bond is used, it is critical that the bond not be too thick ($\lesssim$2 $\mu$m) to avoid ring-up within the epoxy layer that affects the desired thermodynamic state in the sample.  Through careful sample manipulation and construction, several hundred samples can be produced in under a week by talented technicians or scientists. Alternatively, elaborate robotic-controlled target fabrication procedures have been pioneered commercially,\cite{Boehm2017,Spindloe2018,Farrell2022} and bonded or shaped extended tape targets using precision rollers can produce targets that can subsequently be cut or diced.\cite{smith2022}  

In the very near future high repetition rate lasers will be deployed at FEL facilities.\cite{mason2018} In order to take advantage of these new facilities a significant change is needed in target fabrication.  Tape targets  10s of meters in length have been proposed and tested that will enable rapid target alignment and economic production.\cite{smith2022,Prencipe2017}  Of particular interest for their potential for general application and economy are slurry targets where microscopic ($\lesssim$500 nm) sample grains are encased in a medium such as epoxy.\cite{smith2022}  Significant work remains to be done to understand the energy partitioning and thermodynamic state  attained in slurry targets and it is unclear whether Hugoniot states will even be possible.\cite{Kraus2010}    
}

\section{Work to date on femtosecond diffraction on FELs}
\label{S:Work_to_date}

\subsection{Plasticity and Deformation}
\label{SS:Plasticity}
Having given some historical perspective to the field of femtosecond diffraction from dynamically compressed solids, in this section we give a brief overview of some of the experiments that have been performed to date.  Our overarching aim in this paper is to give our perspective on various directions and avenues of study that may be pursued in the future, and the reasons for doing so.  As such, we emphasize that the work outlined here has chosen to be representative of the status of the field, and is by no means comprehensive.

The experiments that have been performed to date in this field broadly fall into two distinct categories: (i) those that interrogate plastic deformation under uniaxial shock compression, and (ii) those that seek to understand the phase diagram of matter at high pressures - although as we shall discuss in further detail below, there are important overlaps between these two areas of endeavour.  

In the first of the above categories, one of the initial experiments performed at LCLS that illustrated the unique capability that the platform affords was a measurement of the ultimate compressive strength of a copper crystal - that is to say the point at which a perfect crystal would yield under uniaxial compression, with the plastic yielding occurring due to the spontaneous generation of so-called homogeneous dislocations owing to the huge shear stresses present. The challenge in demonstrating such an effect is that metallic crystals have pre-existing defects and disclocations, and when subjected to uniaxial compression these defects can alleviate the shear stress, with the associated plastic flow, an effect that is enhanced by dislocation multiplication.  Given that the shear is released in this way, the ultimate strength of the material to uniaxial elastic compression cannot be tested at low strain rates.  However, it had long been conjectured that if the surface of a crystal were instantaneously subjected to uniaxial stress below some theoretical limit, then on some ultrashort timescale the material would still respond purely elastically, as the strain rate would be so high that any pre-existing dislocations would be insufficient to relieve the stress, even as their number multiplied during the short compression. That is to say a simple application of Orowan's equation of plasticity~\cite{Orowan1942}, $\dot{\epsilon_{\rm p}}=Nv|{\bf b}|$, where $\dot{\epsilon_{\rm p}}$ is the plastic strain rate, $N$ the number of mobile dislocations, $v$ their mean velocity (assumed to be subsonic), and ${\bf b}$ their burgers vector, would indicate that the relief of shear strain would be negligible on the relevant pertinent short timescale.

There has long been evidence for the above picture in the following sense.  The elastic wave (the so-called Hugoniot Elastic Limit (HEL)) that is initially seen in samples when they are shocked below the strong shock limit (defined as when the shock velocity of the plastic wave is greater than that of the elastic wave) increases in strength for thinner crystals~\cite{Asay1972}, and for thicknesses in the sub-micron region appears to tend towards what is believed to be the ultimate strength of the sample~\cite{Kanel_2014}.  Indeed, a  prediction was made for copper based on MD simulations by Bringa and co-workers~\cite{Bringa2006}: within the paper describing their work using MD simulations to model shocked single crystals of copper they specifically stated {\it "On the basis of our simulations....we predict that prompt 3D relaxation would not be observed in copper, provided that diffraction measurements could be made at time intervals of a few picoseconds behind the shock front, which is now feasible experimentally"}.  Further MD simulations indicated that copper, when rapidly compressed along (111), would ultimately yield when lattice had been uniaxially compressed by of order 16-20\%~\cite{Dupont2012}.

\begin{figure}
    \centering
           \includegraphics[width=0.95\columnwidth]{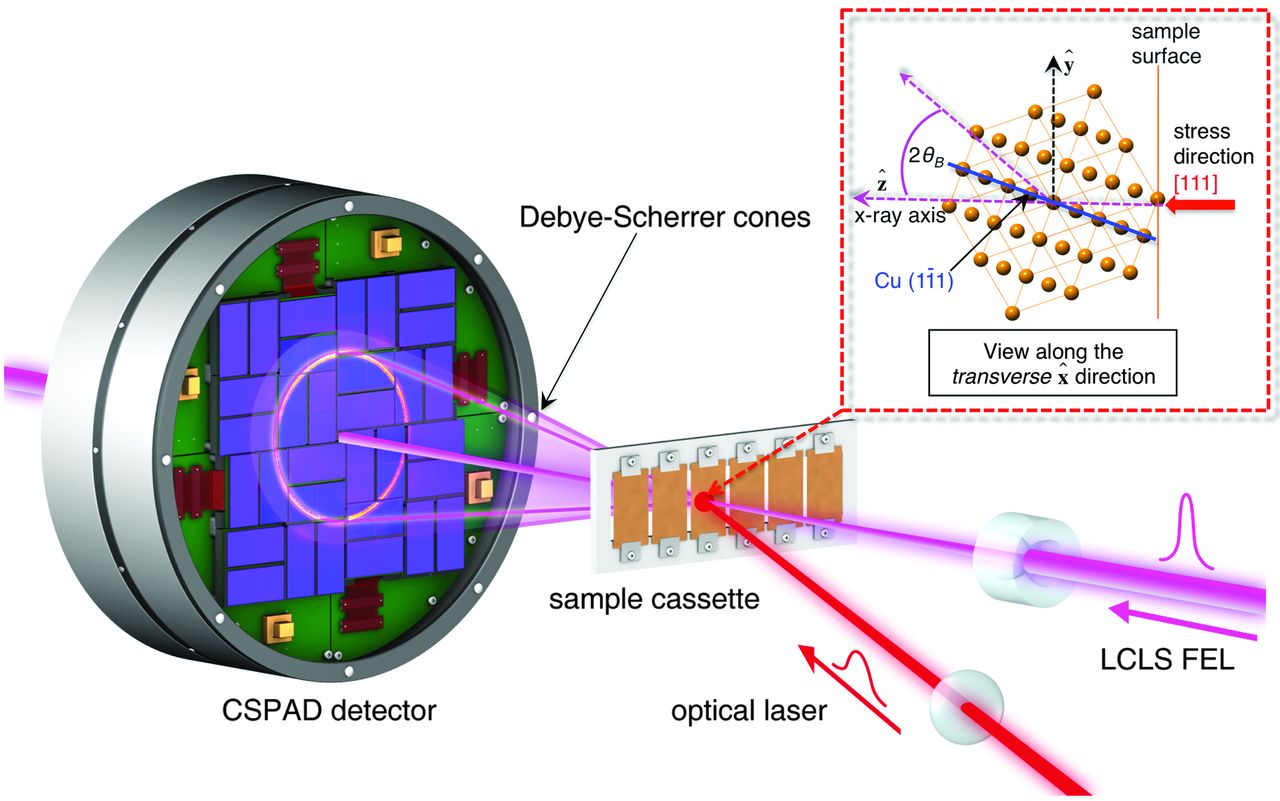}
       \includegraphics[width=0.95\columnwidth]{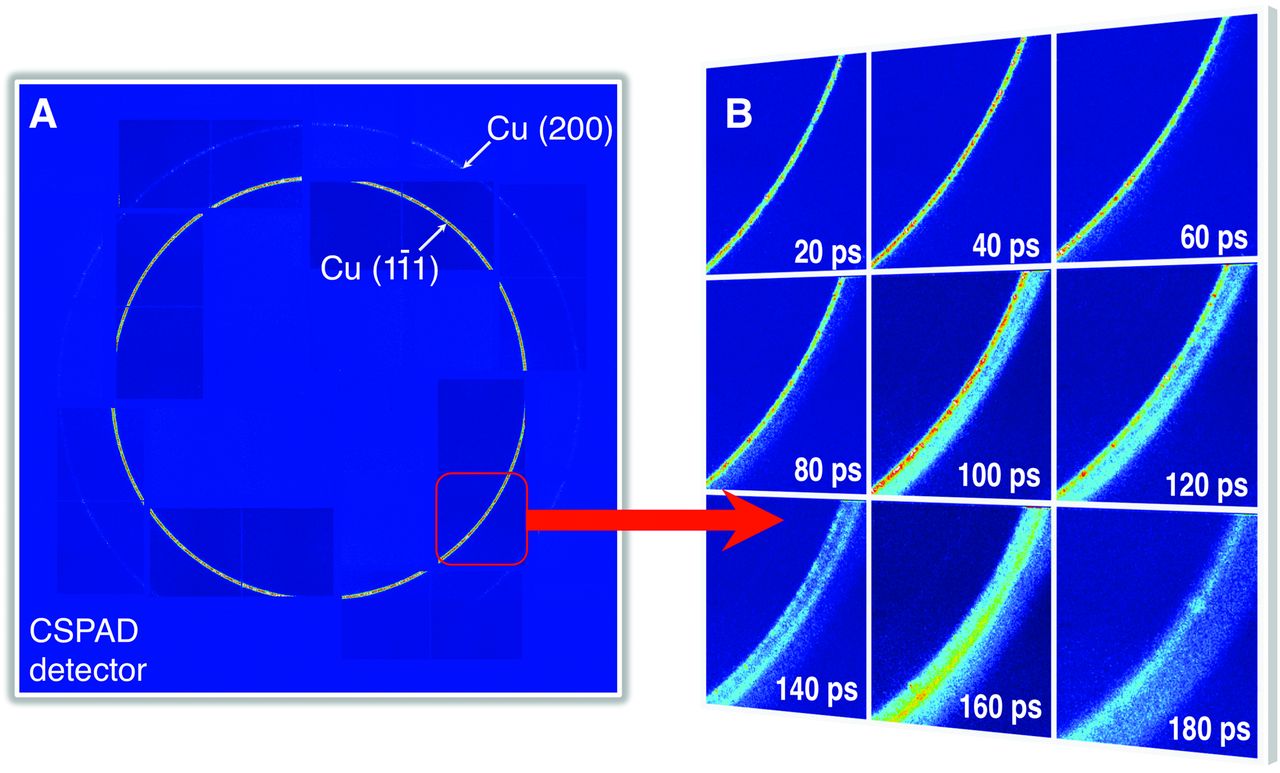}
    \caption{(Upper image) A schematic diagram of the experimental set-up to measure the ultimate compressive strength of copper, with (lower image) raw diffraction data as a function of delay time between the x-ray pulse and the optical laser pulse, as described by Milathianaki {\it et al.}~\cite{Milathianaki2013}}
    \label{F:Milathianaki}
\end{figure}

A schematic diagram of the LCLS experiment~\cite{Milathianaki2013} to test this hypothesis is shown in Fig. \ref{F:Milathianaki}.  Whilst specific to this experiment, the set-up is typical of that used in the majority of dynamic compression experiments performed to date on FELs.   As can be seen from the diagram,  the experiment is in the Debye-Scherrer geometry.  Thin (1 $\mu$m) layers of copper were deposited on Si wafers (in contrast to many experiments, the target did not have a sacrificial ablator on its surface, as is normally the case for target design, as outlined above in \ref{SS:Targets}).  A compression wave, with peak pressure close to 1 Mbar, was launched in them by a laser pulse that had a roughly Gaussian temporal shape with FWHM of 170 psec.  The laser was focused to a spot slightly greater than  250 $\mu$m in diameter.  X-ray pulses from the FEL of energy 8 keV and pulse length under 50-fsec were focused by beryllium lenses to a spot of order 30 $\mu$m diameter, co-centric with the optical laser pulse.  These x-rays were thus diffracted from the shocked region, and could be timed with respect to the shock pulse to better than a picosecond. The diffracted x-rays were detected using an in-vacuum, 2.3-megapixel array detector [the Cornell Stanford Pixel Array Detector (CSPAD)~\cite{Philipp2011}] (for a detailed description of these detectors see elsewhere~\cite{Blaj2015}).

A series of successive shots were then taken with the delays between the optical and x-ray pulses changed at each step by 20 psec, and a fresh target translated into the beams for each laser shot.  Importantly the grains within the copper layers were not randomly oriented.  As they had been grown by deposition, the copper targets were fiber textured - that is to say normal to the target they had sub-$\mu$m grains that were preferentially oriented $\pm 3^{\circ}$ along the [111] direction, but random azimuthally about it.  Diffraction then took place from the [1$\bar{1}$1] planes.

\begin{figure} 
\includegraphics[width=9cm, angle=0]{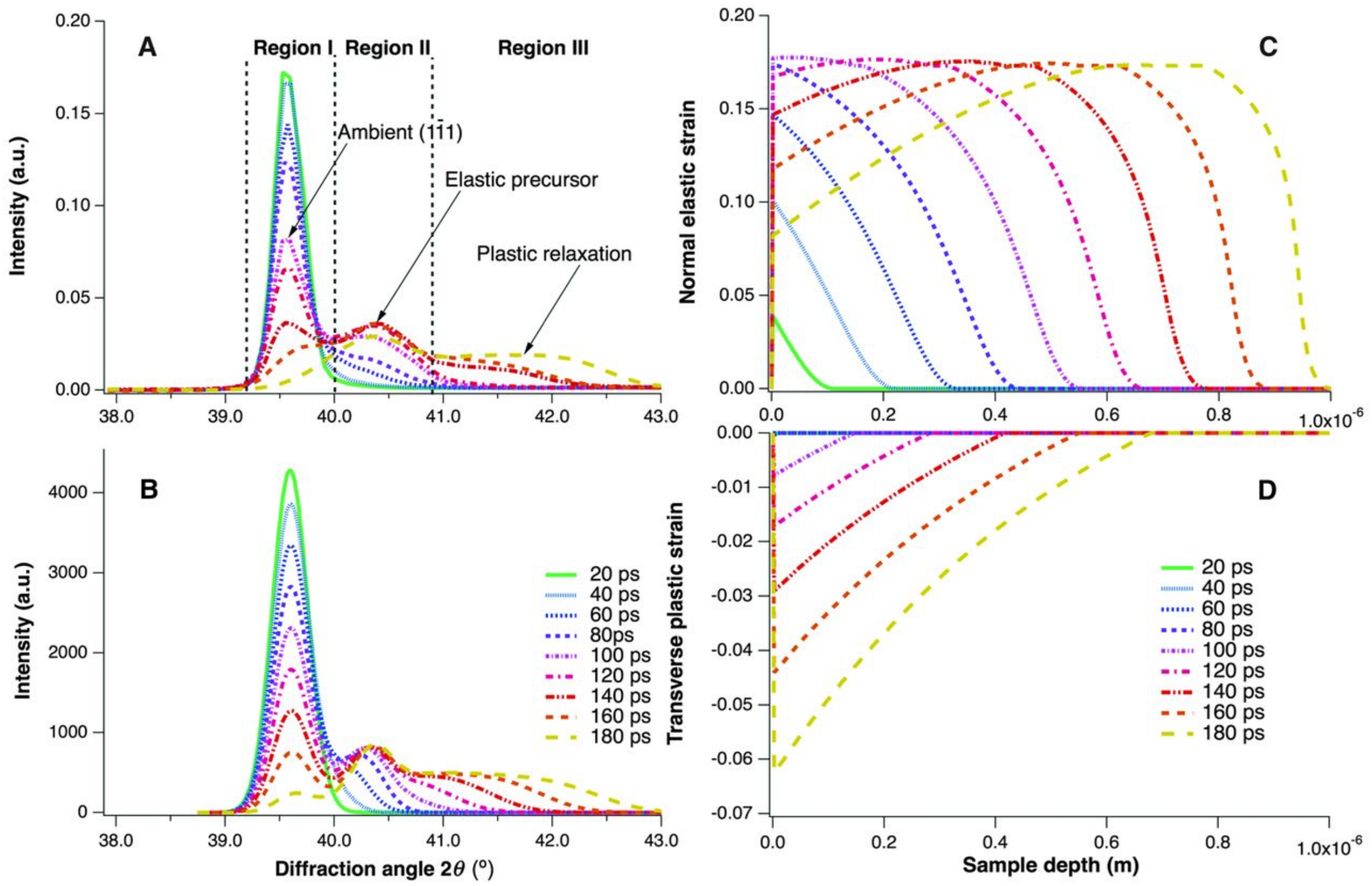}
\caption{\small  Experimental diffraction data from the experiment depicted in Fig. \ref{F:Milathianaki} resulting from the azimuthal integration around the Debye-Scherrer ring. The diffraction profiles are divided into three regions to illustrate the characteristic lattice response: region I, the unstrained lattice; region II, the elastically compressed lattice; and region III, the lattice exhibiting three-dimensional relaxation. (B) Simulated diffraction data resulting from the calculated strain profiles, showing good agreement with experiment. (C) The modelled time-dependent normal elastic strain and (D) transverse plastic strain profiles versus sample depth using the code described in the work of Wark {\it et al.}~\cite{Wark2014}. }
\label{F:Cu-strain}
\end{figure}

The raw diffraction data is shown alongside the experimental setup in Fig.\ref{F:Milathianaki}. The unshocked material scatters x-rays close to 39.5$^{\circ}$, and diminishes in intensity as the wave traverses the sample.  As the surface of the target starts to become compressed, diffraction at higher angles occurs, and in Fig.\ref{F:Milathianaki}B a second diffraction ring of larger radius can clearly be seen to emerge.  At early  time this new peak occurs at around $40.25^{\circ}$, which corresponds to 1-D elastic compression of about 17\%, as predicted.  Note that, given the compression wave is moving slightly faster than the speed of sound, it takes of order 180 psec for the compression wave to traverse the 1 $\mu$m thick sample, and thus early on the x-rays are diffracting from both the compressed region of the target at the surface, and the yet-to-be compressed region below it.  On a timescale of about 60 psec, again in good agreement with original predictions, diffraction at higher angles occurs, consistent with the lattice relaxing in all three dimensions, but with a lower strain in each.  Lineouts of the data are shown alongside the simulated diffraction signals~\cite{Wark2014} in Fig. \ref{F:Cu-strain}.  Excellent agreement between the experimental data and the simulations can be seen.  The simulations use quite a simple model where the plasticity essentially follows Orowan's equation, and the number of dislocations in the system rise copiously once the ultimate strength of the material is exceeded. Thus one of the first dynamic compression experiments performed at LCLS successfully measured the ultimate compressive strength of a single crystal, and confirmed MD predictions made several years prior to its execution.

As noted above, in the experiment of Milathianaki the targets were fiber-textured, and the experiment performed in a Debye-Scherrer transmission geometry.  A further feature of this experiment was that the surface normals of the targets were aligned anti-parallel to the incoming x-ray beam.  A further seminal experiment in this first category of plasticity studies investigated lattice rotation and twinning under shock compression in tantalum crystals.  In many respects the experimental set up was very similar to that shown in Fig \ref{F:Milathianaki}, but with several important differences.  In this experiment~\cite{Wehrenberg2017} tantalum targets were used.  Their thickness ($\sim$6 $\mu$m) was greater than that in the copper experiment.  Furthermore, these targets had a 50 $\mu$m thick kapton ablator glued to their surface.  They were irradiated with an optical laser pulse which had a 5-10 nsec pulse length, and focused to spots varying from 100 to 250 $\mu$m diameter, and enabled dynamic pressures within the Ta target ranging from 10 to 300 GPa to be created.  One effect of the ablator was to ensure that the compression wave that was launched into the Ta had formed into a steeply rising shock. Finally, unlike in Fig. \ref{F:Milathianaki}, the normal to the target surface was not anti-parallel to the direction of propagation of the FEL x-ray beam: the target was tilted so that its surface normal made an angle to the x-ray beam axis. Given that the targets were again fiber-textured (in this case with grains having the normals to their (110) faces along the surface normal) symmetry is broken, and because of this symmetry breaking complete Debye-Scherrer rings are not observed, but arcs of diffraction are seen, the azimuthal position of which encodes the angle that the lattice planes make to the surface normal.  Thus, by monitoring the position of the arcs, both compressive elastic strains, and the angles that distinct lattice planes make to the surface normal, can be interrogated~\cite{McGonegle2015}.  

\begin{figure}
    \centering
        \includegraphics[width=0.95\columnwidth]{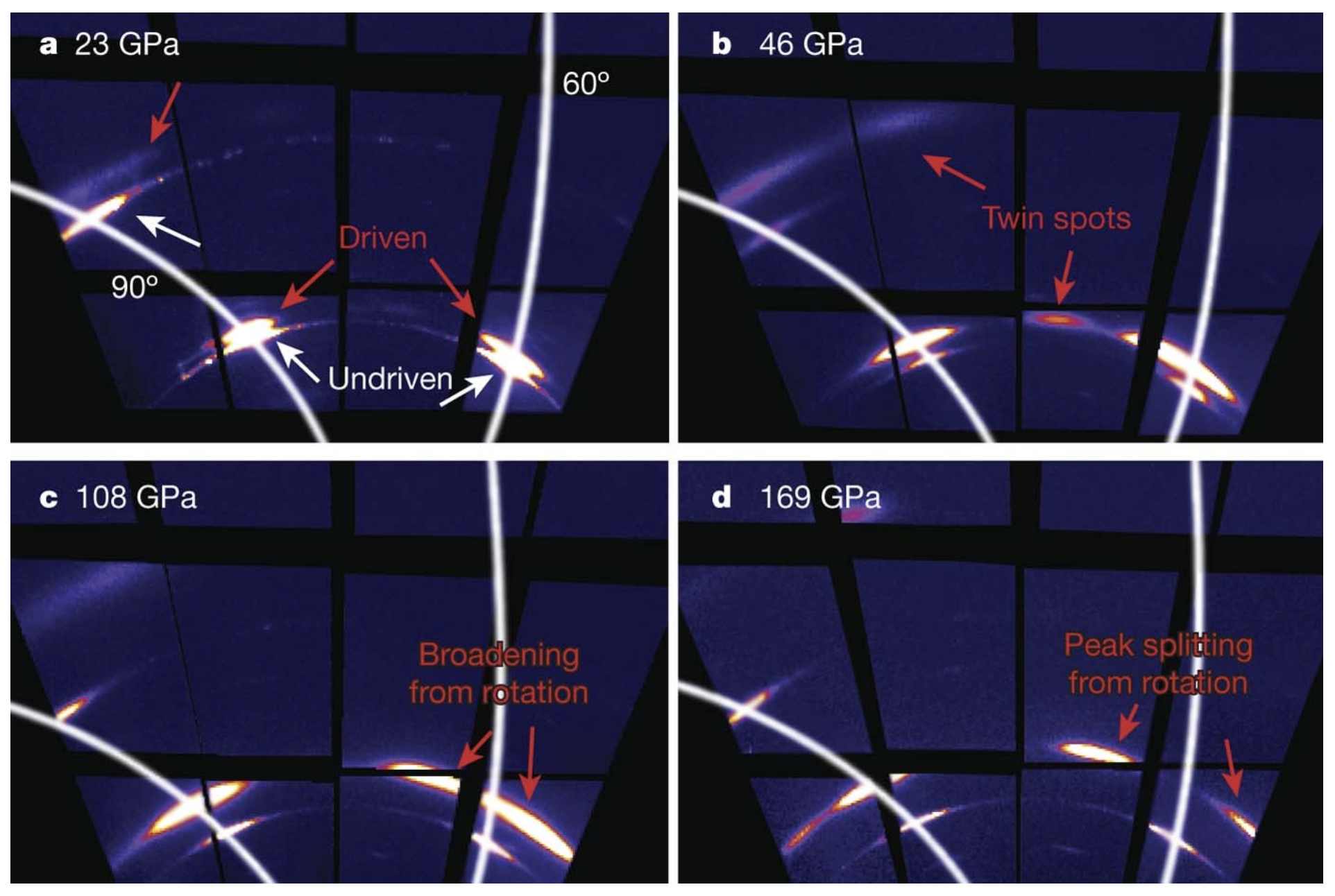}
    \caption{Diffraction patterns from tantalum shocked to the pressures shown.  The spots corresponding to the driven portion of the sample (highlighted by red arrows) appear at higher diffraction angles than those corresponding to the undriven portion (highlighted by white arrows), appearing as a shift upwards in this view. Lines of constant $\chi$ (angle between the lattice plane and sample normal) are shown at 60$^\circ$ and 90$^\circ$ to illustrate the reorientation of the texture spots in driven tantalum. ({\bf a}), At 23 GPa, no reorientation is observed, because the diffraction spots for the driven and undriven sample have the same azimuthal location. ({\bf b}) At 46 GPa, new texture spots arise that correspond to twinning across a {112} mirror plane and an additional 2.5$^\circ$ rotation. ({\bf c, d}) At higher pressures (108 GPa, {\bf c}; 169 GPa, {\bf d}), large lattice rotations are observed, causing the $\chi$ = 60$^\circ$ (110) spot (highlighted by red arrows) to broaden and then split. Figure taken from the work of Wehrenberg {\it et al.} ~\cite{Wehrenberg2017}}
    \label{F:Wehrenberg-diffraction}
\end{figure}

\begin{figure}
    \centering
        \includegraphics[width=0.95\columnwidth]{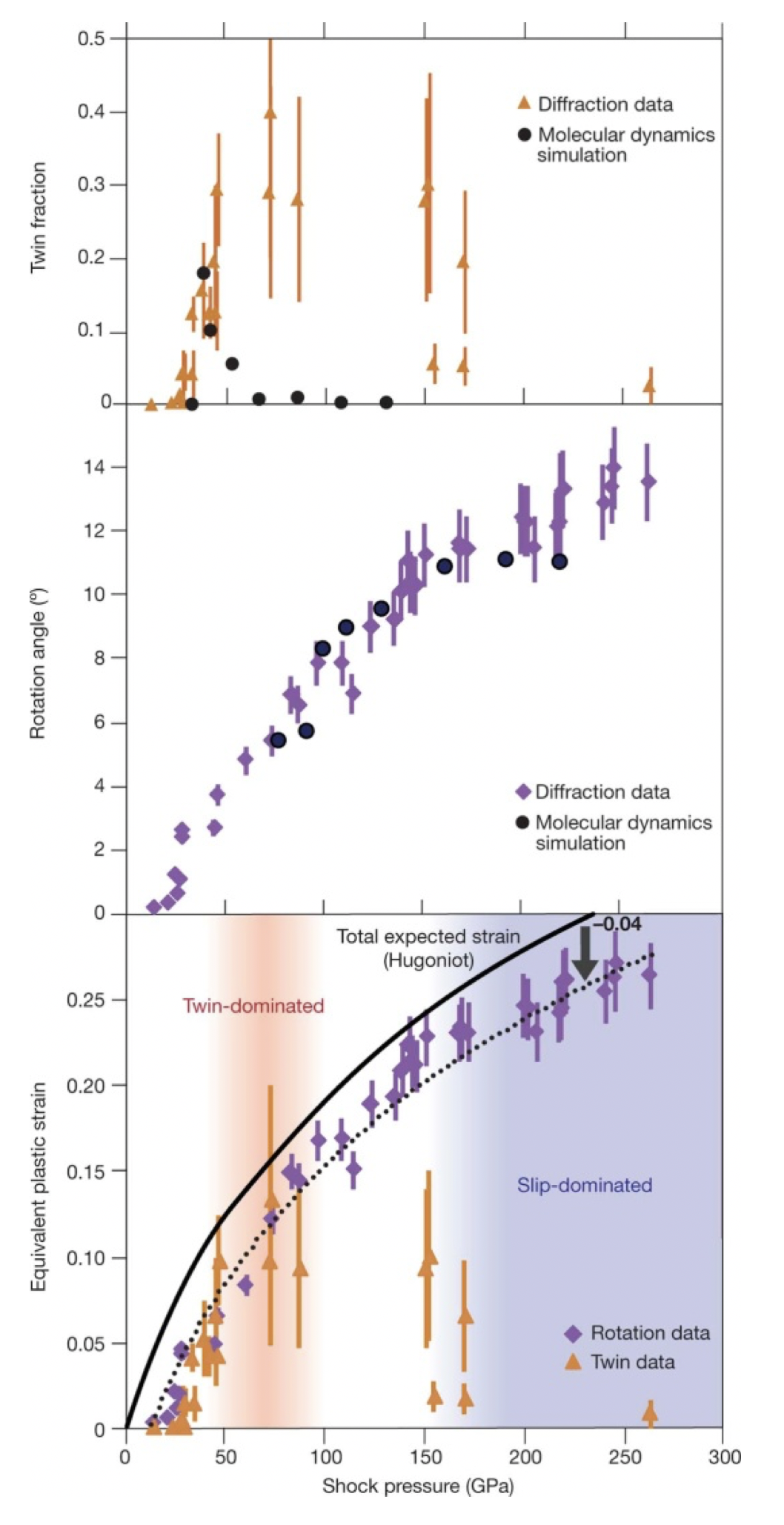}
    \caption{Twin fraction, rotation angle and plastic strain computed from the diffraction data such as that shown in Fig. \ref{F:Wehrenberg-diffraction}. Upper plot: twin fraction as a function of shock pressure, measured from diffraction
data by the ratio of texture spots (orange triangles) and from molecular dynamics simulations (black circles). Middle plot: lattice rotation angle from x-ray diffraction data (purple diamonds) and from molecular dynamics simulations (black circles). Lower plot:  equivalent plastic strain inferred from lattice rotation (purple diamonds) and twin fraction (orange triangles). The equivalent plastic strain for a material subject to hydrodynamic shock (‘on the Hugoniot’) is also plotted (black curve). Figure taken from the work of Wehrenberg {\it et al.} ~\cite{Wehrenberg2017}}
    \label{F:Wehrenberg-rotation}
\end{figure}

The key results from this experiment can be seen in Figs. \ref{F:Wehrenberg-diffraction} and \ref{F:Wehrenberg-rotation}.  The first of these, Fig. \ref{F:Wehrenberg-diffraction}, shows the observed diffraction patterns for 4 shots at 4 different pressures (as determined by the VISAR measurements).  Diffraction from certain members of two families of planes,  \{110\} and \{200\}, can be seen at certain azimuthal angles.  Also shown are the angles, $\chi$, that the specific planes make with respect to the target normal - in this case we can observe two members of the \{110\} family of planes, one that is perpendicular to the surface normal, and another that makes an angle of 60$^{\circ}$ with respect to it. As diffraction is taking place when the shock has only traversed part of the crystal, diffraction takes place both from the shocked region, and the unshocked regions of the target, and this can be seen by noting the shift in diffraction angle, and the expansion of the Debye-Scherrer ring, which increases with shock pressure, and gives a measure of the change in the lattice spacing.  However, what can also be observed, and is most noticeable at the highest pressure of 169 GPa, is that the diffraction coming from the plane that makes an angle of 60$^{\circ}$ to the surface normal splits in two, indicating that these planes have rotated significantly.  

Analysis of the diffraction patterns over a wide range of pressures showed that the lattice planes rotated under compression, rotating by up to 14$^{\circ}$ at shock pressures of 250 GPa. Deformation of the sample due to twinning could also be directly observed, as the twins could be identified owing to their differing position azimuthally in the diffraction pattern, and the fraction of the sample that had twinned could be deduced from the associated diffraction intensity.  The experiments found a peak in the twin fraction at pressures of order 75 GPa, as shown in Fig. \ref{F:Wehrenberg-rotation}, where the rotation angle as a function of shock pressure is also shown.  Reasonable agreement with MD simulations was found for the observed rotation angles, though the experiments were undertaken with fibre-textured polycrystalline material, and the MD simulations were of single crystals oriented along (110).

In the original paper describing this work the rotation angle was used to predict the total amount of plastic strain using a model that assumed two things: firstly, that the plasticity at each point in the crystal was due to the activation of a single slip plane, and secondly, that the rotation angle and plasticity was related in a way originally determined by Schmid~\cite{Schmid1926,Schmid1935}.  If we denote by $\phi_0$ and $\phi$ the angles made by the slip plane normal with the compression direction before and after compression, then the rotation angle $\omega_{\text{Schmid}} = \phi_0 - \phi$ is given by 
	\begin{equation} \label{eq:schmid_rotation}
	\omega_{\text{Schmid}} = \phi_0 - \arccos\left(\frac{\cos\phi_0}{\ell/\ell_0}\right).
	\end{equation}
where the quantity $\ell/\ell_0$ coincides with the total engineering strain along the loading direction.

Under these assumptions, the plastic strain inferred from the rotation angles could be compared with that predicted to be present on the Ta Hugoniot (assuming complete loss of strength).  It can be seen from the lower plot of Fig. \ref{F:Wehrenberg-rotation} that within this framework the rotation measurement underestimates the plastic strain by about  4\% (dotted curve), and that twinning accounts for the majority of the Hugoniot plastic strain in the 40–80 GPa range (pink shaded region); however, twinning-induced strain falls well below the total plastic strain above 150 GPa (blue shaded region), indicating that the response is slip-dominated in this regime.

The causes for the discrepancy between the plasticity inferred from rotation, and that expected on the Hugoniot, could be thought to be due to either more than one slip system being activated, or some other discrepancy in the model (especially as the MD data seems to predict the rotations reasonably well).  Indeed, both of these issues are pertinent, and we will return to them in  section \ref{SS:Strength}, where we discuss how measurements such as this are influencing the direction in which plasticity studies on FELS are taking, and our perspectives on future directions being taken in the field.

This same experiment also showed that the lattice rotations and twinning imparted by the shock were largely reversed when the shock broke out from the rear free surface of the target~\cite{Sliwa2018}.  This is a particularly important result for the field of shock physics as a whole: it had long been known that very high dislocation densities must be present at and behind the shock front in order to explain the plastic strain rates observed (again as dictated by Orowan's equation), but such very high defect densities are not seen in recovered samples.  Whilst previous MD simulations had indicated that dislocation annihilation could occur upon shock release and rarefaction, the LCLS data, demonstrating de-twinning and the rotation of the lattice back towards its original position, provides some of the first experimental insight into such an effect happening in real time.  Whilst the fact that the lattice rotated largely back to its original position is not of itself full proof of defect annihilation, it is consistent with such a microstructural effect, and in the MD simulations the lattice reorientation and defect density reduction were seen together.

As well as observation, upon release, of a reversal in the lattice rotation, these experiments also showed that whilst the rarefaction wave travelling back into the sample did so at an every decreasing strain-rate (producing a classic rarefaction fan), close the surface the plastic strain rates were still so high that significant plastic work was being performed~\cite{Heighway2019}.  This plastic work led to a heating upon release of the surface of sample, counteracting the cooling brought about by the thermo-elastic effect.  It is only the latter of these that is normally taken into account in a textbook description of shock release from a free surface, where the release process is assumed to be isentropic.  These experiments showed that this is an oversimplification of the processes that are taking place within the first few microns of the surface, and on the nanosecond timescale - i.e. precisely the time and length scales of typical laser shock experiments.

In conclusion of this section, we note that the experiments discussed above have investigated classic face-centred and body-centred cubic metals, and even in these cases we are only just starting to discover how materials act at the lattice level to such rapid compressions.  Other types of materials are clearly also of interest, and particular attention is being paid to the response of diamond above its elastic limit~\cite{MacDonald2020}, given that it is often used as a confining material in ramp compression experiments.

\subsection{Shock-Induced Phase Transitions}

In the previous subsection we have described some of the experiments that have taken place to date using femtosecond diffraction to directly study plasticity and deformation.  The second area that has received significant attention is that of the observation of shock-induced phase transitions.  Within section \ref{SS:Some-History-JSW} we described how in many ways the whole field of the determination of the structure of crystalline matter was vindicated by the $\alpha$-$\epsilon$ transition in iron being deduced by shock wave measurements that were later confirmed by static measurements, with good agreement regarding the transition pressure being found.  However, it is clear that there is by no means always such a simple relationship between what is seen (or predicted to be seen) under long term static conditions, and what is seen dynamically - again in section \ref{SS:Some-History-JSW} we described how diamond is predicted to transform to a BC8 structure above 1 TPa, but appears to remain in the cubic diamond phase in the dynamic experiments performed to 2 TPa \cite{Lazicki2021}.  As we will indicate in section \ref{SS:Phase_Boundaries}, understanding such discrepancies between static and dynamic measurements will be necessary in the context of relating the two - i.e. making inferences about the long-term behavior of materials within planets, and how they behave on nanosecond timescales when generated by laser compression.

One material that has received considerable attention over many years is elemental silicon.  Indeed, as this material can be made in an almost defect-free single-crystal form, one might initially envisage that its response to rapid compression would be relatively straight-forward to predict.  In practice the converse has been true.   Studies of shock-compressed single crystals of silicon date back over many decades, to the seminal work of Gust and Royce~\cite{Gust-Royce1971}, where the multiple waves observed in samples shocked along various crystallographic directions were interpreted in terms of a plastic response, followed by a phase transition, with the Hugoniot Elastic Limit (HEL) being observed to vary from 5.9 GPa (along $\langle110\rangle$) to 5.4 GPa  ($\langle111\rangle$) and 9.2 GPa ($\langle100\rangle$).  As noted in section \ref{SS:Some-History-JSW}, silicon was also the first material to be studied on nanosecond timescales using a laser-plasma x-ray source~\cite{Wark1987,Wark1989}. Such nanosecond laser-based experiments found the response remained elastic to much higher pressures than those seen in the $\mu$s explosively-driven experiments of Gust and Royce, and also observed two elastic waves in silicon shocked along $\langle100\rangle$, which were at the time difficult to explain~\cite{Loveridge-Smith2001}, and were followed by a significant number of studies of the material on different platforms~\cite{Turneaure2007,Turneaure2007b,Smith2012}.
On the basis of MD simulations it was suggested that the first wave, identified as an elastic-plastic transition by Gust and Royce, was in fact the onset of a phase transition to the tetragonal $\beta$-Sn or the orthorhombic $Imma$ phase~\cite{Mogni2014}. The response of the crystal is termed inelastic, rather than plastic, in the sense that conventional plasticity under uniaxial strain leads to a situation where there are more unit cells of the material per unit length transverse to the shock direction.  However, Mogni {\it et al}. posited that the relief of the shear stress due to the phase transition did not alter the number of (primitive) unit cells in this transverse direction - the new phase has a larger lattice vector perpendicular to the shock than that of the ambient lattice, and thus its generation within the material leads to a reduction in the lattice spacing perpendicular to the shock of the unconverted silicon, thus letting it relax towards the hydrostat (given the lattice spacing along the shock direction is already reduced).

This picture was largely confirmed in a series of laser-plasma based x-ray diffraction experiments that agreed qualitatively with the MD simulations~\cite{Higginbotham2016}.  Furthermore, these experiments finally provided an explanation for the two elastic waves seen in the nanosecond laser experiments, but apparently absent in dynamic experiments on longer timescales: the double elastic wave structure was found to be consistent with a phase transformation taking place over a few nanoseconds, and with the transformation leading to a volume collapse.  Before the transformation takes place, the elastic response is large (if the applied stress is large), but the delayed volume collapse upon transformation, which starts at the irradiated surface of the crystal, leads to a rarefaction wave traveling forward into the elastically compressed material, lowering the elastic compression, and eventually outrunning the shock front of the large elastic response, and thus leading to a single elastic wave at lower stresses.  It is for this reason that the very large elastic response was not seen in longer time-scale experiments.

\begin{figure}
    \centering
        \includegraphics[width=0.95\columnwidth]{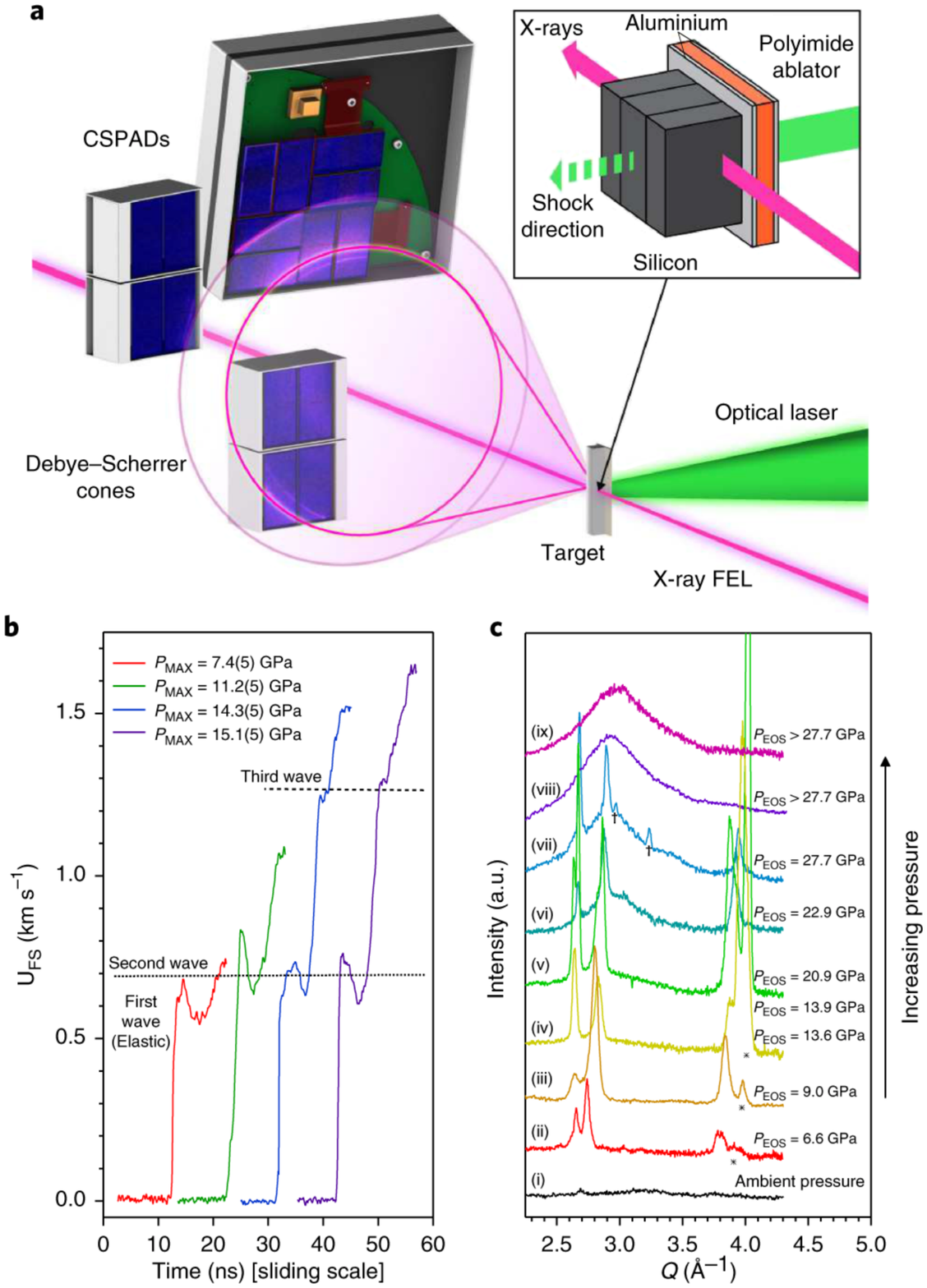}
    \caption{(a) The transverse probe set up used by McBride and co-workers whereby the compression laser was perpendicular to the x-ray beam. (b), VISAR lineouts showing the free surface velocity (UFS). Dashed lines indicate the onset of the second and third wave. PMAX denotes the maximum pressure reached as inferred from VISAR. (c) Azimuthally integrated one-dimensional diffraction patterns as a function of increasing laser intensity, and hence increasing pressure, as determined from the equation of state (PEOS) in both the transverse: (i), (ii), (iv), (v), (ix) and collinear (iii), (vi), (vii), (viii) configurations. Peaks marked with the * symbol belong to the compressed cubic diamond phase. Peaks marked with the $\dagger$  symbol cannot be described by the cubic diamond, $\beta$-tin, {\it Imma}, or simple hexagonal phases.
   Figure taken from the work of McBride {\it et al.} ~\cite{McBride2019}}
    \label{F:McBride-Silicon}
\end{figure}

The utility and versatility of x-ray FELs in providing novel information is evident in the way that such experiments have been able to confirm much of the picture painted above.  Given the few nanosecond time-scale of the phase transformation, corresponding to length scales of tens of microns, in laser based experiments both the elastic waves, and the subsequent regions of the crystal undergoing phase transitions due to the rapid compression, co-exist.  It has been argued that this co-existence complicates the diffraction pattern, as the diagnostic x-ray beam is diffracted from a range of different phases, all of which have strain gradients along the shock direction associated with them.  To overcome this problem, McBride and co-workers developed a geometry, shown in Fig. \ref{F:McBride-Silicon}, where the focused x-ray beam is incident on the target perpendicular to the shock propagation direction~\cite{McBride2019} which ensures that in the majority of cases a single wave within the multi-wave profile is probed, aiding greatly in the data interpretation.

Whilst referring the interested reader to the original paper for more details, the conclusion of this study was that the wave following the elastic wave was indeed due to a phase transition.  Importantly, the onset of the transition to $\beta$-tin was seen at a longitudinal stress as low as 5.4 GPa, consistent with the HELs seen in the original Gust and Royce work.  Of further note is that this figure should be compared with the pressure of 11.7 GPa at which the same transition is seen in static DAC experiments~\cite{McMahon1994}.  The fact that both strain rate and high shear stresses could have a large influence on the transition pressure had been noted previously in nano-indentation experiments at loading rates of order 0.1 GPa s$^{-1}$, where some evidence for the transition occurring around 8 GPa was found for silicon loaded along [111]~\cite{Gupta1980}.

Additional evidence for the mechanism underlying the transition has recently been put forward by experiments at LCLS utilising single crystals ~\cite{Pandolfi2021}.  This work supports the picture put forward by Mogni, but is at odds with models for the transformation mechanisms put forward by Turneaure and co-workers in shock experiments, but on longer timescales than the nanosecond laser-ablation experiments~\cite{Turneaure2016}. It is believed that this indicates that even once the transformation has taken place, relieving some of the shear stress, further relaxation can alter the orientational relationships between the initial and final phases.

In any event, the series of experiments cited above leads to an important conclusion: owing to the high shear stresses that are present in dynamic compression experiments, or the timescale of the kinetics of the transition, or indeed both, we cannot automatically assume that the phase diagram of matter that we might deduce from dynamic compression experiments is the same as that which pertains under static conditions, and this fact must be borne in mind when making claims that attempt to relate dynamic experiments to fields where the relevant timescales are considerably different - e.g. the state of matter within the interior of planets.

The above statement has also been shown to be pertinent in another class of materials.  We recall the words of Bridgman quoted in section \ref{SS:Some-History-JSW} where he questioned whether lattice rearrangement could indeed take place under the short timescales of shock compression. Even though such modifications can indeed be made, it is still of interest to question whether quite complex new phases can form on nanosecond timescales.
It is in this context that we note that a major and unexpected discovery in static high-pressure science at the turn of the millennium was the discovery of both modulated and composite incommensurate phases in elements such as barium \cite{Nelmes1999}, bismuth \cite{McMahon2000a}, iodine\cite{Takemura2003}, and tellurium \cite{Hejny2003}. Similar phases were subsequently found in other Group 1 \cite{Lundegaard2009,McMahon2006d,McMahon2001a}, 2 \cite{McMahon2000a}, 15 \cite{Degtyareva2004b}, 16 \cite{Hejny2003,Hejny2005,Fujihisa2007} and 17 elements \cite{Kume2005,DalladaySimpson2019}, as well as in Sc \cite{Fujihisa2005} and Eu \cite{Husband2012}.

The most exotic of these structures are the composite host-guest structures that comprise a tetragonal or monoclinic ``host" structure with 1-D channels running along the $c$-axis. Within these channels are chains of ``guest" atoms, which crystallize in structures with a $c$-axis incommensurate with that of the host structure. The diffraction patterns from such structures comprise diffraction peaks that come from either the host structure alone, from the guest structure alone, or from both. Such structure can be described within the formalism of 4D
superspace where diffraction peaks are indexed using using four Miller indices ($hklm$) rather than the traditional 3 indices \cite{VanSmallen1995}. Reflections from the host component of the basic composite structure then have
indices ($hkl0$), those from the guest have indices ($hk0m$), and the ($hk00$) reflections are common to both host and guest. In both Rb and K, a ``chain melted" phase of the host-guest structures was observed in DAC experiments where the ($hk0m$) peaks from the guest structure disappeared on either pressure decrease \cite{McMahon2004b} or temperature increase \cite{McBride2015}. In such a structure the atoms within the guest chains have no long-range order, and might therefore be regarded as 1D chains of liquid within the crystalline host structure.

\begin{figure}[!htb]
\begin{center}
\includegraphics[width=0.9\columnwidth]{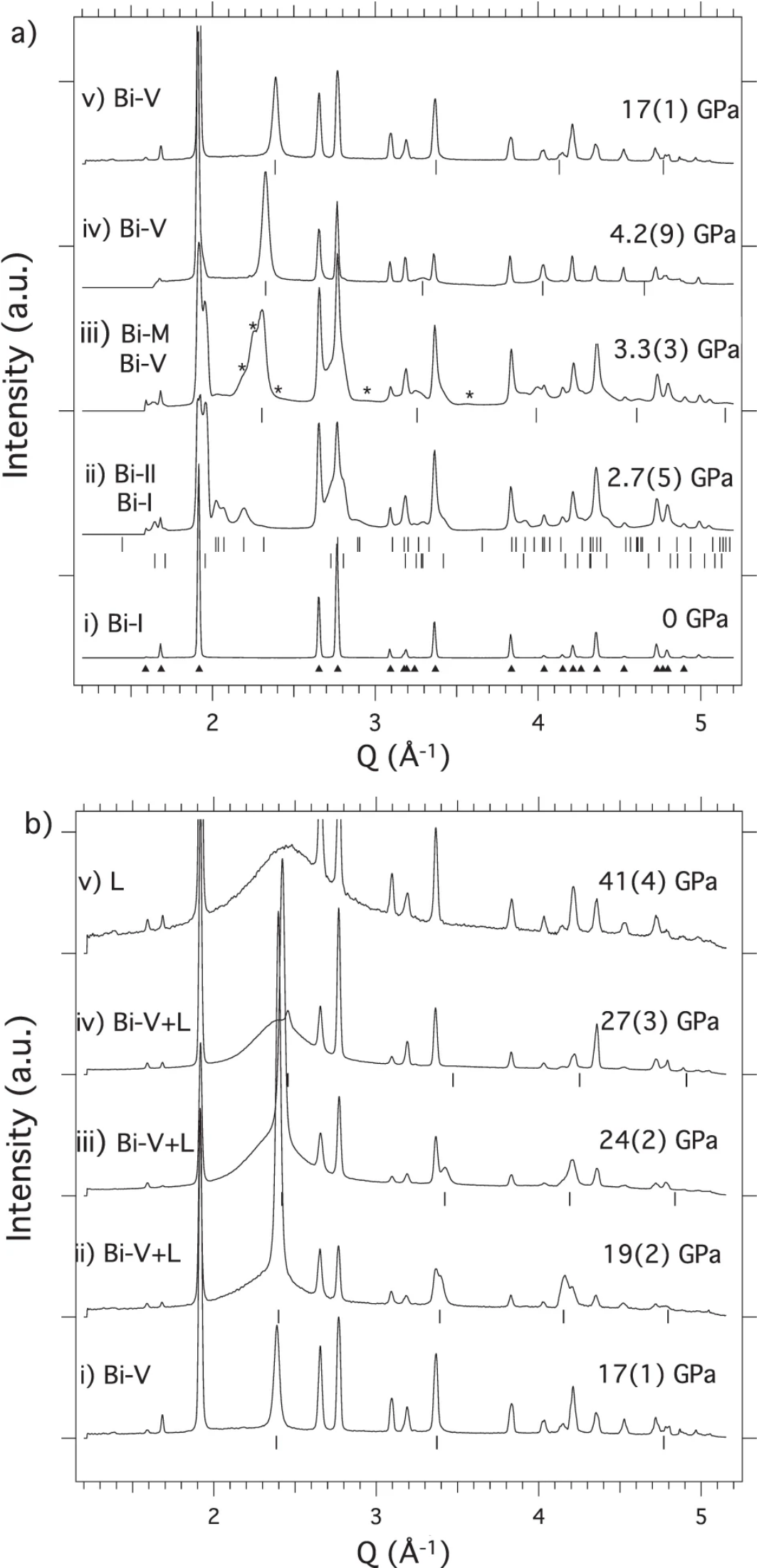}
\caption{Diffraction profiles from the different solid and liquid (L) phases of Bi seen on shock compression, as obtained on the MEC beamline at the LCLS\cite{Gorman2018}. Bi-I (profile a(i)) first transforms to Bi-II (profile (ii)) at 2.5 GPa (upper tick marks). Rather than then transforming to incommensurate Bi-III, a
transition to a metastable phase (Bi-M, and whose peaks are marked with asterisks), not seen in static compression studies, is then observed at 3 GPa (profile a(iii)). Bi-M is always accompanied by diffraction peaks from the higher pressure bcc Bi-V phase, the locations of which are indicated
by tick marks. Above 4 GPa only peaks from Bi-V are observed (profiles a(iv,v)). In static compression studies, Bi-V is seen only above 7 GPa. Scattering from liquid-Bi first appears at 19 GPa (profile b(ii)) and the liquid phase-fraction increased with pressure (profiles b(iii,iv)) until only diffraction from the liquid phase is observed above 27 GPa (profile b(v)).}
\label{fig:Bi}       
\end{center}
\end{figure} 

When FEL diffraction experiments began at the LCLS an obvious question to address was whether such complex structures would form  on the nanosecond timescales of dynamic compression experiments. As the incommensurate phase in Bi, Bi-III, is stable at pressures as low as $\sim$3 GPa in DAC experiments \cite{McMahon2000a}, and is easily accessible along the Hugoniot, such experiments were expected to be straightforward. However, Bi is well known to have phase boundaries that are different in static and dynamic experiments, and that are strain dependent \cite{Smith2008}.

\begin{figure}[!htb]
\begin{center}
\includegraphics[width=0.95\columnwidth]{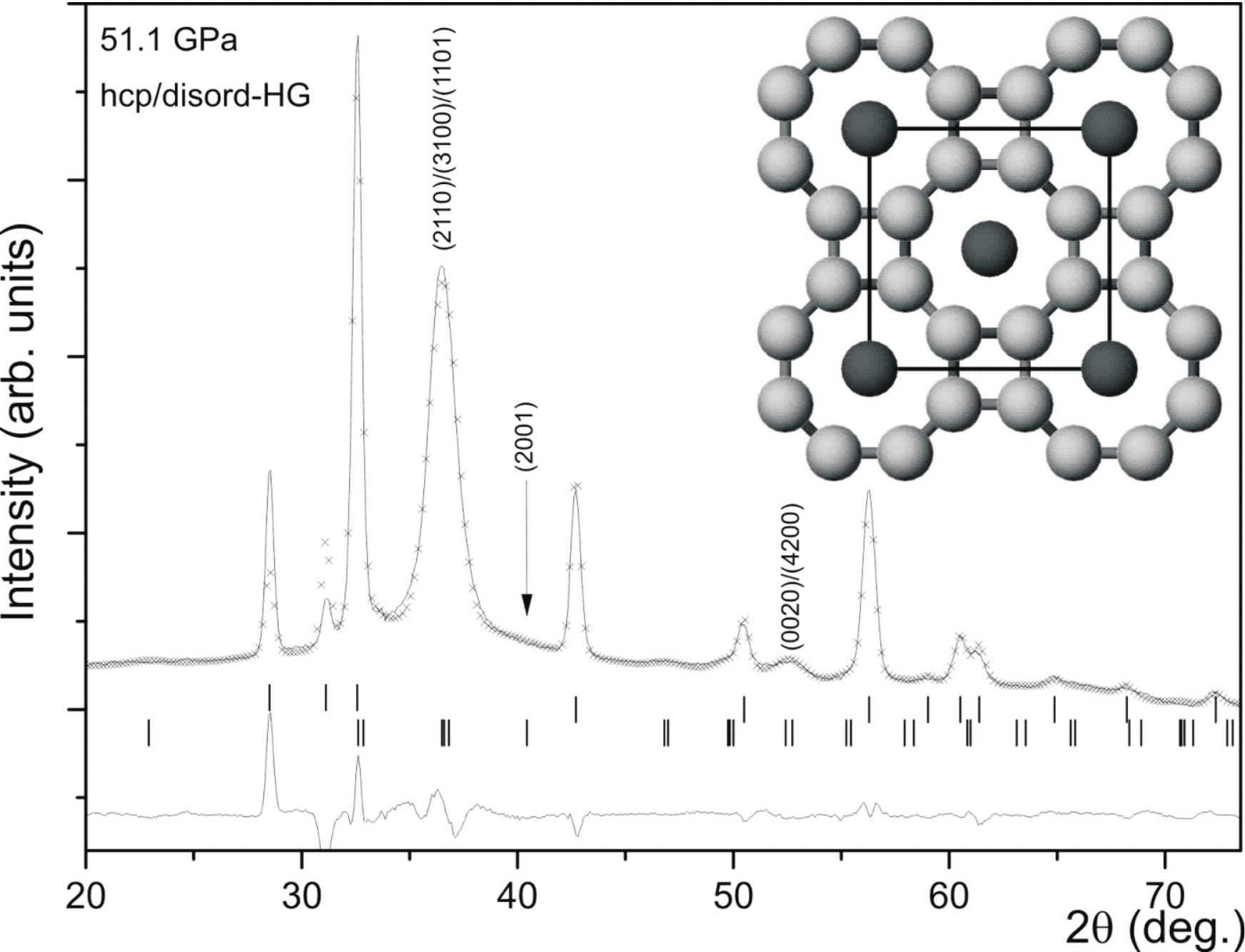}
\caption{A two-phase hcp:host-guest Rietveld fit to a diffraction profile obtained from shocked Sc at 51.1 GPa, with the most intense peaks indexed. The data were obtained at the MEC beamline at the LCLS \cite{Briggs2017}. The best fit to the profile is obtained when the 1-D chains running along the $c$axis of the incommensurate host-guest structure (shown as darker atoms in the structural figure in the inset) are disordered and liquid-like. If the chains were ordered, then the (2001) reflection at $\sim$40$^{\circ}$, marked with an arrow, would be clearly visible.}
\label{fig:sc}    
\end{center}
\end{figure}

A detailed series of experiments at LCLS revealed that while the Bi-I$\rightarrow$Bi-II transition is observed under shock compression in Bi 
(Figure \ref{fig:Bi}a(ii)), the incommensurate Bi-III phase is \emph{not} seen on further compression \cite{Gorman2018} (Figure \ref{fig:Bi}), although the equivalent incommensurate phase \emph{is} seen to form in Sb \cite{Coleman2019} on nanosecond timescales. While the guest chains in Sb were found to form the same crystalline structure within the host as was previously seen in static compression experiments, this was \emph{not} the case in dynamically compressed Sc, where the chains of guest atoms were found to have no long range order, as determined by the absence of the ($hk0m$) ``guest-only" diffraction peaks \cite{Briggs2017} - as illustrated in Figure \ref{fig:sc}. 

While the host-guest Bi-III phase was not observed in Bi,  a new high-pressure metastable phase, Bi-M, the structure of which remains unknown, \emph{was} observed (see Figure \ref{fig:Bi}a(iii)). Also, the higher-pressure Bi-V phase was seen at pressures as low as 3.3 GPa (Figure \ref{fig:Bi}a(iii)) while it is not seen below 7 GPa in static compression studies \cite{McMahon2000a}. On further compression, Bi-V was observed to melt at 19(2) GPa (see Figure \ref{fig:Bi}(b(i)), and the diffraction data from shock-melted Bi were also of sufficient quality (see Figure \ref{fig:Bi}b(v)) to follow the structural changes in liquid-Bi for the first time \cite{Gorman2018}.

These experiments perfectly demonstrated the quality of the diffraction data that are available from dynamic compression experiments at FELs, such that multi-phase profile refinement of incommensurate structures, and the extraction of coordination changes in liquids, are possible. More recent experiments have extended such studies to 80 GPa and 4000 K in Sn \cite{Briggs2019}.

The key result from the Bi studies at LCLS thus reveals that on nanosecond timescales not only are the boundaries between different phases potentially found at different regions of the phase diagram, but it is possible that phases seen statically are completely absent in the dynamic case, and vice versa. This stresses the importance of knowing the crystal structure along the compression path, and determining exactly where it changes. It also raises the possibility that the phases seen under extreme P-T conditions in short-lived laser compression experiments are not necessarily the equilibrium phases existing under the same P-T conditions within planets: a fact that reinforces the weight of our view that a degree of circumspection is advised when attempting to directly relate dynamic experiments to static contexts.  We are only starting to make progress in this area, where the results from static compression, dynamic DAC \cite{Husband2021}, impactor, and laser compression experiments will be required in order to probe sample behaviour over strain rates spanning 9 orders of magnitude.

\section{Perspectives on Future Developments}
\label{S:Future}

In the previous sections we have outlined some of the relevant history of the field of dynamic compression combined with x-ray diffraction techniques, and given some examples of the sorts of experiments that have taken place at FEL facilities in recent years.  We now turn our attention to providing our perspective on what we believe to be important lines of development in the near to medium term.  It will be evident that from the comments we have made thus far that we consider an understanding of the potential differences between the response of matter on nanosecond timescales to those that pertain on geological timescales to be of import, at least if claims of relevance of the former to the latter are to be made with any coherent degree of credence.  Broadly speaking the themes that we see as being ripe for investigation (and in many cases studies have already started) are the following:
\begin{itemize}
    \item {An improved understanding of the boundaries and differences between shock and so-called quasi-isentropic compression.}
    \item{Investigations, across the length-scales, of the mechanisms that underpin the plasticity that invariably accompanies uniaxial compression of solids, and leads to substantial heating, and is associated with the dynamic strength of the material.}
    \item{The development of techniques to measure directly temperature during dynamic compression.}
    \item{An improved understanding of under what conditions phase boundaries seen under dynamic compression differ from those seen on longer timescales.}
    \item{The development of the study of liquids at high pressures.}
    \item{The use of dynamic compression in novel material synthesis.}
    \item{The use of small angle x-ray scattering (SAXS) and line-profile analysis to monitor crystallite sizes, voids, and defect densities.}
        \item{The development of experiments that use convergent, rather than uniaxial, geometries.}
    \item{The exploration of x-ray techniques that complement diffraction such that they give information about the electronic structure of matter, as well as its crystalline structure. }
    \item{Using advances in high-power optical laser technology to increase the repetition rate at which material can be studied, leading both to improved statistics, but also increased throughput and associated access for users. }
\end{itemize}

We provide our views on the above issues below.

\subsection{Shock and Ramp Compression}
\label{SS:Shock-Ramp-JSW}

As outlined in section \ref{S:Intro}, initial dynamic studies of matter have concentrated on shock-loading of materials.  Whilst there are very good reasons to study matter under shock compression, compressing a material by a single shock does not allow the study of solids at the highest pressures, and in this regime will generally not be directly relevant to the study of matter within planetary cores.  This is because the shock process itself is highly entropic, with the material behind the shock front being heated to very high temperatures.  Indeed, even though, as a rule of thumb most metals have a melting point that increases as a function of pressure, we find that the Hugoniot (the locus of states that can be reached via shock compression) typically crosses the melt curve in the region of 100 GPa or so, as we have shown in our depiction of a generic material in Fig. \ref{F:Schematic} (as examples of real materials, Pb starts to melt under shock loading at just above 50 GPa~\cite{Song2010}, and Ta at 300 GPa~\cite{Chengda2009}).  Thus, in order to study solid state matter at high pressure, compression via a single shock may not be appropriate.  Indeed, as a reasonable approximation, if one follows the compression path towards a planet whose interior is believed to be convecting, the temperature gradient is constrained to be close to isentropic \cite{Stacey2005}, although it is acknowledged that in many cases a fully adiabatic model is overly simplistic~\cite{Helled2017}.  Very crude arguments show that in the vast majority of cases, the temperature rises along an  isentrope more gradually than the melt curve~\cite{Wark2020}.  Therefore, to compress materials to the highest pressures, whilst keeping them solids, requires us to compress them quasi-isentropically, rather than via shock compression.

We take care to specifically use the term isentropic, rather than adiabatic compression: both the rapid compression of a shock, and a far slower isentropic compression, are adiabatic, in that only compressive work is done on the system, and the external driving force does not of itself directly supply heat.   However, the shock adiabat results in significantly more heating than compressing more slowly along the isentropic adiabat.

The above realisation immediately leads to the question of how slowly must we compress matter if we are to regard the process as close to isentropic?  In the absence of plastic flow, the difference between a shock and dynamic compression that is close to an isentrope is simple to understand.  In this case uniaxial compression can be visualised in terms of a single chain of atoms.  In the case of an elastic shock, the shock-front is at the atomic scale, such that an atom in front of the shock has no interaction with it until its neighbour is accelerated towards it, imparting considerable kinetic energy in the compression process - i.e. it 'overshoots', with atoms at the shock front itself for a time shorter than a single oscillation having a closer interatomic distance than the mean distance between atoms behind the shock front.  This simple picture of the difference between an elastic ramp and an elastic shock has been confirmed by MD simulations.  \cite{Higginbotham2012}.

However, once the material starts to yield owing to the high shear stresses induced by the compression, the situation becomes more complex, and the demarcation between a shock and a ramp is less well understood.  In this context it is important to note that for the vast majority of experiments performed to date using optical laser ablation to compress materials for diffraction studies, the compression process, whether ramped or as a shock, is uniaxial in strain.  That is to say the total deformation of the material (deformation being comprised of elastic and plastic components) only changes along the compression direction.  This requires that the so-called deformation gradient matrix, $F$, be diagonal:  
\begin{equation}
    F = F_e F_p = {\rm diag}(1,1, \eta) \quad, \label{E:deformation}
\end{equation}
where $F_e$ and $F_p$ are the elastic and plastic deformation gradient matrices respectively, and $\eta= V/ V_0$ is the volumetric compression experienced by the crystal.

Once uniaxial compression of the material exceeds the yield point of the material (in the case of a shock, this is known as the Hugoniot Elastic Limit, or HEL), then the material will start to plastically flow to relieve the shear stress.  Importantly, the plastic work performed (dictated by the time-integrated product of the relevant plastic stresses and strains) will give rise to heating of the material.  Strictly speaking, therefore, it is impossible to uniaxially compress a material to the ultra-high pressures (above the HEL) and keep the material on an isentrope - there will always be some degree of plastic work performed, and thus heat dissipated in the system,~\cite{Swift2005} and this is why the ramp compression path depicted in Fig. \ref{F:Schematic} lies between the isentrope and the Hugoniot.  In the case of a shock, the amount of work that must be dissipated is, to a good approximation, dictated by the integral of $P dV$ between the Rayleigh line (this is the straight line connecting the initial to final state, as shown in Fig. \ref{F:Compression}) and the cold compression curve.  Note this is only an approximation, as this neglects any residual strength that the material may still exhibit on the Hugoniot.  However, and importantly, within this approximation we see from the Hugoniot relations that if two shock parameters are measured, we know the increase in energy of the shock compressed material, and thus, when combined with knowledge of the cold compression curve, we can make reasonable estimates of the temperature.

For ramp compression, however, the situation is very different.  Unless we know the integral of the plastic work done in relieving the shear strains built up during uniaxial compression, we have no information whereby we can plausibly estimate the temperature rise.  Thus, crucially, we find that we need to understand strength at high pressure, and at high strain rates if we are to predict where in temperature space, between the isentrope and the Hugoniot, we may lie.  
Indeed, this leads us to understand one of the major problems facing this field: uniaxial compression is favored (in large part as we then have tried and tested methods to determine the longitudinal stress via VISAR), but this inevitably means that plastic flow takes place, but this in turn means that plastic work is dissipated as heat.  Quite how much plastic heating occurs depends on the time history of the plastic stresses and strains - a history that itself will depend on the time-dependent strength of the material.  In the absence of knowledge of this plastic work history, we have no means to reliably estimate the final temperature of the material, and thus do not know exactly where we lie in a ($P,V,T$) phase diagram.  We return to this point in section \ref{SS:Temperature}, where we discuss current efforts to develop methods for temperature measurements, and section \ref{SS:Strength} where we indicate possible directions of enquiry that may lead to an improved study of strength at these ultra-high strain-rates.

What we can assume, however, is that if we wish to keep the degree of plastic work, and thus heating, to a minimum, the material must be compressed on a time-scale that is long compared with that taken for an element of the same material to be compressed by a shock of the same strength.  This in turn leads us back to our original discussion of the underlying differences between shock and ramp compression, but now in the presence of plasticity. It is at this juncture that we appeal to the empirical result, known as the Swegle-Grady relation \cite{Grady1981, Swegle1985, Grady2015}, that when a plot is made the of Hugoniot stress, $\sigma_{\rm H}$, behind the shock versus the maximum strain rate, $\dot{\epsilon}_m$ within the front for a given material, then for a vast array of different materials we find the relation $\dot{\epsilon}_m = A \sigma^4$, where $A$ is a constant that depends on the material.  Furthermore, for a wide range of materials $\dot{\epsilon}_m$ is found to be in the region of 10$^9$ s$^{-1}$ once Hugoniot stresses reach values of order 10 to 100 GPa.  As the peak strains themselves in this region are of order tens of percent, we readily deduce that even in the presence of plasticity the shock rise time in this pressure range is short compared with a nanosecond, and surmise that therefore, if the material is compressed on nanosecond or longer timescales, then the material will not undergo the same degree of heating as it does under shock compression. 

There is compelling evidence that underpins the above picture.  Firstly, the diffraction experiments that have been performed using laser-plasma x-ray sources on the NIF and Omega facilities have readily demonstrated that, when ramp-compressed on the time-scale of several nanoseconds, the material remains solid into the TPa regime~\cite{Rygg2020,Lazicki2021}.  Secondly, MD simulations also indicate that ramp compression, even on rapid timescales (but slower than shock compression) keeps the material cool.  In these simulations single crystals of Cu were compressed up to 250 GPa with a ramp of rise time of 300 psec, and it was shown that the temperature could be kept well below that produced by a shock, at least for the short period of time before the ramp steepened up to the strain rate determined by the S-G relation.~\cite{Higginbotham2011} This meant that only a few tens of nm of material at the surface of the crystal remained solid, but the basic concept was proven.

\begin{figure}
        \includegraphics[angle=270,width=0.95\columnwidth]{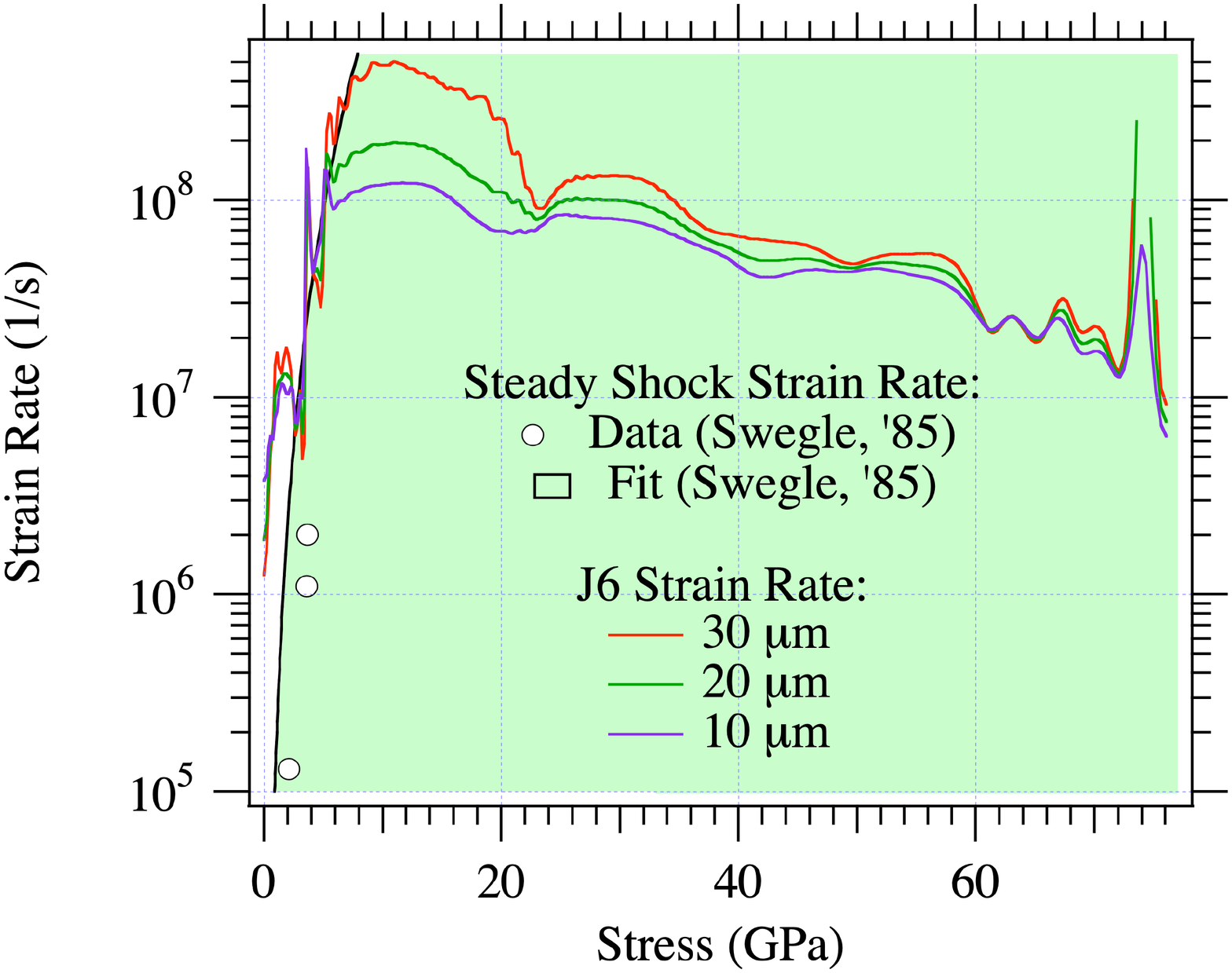}
 
    \caption {A plot of strain-rate (as defined in the text) as a function of stress for ramp-compression waves in laser-shocked Al targets of the three thicknesses shown.  Data taken with permission from the experiment of Smith {\it et al}~\cite{Smith2007}}
    \label{F:Strain-Rate}
\end{figure}

Further insight into when compression is in the ramp rather than shock regime can be gained from monitoring the rate of compression from a VISAR signal (if afforded by the temporal resolution of the instrument). The local Lagrangian sound speed, $C_L$ is a function of the particle velocity, $U_P$, and we can define the strain rate as $\dot \epsilon =[1/C_L(U_P)]*[dU_P/dt]$~\cite{Smith2007}.  The expectation is that a compression wave is in the ramp regime if the strain rate at a specific stress increases with samples of thickness for the same compression rate on the front surface - that is to say that the wave steepens as it propagates, and if the target were thick enough, would lead to shock formation.  We illustrate this concept in Fig. \ref{F:Strain-Rate}, where we show an analysis of data obtained in an experiment where thin foils of Al, of thicknesses of 10, 20 and 30 $\mu$m, were ramp compressed up to 80 GPa~\cite{Smith2007} The foils were mounted on a LiF window, which was well impedance-matched to the Al, and the response of the rear surface of the Al monitored via VISAR, providing a time-dependent particle velocity of the rear surface of the Al via an iterative characteristic method~\cite{Rothman2005}.  Fig. \ref{F:Strain-Rate} shows the local strain-rate (as defined above) as a function of stress for the ramp waves in the three foils.  Also shown in open circles is the maximum steady-shock strain rate measured by Swegle and Grady, and the green shaded area represents regions of stress and strain-rate where we would expect quasi-isentropic behavior to hold, with the demarcation line between the shaded and unshaded areas being determined by the S-G relation.  It can be seen that in these experiments, at the higher stresses above approximately 10 GPa, the strain rates are below those expected in a shock, indicating true ramp-compression. It is interesting to note that the experiments from which these data are taken indicate that compression along the quasi-isentrope revealed a stiffer response than models consistent with static and shock results, highlighting once more the importance of understanding strain-rate dependent strength.  Indeed, if the rate-dependent strength were fully understood, and the total plastic work could be deduced, then accurate predictions of the temperature of ramp compressed matter could be made~\cite{Fratanduono2015}.

We further note that owing to the non-linear response of materials, care is required in designing the pressure (and thus laser) profiles required for ramp compression, to ensure that portions of the ramp do not steepen up into a shock during the process.  Examples of how this might be done for several ablators and materials has been undertaken by a study of the flow characteristics of materials under compression (i.e. as a function of their strength)~\cite{Swift2008}.

\subsection{Strength and Plasticity}
\label{SS:Strength}

A recurrent theme throughout our discussions has been the fact that in order to keep matter solid at high pressures, the rate of dynamic compression should be such that the thermodynamic path followed is close to that of an isentrope, rather than the sample being shocked. Given (at least to date) most dynamic studies employ uniaxial compression, such that the total strain (plastic plus elastic) transverse to the compression direction is zero, the sample potentially experiences high levels of shear stress, which must be relieved either via plastic flow, or via the process of polymorphic phase transitions, or both.  The plastic work experienced during the relief of this shear stress will, to a large part, dominate the temperature rise of the sample.  Some plastic work on the sample will also be expended in the generation of defects, with the fraction converted to heat being known as the Taylor-Quinney factor~\cite{Taylor1934}, with the thermal component thought to dominate, even at high strain rates and high defect densities~\cite{Mason1994}, and as seen in MD simulations~\cite{Heighway2019}.  The temperature rise during dynamic compression is thus intimately linked to material strength.

The incorporation of strength models into hydrodynamic descriptions of material response under dynamic compression has a long history.  The main thrust of early implementations has been to determine expressions for the yield strength, $Y$, in the von Mises sense, and of the shear modulus $G$, as a function of the relevant parameters, i.e. strain, strain-rate, temperature and pressure.  Two well-known models of this form are those due to Steinberg and Guinan~\cite{Steinberg1980}, and Preston, Tonks and Wallace~\cite{Preston2003}.

Ultimately, one would like to understand the physics of plasticity across a multitude of scales.  The motion of one atom with respect to another, and thus the stresses required for plasticity, is ultimately linked to the strength of atomic bonds, which break during dislocation motion.  At another level, the velocity at which dislocations move as a function of shear stress plays a significant role, as does their number, given that the plastic strain rate is governed by Orowan's equation.  Dislocations can also interact with each other, leading to hardening effects.
This approach, whereby information about bonding at the atomic level is determined via, for example, density functional theory, and dislocation mobility by classical molecular dynamics calculations and so forth,  is known as the multiscale approach, and a review of such methods has been provided by Rudd~\cite{Rudd2014}, and one detailed form of this approach has been constructed for tantalum, and has become known as the Livermore Multiscale Model~\cite{Barton2013}, discussed further below.

We note that if a given strength model can be trusted, then a calculation of the plastic work expended under ramp compression can be made, and thus the temperature rise predicted.  For example, in work on the NIF where Ta was ramp-compressed to approximately 600 GPa, the wave profiles determined from VISAR and the iterative Lagrangian analysis were compared with those from Lasnex hydrocode simulutions, where the hydrocode incorporated the Steinberg-Guinan model~\cite{Fratanduono2015}, good agreement between predicted and observed wave profiles enabled the temperature rises to be deduced.  In these experiments, where, after an initial small shock, the Ta target was compressed on a timescale of order 10 nsec. As the ramp wave steepens during the compression, different parts of the target experience different compression rates, and thus have their temperatures raised by differing amounts.  As an example, for compression to 600 GPa the code indicated that shock compression would take the material above the melting point, and to a temperature of 24,000 K, whereas under ramp compression, at a position 63 $\mu$m from the surface at which pressure is applied, the temperature  would have been 3850 K, keeping the material solid.

As said, a more sophisticated model than the Steinberg-Guinan model referred to above has been developed by workers at LLNL, and is known as the
 Livermore Multiscale Model (LMS)~\cite{Barton2013}. It is designed to bridge various length scales by, for example, obtaining data about dislocation mobility from molecular dynamics simulations, and then using an analytic form based on this data as input to the model.  In this model, thus far applied to tantalum, a number of slip systems $\alpha$ are taken into account, and the rate of an amount of slip on each system, $\dot{\gamma}^{\alpha}$ assumed to follow Orowan's equation:  $\dot{\gamma}^{\alpha} =\rho^{\alpha}b^{\alpha}v^{\alpha}$, where $\rho^{\alpha}$ is the mobile dislocation density, $b^{\alpha}$ the magnitude of the dislocation density, and $v^{\alpha}$ the average velocity of mobile dislocations on the relevant slip system, with this velocity being a function of the effective resolved shear stress, $\tau^{\alpha}_{\rm eff}$, with this relationship being fitted to MD simulations of screw dislocation mobility.  The multiplication and annihilation of dislocations is taken into account by assuming
\begin{equation}
    \dot{\rho}^{\alpha} = R \left (  1 - \frac{\rho^{\alpha}}{\rho^{\alpha}_{\rm sat}(\dot{\gamma}^{\alpha})}   \right ) \quad ,
\end{equation}
where $R$ is specific to the material, and $\rho^{\alpha}_{\rm sat}$ a saturation dislocation density that depends on the strain rate, and is constrained by larger scale dislocation dynamics models.
The model has been used in hydrodynamic simulations  with some success to predict the growth rate of the Rayleigh-Taylor hydrodynamic instability in laser-driven targets~\cite{Remington2012,Remington2019}.

Ideally, as noted above, one would like to be able to use such models to predict temperature rises in ramp compressed targets, and compare those calculated with values deduced experimentally.  However, as we discuss further in section \ref{SS:Temperature}, accurate temperature measurements are proving challenging, and remain an area for further investigation and development.  That said, direct links between models such as the LMS model and experimental results can be made, in that as the model takes into account specific slip systems and the degree of plasticity active on each of them, it as such encodes information on the way in which plastic flow changes the texture of the target.  It is in this context that recently a version of the LMS model has been incorporated into a crystal plasticity finite element code by Avraam and co-workers~\cite{avraam2021},  and then used to predict the lattice rotation observed in the Ta experiments of Wehrenberg discussed in section \ref{SS:Plasticity}.  As the finite element code keeps track of the orientation of lattice planes with a spatial resolution much smaller than the size of a grain, but can incorporate multiple grains, simulated diffraction patterns can be generated that can be compared directly with the experimental data.  

\begin{figure}
        \includegraphics[angle=270,width=0.99\columnwidth]{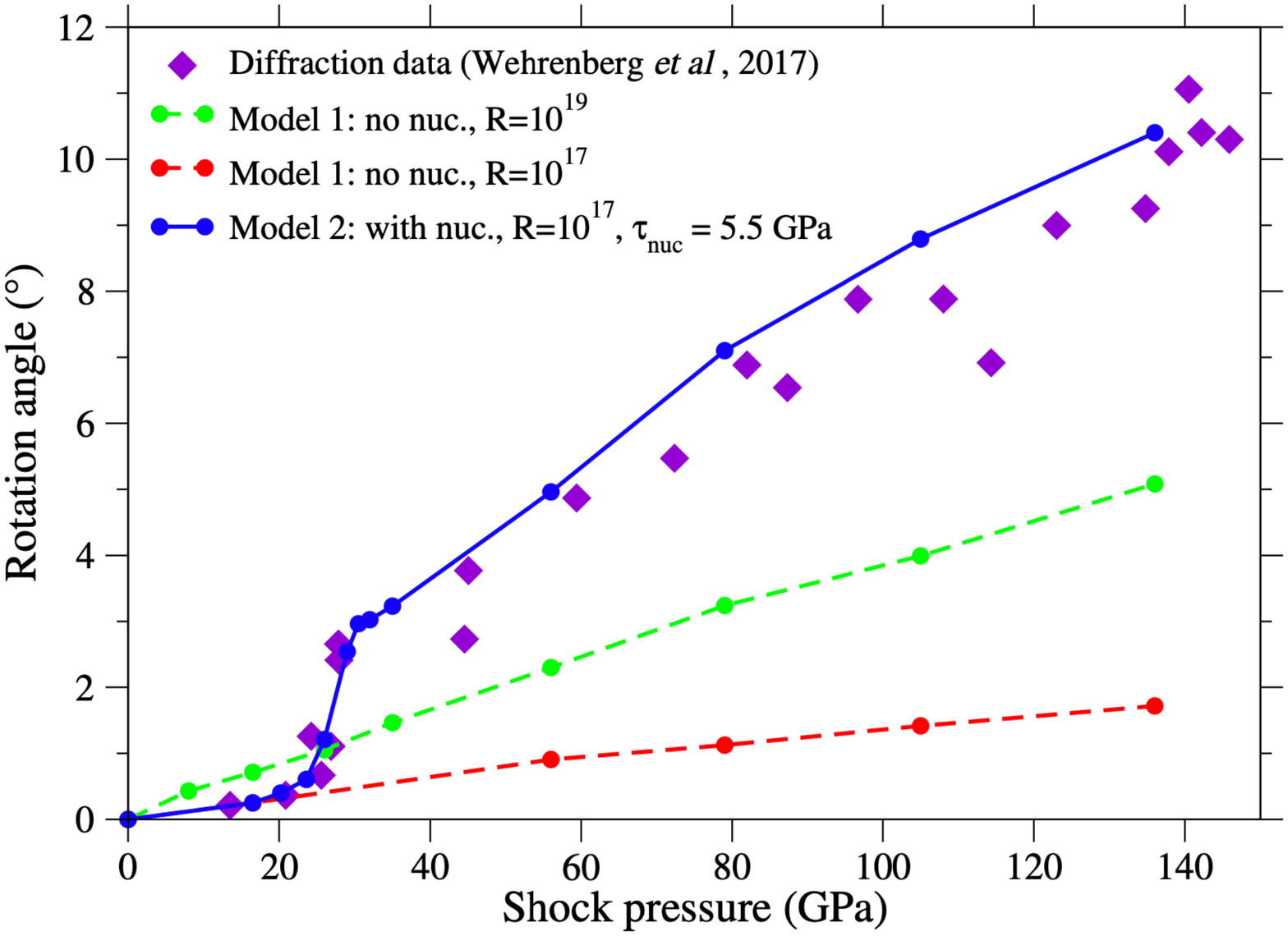}
    \caption {The predictions of lattice rotation from the crystal plasticity finite element modelling of Avraam {\it et al.}~\cite{avraam2021} . Model 2 is able to capture the low and high pressure regimes. Model 1 is based on the LMS model and accounts for dislocation multiplication and annihilation, with multiplication rate controlled by parameter R. Model 2 additionally incorporates a nucleation process, with nucleation threshold T$_{nuc}$ = 5.5 GPa, as described in the text.}
    \label{F:Avraam}
\end{figure}

In order for the predicted lattice rotations to match the experimental data, Avraam {\it et al.} needed to add a further term to the LMS model in order to take into account the homogeneous nucleation of dislocations.  When this term was included, excellent agreement was found, as can be seen in Fig. \ref{F:Avraam}.  In model 2 shown in the figure, nucleation of dislocations occurs when a threshold stress of 5.5 GPa is reached at the shock front.  It can be seen that this model correctly predicts the `knee' in the graph of the rotation angle versus shock pressure, occurring around 27 GPa.  What the authors find is that in the lower pressure regime the kinetics are relatively slow resulting in two competing slip systems acting in a balanced way, giving rise to a relatively small net lattice rotation, whereas above the nucleation threshold the kinetics are faster, with one slip system dominating, leading to a larger net lattice rotation, though significant plasticity occurs in both regions.

It is also worth noting that even prior to the work of Avraam, it had been pointed out that the models used originally to model the Wehrenberg data were somewhat deficient, in that they assumed that only a single slip system was active in each grain, and relied on the classic Schmid model~\cite{Schmid1926,Schmid1935} for lattice rotation discussed in section \ref{SS:Plasticity}.  This model assumes that  line elements of the material parallel to the compression axis remain conserved.  In previous laser-plasma shock experiments that viewed rotation in single crystals~\cite{Suggit2012} the Taylor model was used~\cite{Taylor1934}, where line elements perpendicular to the compression access are conserved.  However, it can easily be shown that the situation is more complex for single slip when the deformation of the crystal takes place under the boundary conditions of uniaxial compression, and Heighway and Wark have shown that neither of these classic models correctly predicts the degree of rotation for a single crystal subject to uniaxial compression and undergoing single slip~\cite{Heighway2021}; furthermore, there is good evidence from MD simulations that multiple slip systems are active in these experiments~\cite{Heighway2022}.  Of course, in real polycrystalline materials, such as those observed in the Wehrenberg experiments, whilst the condition of uniaxial strain must hold globally, it need not hold locally owing to grain-grain interactions~\cite{Heighway2019b}.  It is here that the finite element crystal plasticity model shows its advantages, in that it can model a large number of realistically sized grains, something that is at present still beyond the computational capabilities of MD simulations.

In our view the work referenced above is only just beginning to give us an insight into deformation processes that give rise to temperature rises under ramped dynamic compression.  Where some insight has been gained thus far it is mainly in simple metallic systems.  Little attention has yet been given to the underlying plasticity mechanisms that come into play in more complex materials, and perhaps particularly materials that are covalently bonded, where different processes may be at play.  For example, in the ramp compression of diamond targets to 2 TPa, discussed in section~\ref{SS:Some-History-JSW}, the authors note that the observation of solid diamond at these pressures implies that the material is less strong than they initially thought.  Whilst some work has been performed to investigate this, it still largely relies on the input of empirical parameters~\cite{Gonzalez2021}, and possibilities such as a low Taylor-Quinney factor have also been suggested\cite{Swift2022}. 

In conclusion of this section, therefore, we would maintain that an increased understanding of plasticity at the lattice level is not only necessary from the perspective of those interested in the fundamentals of rapid deformation for its own sake, but is a necessary prerequisite for providing a capability of predicting the temperature rises to which materials will be subject under uniaxial loading at different strain rates.

\subsection{Temperature Measurements}
\label{SS:Temperature} 

Under dynamic compression, the density of the target can be determined from the diffraction pattern, and the pressure (or more accurately, the component of the stress along the shock propagation direction) can be deduced from VISAR.  However, as mentioned in several places above, accurate measures of the temperature have yet to be developed.  This problem is particularly acute in the case of ramp compression, where the degree of plastic work experienced by the sample is hard to quantify.  In contrast, in the case of shock compression, one has some measure of the total internal energy imparted to the sample by the shock, as this is the difference between the Rayleigh line and the low temperature isotherm, as shown in Fig. \ref{F:Compression}.  Indeed, such a knowledge of the energy budget, along with other experimental parameters, has recently been used to constrain the melting temperature of Ta under shock compression~\cite{Kraus2021}

One method that has been pursued for many years in the field of shock physics is optical pyrometery~\cite{Hereil2000}, whereby the thermal emission from the rear surface of a sample is recorded (often through a window that is transparent). However, it has been shown that these experiments are hampered by small optical depths and thermal conduction effects, such that the surface temperature observed  may not be representative of the bulk temperature~\cite{Brantley2021}.  The technique requires not only very accurate calibration, but also a knowledge of the wavelength-dependent emissivity of the sample, which itself may change upon compression.  That said, at high enough temperatures, above several thousand Kelvin, a combination of reflectivity measurements with pyrometry can allow good temperature measurements to be made at the relatively high temperatures found along the Hugoniot~\cite{Millot2018}, though the technique is less suitable for ramp compression, where the sample is kept much cooler and thermal diffusivity effects are always present.

It has been proposed that temperature could in principle be deduced via elastic scattering by monitoring the relative intensity of Bragg peaks and utilising the Debye-Waller (DW) effect.  This effect exhibits itself in a relative reduction of the intensity of the diffracted x-rays as a function of the reciprocal lattice vector associated with the Bragg scattering, such that, in a simplified form of the DW effect for a material with a monatomic basis,  the scattered intensity is given by
\begin{equation}
    I \propto I_0 \exp \left ( -2 B \sin^2 \theta / \lambda^2 \right )  \quad , \label{E:DW}
\end{equation}
where $ B = 8 \pi^2 \langle u^2 \rangle$ with $ \langle u^2 \rangle$ being the mean squared displacement of the atom from its average position.
The physics of the effect is that thermal motion implies an rms deviation of an atom from its lattice point, and thus the x-rays are not scattering from a perfect lattice. When the momentum transfer of scattered x-rays increases (as it does within increasing reciprocal lattice vector, or $\sin \theta / \lambda$), a given displacement of an atom from its ideal position induces a larger random phase into each scattered x-ray, reducing the overall observed intensity.  The displacement of the atoms from their perfect positions is a function of both the temperature, and the Debye temperature, $\theta_{\rm D}$ of the material (given that the Debye temperature is related to the speed of sound, and hence bond stiffness).  This in turn means that the exponent in eqn. \ref{E:DW} is proportional to $G^2_{hkl} T / \theta_{\rm D}^2$, where $G_{hkl}$ is the reciprocal lattice vector of the relevant Bragg reflection.

Very few studies of how the Debye-Waller effect might exhibit itself in dynamic compression experiments have been undertaken using femtosecond x-ray diffraction, although some initial computational analysis has been embarked upon~\cite{Murphy2007,Murphy2008}.  Some work has been performed in using the same phenomenon in absorption spectroscopy experiments, given that the same physical effect occurs in Extended X-Ray Absorption Fine Structure (EXAFS). In such experiments the absorption coefficient above an x-ray absorption edge is modulated due to scattering of the photo-ejected electrons from surrounding atoms, and their deviation from perfect lattice positions exhibits itself in a similar temperature dependent damping of the modulations in the absorption coefficient.  This technique has been used by to deduce the temperature of iron ramp compressed to 560 GPa~\cite{Ping2013,Ping2016}, although the relative errors in the measurement in those experiments were quite high, often exceeding $\pm 20 \%$.

It could be argued that using the DW approach in elastic Bragg scattering to measure temperature may be hampered by texturing effects: the texture of a target will evolve under compression, as lattice planes can rotate under compression due to slip.  How this occurs under uniaxial conditions is only just starting to be investigated~\cite{Heighway2021}, but if the degree of texturing can be understood, and indeed retrieved from the diffraction data, the non-random orientation of the grains within the sample could then be accounted for any DW analysis.  We note that texture should not be such an issue in EXAFS measurements, as here the DW effect is occurring at the local level, and the absorption is independent of orientation.  However, in both cases the extraction of temperature relies on accurately knowing the effective Debye temperature of the system, which is itself a function of compression -- that is because what is actually deduced from the relative intensity of the Bragg peaks is $G_{hkl}^2 T/ \theta^2_{\rm D}$, rather than the absolute temperature itself.  However,  information about $\theta_{\rm D}$ is implicit in the knowledge of the pressure-density curve (given that the Debye temperature is also related to the local speed of sound).  That said, several static studies on highly defective nanocrystalline material have shown that the defects themselves can modify the effective Debye temperature by some considerable margin, further complicating this approach to temperature measurement~\cite{Scardi2017}, and this has been confirmed in the case of materials under shock compression~\cite{Sharma2021}.  Whilst further investigations are required, using the
relative intensity of Bragg peaks to provide temperature information remains an attractive prospect, as such measurements could in principle be made on a single-shot basis.

We mention in passing one potentially interesting aspect of the DW effect in the context of ramp compression.  Whilst the temperature and length of the reciprocal lattice vector will increase under ramp compression, in general so will the Debye temperature, owing to the curvature of the $PV$ hydrostat.  It is thus pertinent to ask which effect dominates - because if the increasing $\theta_{\rm D}$ is most important, it is possible that higher order diffraction peaks will actually get relatively brighter under compression.  Indeed, it can be shown that were the compression perfectly isentropic (which it will not be, owing to plastic work), then in a simple Gr{\"u}neisen model of a solid, this effect will occur for most solids, as it merely requires a Gr{\"u}neisen parameter greater than 2/3~\cite{Murphy2008,Wark2020}.

In the context of the difficulties in extracting temperatures using the methods discussed above, a novel method (for dynamically compressed solids) that has been proposed is that of the inelastic scattering of x-rays from phonons.  As well as the elastic scattering observed when x-rays meet the Bragg condition, incident x-rays of energy $\hbar \omega_i$ and of wavevector ${\bf k}_i$ can scatter from phonons of wavevector $\bf q$  and energy $\hbar \omega_{\rm ph}$, with the scattered x-ray photon having wavevector ${\bf k}_s = {\bf k}_i +{\bf G}_{hkl} \pm {\bf q}$, and frequency $\omega_s = \omega_i \pm \omega_{\rm ph}$, with the $\pm$ denoting the absorption or creation of a phonon during the process.  We note that the most energetic phonons in the sample will have energies of order $k_{\rm B} \theta_{D}$, which, given typical Debye temperatures are of the same order of magnitude as room temperature, implies phonon energies of up to a few 10s of meV.  In contrast, the incident x-rays have energies of many keV, and thus the change in energy of the x-ray upon scattering is extremely small, and the spectral resolution required for such a measurement must be high, with $\Delta E / E$ better than $10^{-5}$ by some margin.  

However, given that the relative intensity of the Stokes and anti-Stokes peaks of inelastically scattered x-rays (i.e. photons that during the scattering process have either created or annihilated a phonon) are determined by the principle of detailed balance for a system in thermodynamic equilibrium\cite{Vieillefosse1975,Burkel2000,Sinha2001,Scopigno2005}, if such a technique could be employed during dynamic compression it would give an absolute temperature measurement, and not be reliant, for example, on knowledge of $\theta_{\rm D}$~\cite{McBride2018, Descamps2020}. However, the transient nature of the compression process means the crystal is only in the state of interest for a few nanoseconds at most. Owing in part to the very high resolutions required, inelastic x-ray scattering measurements are `photon-hungry'\cite{Sinha2001}, and it might seem that the prospect of retrieving information about the phonon modes on such timescales is unfeasible. Nevertheless, despite the considerable challenges that such measurements pose, progress is being made towards the development of inelastic scattering techniques on these short timescales. As noted above, FELs have an enormous peak spectral brightness. Indeed, recent work on both the Linac Coherent Light Source (LCLS)\cite{McBride2018} and at the European XFEL (EuXFEL)\cite{Descamps2020,Wollenweber2021} has demonstrated inelastic scattering from static, unshocked samples of almost perfect diamond single crystals. The temperature of samples at both ambient and elevated levels could be determined with an error of less than $8 \%$. In these experiments the photon numbers were low and just a few photons were collected for every FEL shot, but this signal level can be increased significantly by seeding of the FEL\cite{Feldhaus1997,Saldin2001}. 

These initial experiments at LCLS and EuXFEL were conducted at an x-ray energy of 7.49 keV, utilising a Si (533) analyser crystal.  The resolution of the setup was approximately 44 meV.  As with inelastic scattering measurements performed at synchrotrons, this can in principle be improved further (at the expense of a further reduction in photon numbers) by, for example, using a higher x-ray energy and, given the analyser operates with a Bragg angle of close to $\pi/2$ in order to maximise dispersion, an analyser crystal with a smaller lattice spacing~\cite{Baron2009,Burkel2000}.  For example, resolutions better than 3 meV have been achieved at synchrotrons using the Si (999) reflection at 17.8keV~\cite{Masciovecchio1996}.

Whilst the static measurements performed at FELs cited above shows promise, there are other complications that may need to be born in mind.  In these inelastic scattering measurements, the x-rays are collected between the Bragg peaks.  However, there will still be some elastic scattering present at this position due to defects in the sample, and the dynamic compression process itself will introduce significant defects owing to the plastic deformation processes incurred. To date little consideration has been given to this problem -- it could be that the intensity of this quasi-elastic scattering could overwhelm any inelastic signal, unless the resolution of the set up is sufficient.  Preliminary computation work has started in this area, and MD simulations of shocked single-crystals of Cu have demonstrated the effects of the copious stacking faults on the observed signals~\cite{Karnbach2021}.

Even with seeding it is likely that inelastic scattering experiments will need multiple shots to accumulate sufficient signal for accurate temperature measurements. Thus this technique will require combination with a high-repetition-rate high-energy nanosecond optical laser for compressing samples if experimental data are to be acquired on realistic experimental timescales. Indeed, this is one of the motivations for the building and placement of the DiPOLE laser\cite{Mason2017} at the EuXFEL, DiPOLE being a 100~J diode-pumped laser capable of delivering pulses of a few nanoseconds duration at 10~Hz.  As noted above, it may be that useful temperature information could be extracted more easily by use of the DW effect, which only requires a single shot, but in any event inelastic scattering experiments could prove extremely useful in that they should allow extraction of some information about phonon dispersion curves under compression (which themselves could provide information on the effective value of $\theta_{\rm D}$).  It has also been pointed out that very high inelastic scattering, if employed away from a forward scattering geometry, would be sensitive to the bulk (particle) velocity of ramp compressed material, in that the scattered signal will be Doppler shifted by a degree detectable with these high resolutions~\cite{Karnbach2021}.  Whilst the accuracy of such a particle velocity measurement cannot compete with VISAR, it would comprise a near instantaneous measurement within the material, rather than relying on measurement of a surface or interface velocity.

Alongside an understanding of the plasticity that gives rise to the major part of the temperature increase under uniaxial compression, we consider the development of techniques to measure such temperature rises accurately {\it in situ} to be an area ripe for further development.  As we have noted above, each technique proposed to date appears to have its advantages and disadvantages, but as of yet no clear favored candidate has emerged.

\subsection{Phase boundaries under dynamic compression}
\label{SS:Phase_Boundaries}
As noted previously, it has long been known that the locations of the phase boundaries between different high-pressure phases can be dependent on the rate of compression \cite{Smith2008}. However, early structural studies on Bi and Sb at the LCLS \cite{Gorman2018,Coleman2019}, which have directly determined the different structural forms obtained on shock compression, have shown that not only do the phase boundaries move, but also that one can obtain structures under dynamic compression that are not been seen in longer-timescale DAC experiments (see Figure \ref{fig:Bi}). Similarly, early FEL experiments have also shown phases that are seen in static compression studies are sometimes not seen on dynamic compression. One example is in H$_2$O, where ice-VI, the tetragonal form of H$_2$O ice seen above 1 GPa on the slow compression of liquid H$_2$O at 300 K, is not seen on compression on either millisecond \cite{Lee2006} or nanosecond \cite{Gleason2017} timescales, but rather that super-compressed H$_2$O crystallises directly into the higher-pressure ice-VII at $\sim$ 2GPa. Further ramp compression experiments over 100s of nsec at the Z-machine \cite{Dolan2007} have shown homogeneous crystallisation of the liquid directly into ice VII at 7 GPa, while nanosecond isentropic compression at Omega reported the liquid-ice-VII transition to occur at at least  8 GPa \cite{Marshall2021}.
The ability to determine the actual crystal structures observed under dynamic compression is providing important new light on the behaviour such materials. However, we have yet to determine when, why and how the changes in phase boundary locations and structures are obtained in materials under uniaxial compression and this is a topic under active investigation. Such changes might be driven by shear effects, or they might be a kinetic effect that is related to the timescale of the experiment. The observation that ice-VI is “missing” in both nanosecond and millisecond timescale experiments provides invaluable input to our current understanding. In any case, great care must obviously be taken in assuming that the structures and physical properties (such as melting temperature) obtained from short-lived experiments can be applied without caveats to our understanding of geoplanetary bodies, the internal structures of which have evolved on geological timescales.

It is likely that effects of material strength and shear will be relevant to developing our understanding in this area. Such effects might be mitigated by the use novel target designs - e.g. by the use of non-planar compression (see below), or by embedding materials of interest in a ‘weak’ matrix \cite{smith2022}, much as DACs use a pressure transmitting medium in order to obtain more hydrostatic conditions.
Further insight will be gained from more systematic studies of P-T space in order to map phase boundaries over extended ranges. At present, phase transitions have been determined mostly along either the Hugoniot or a (quasi)-isentrope (see Fig. \ref{F:Schematic}). Locating phase transitions at other locations (such as location ‘A’ in Fig. \ref{F:Schematic}) will require both inventive use of pulse shaping in order to move away from the Hugoniot and isentrope, and a solution to the problem of measuring the sample temperature, an issue we discussed in the previous section.

\subsection{Prospects for the study of the structure of liquids at ultrahigh pressures}
\label{SS:liquids}

The study of liquid phases and liquid-liquid phase transitions has long been a topic of active interest in the static compression community. While first-order transitions are commonplace in solids, they are very much rarer in liquids. The clearest such transition at high-pressures has been seen in phosphorous, where there is a transition from a molecular-liquid to polymeric form at $\sim$1 GPa and $\sim$1400 K \cite{Katayama2000}. While there have been numerous other claims for liquid-liquid transitions in a range of other materials at high pressure \cite{Brazhkin1997}, both the evidence for, and nature of, the transitions is less clear than in the case of phosphorous.

When trying to determine the existence and nature of such transitions using diffraction, the key issues in static compression experiments are the limited sample volume, the structured background arising from Compton scattering from the diamond anvils, and the limited $Q$-range of the diffraction data, which is constrained by the pressure cell housing and the choice of x-ray wavelength\cite{eggert2002}. As a result, x-ray data are typically only collected to a maximum-Q of 8-10 \AA$^{-1}$, and sometime much lower \cite{Cadien2013}, as opposed to the ranges of 25 \AA$^{-1}$  used in ambient-pressure x-ray studies, and values of 30-50 \AA$^{-1}$ obtained in neutron studies \cite{Keen2020}.

Studies of liquid phases of a number of materials have already been made at the LCLS to P-T conditions of 190 GPa and 10,000 K \cite{Gorman2018,Hartley2018,Briggs2019,Morard2020}, and changes in coordination have been reported in Bi \cite{Gorman2018} and Sn \cite{Briggs2019}. Such studies have revealed some of the inherent advantages of using dynamic compression techniques at FELs to conduct such experiments. Firstly, the accessible P-T conditions are greatly extended over those available in a laser-heated DAC, and the timescale of the FEL experiments precludes reactions between the sample and its surrounding, even at these extreme temperatures. Secondly, the volume of the dynamically-compressed sample is much greater than that compressed in a DAC, and the incident x-ray beam is also typically larger, so the diffraction signal is much greater. Thirdly, the scattering from the sample is not constrained by the target environment. And finally, the absence of the diamond anvils, and the vacuum in which dynamic compression experiments are conducted, means that anvil and air scattering are absent. Only scattering from the low-Z ablator, if present, and x-rays resulting from the ablator plasma, provide additional background. Despite these advantages, the $Q$-range of the liquid data that has been collected at the LCLS to date is very similar to that collected from samples in DACs as a result of the current limitations in the lower x-ray wavelengths at FELs.

Such limitations will be less constraining in the future, with the availability of higher-energy radiation at both FELs such as EuXFEL and LCLS-HE, and synchrotrons such as PETRA-IV. On the HED beamline at EuXFEL a $Q$-range up to 12 \AA$^{-1}$ can be obtained using a 20 keV x-ray energy, which, while an improvement, is still well short of the range accessible at ambient conditions of 25-30 \AA$^{-1}$ \cite{Keen2020}.

While the use of higher energy x-ray harmonics from the FEL and synchrotron undulators would provide access to an extended $Q$-range, this would be accompanied by a decrease in flux and scattering intensity (which scales as $\lambda^3$). However, for molten phases, especially hot molten phases accessible along a Hugoniot, there is probably little point in extending the $Q$-range to 25 \AA$^{-1}$ and above as there will no structural features in the data at these momentum transfers. But data at high
values of $Q$ do aid data normalisation as they help in scaling the scattering data to follow the average total scattering curve of the sample. This means that the high-$Q$ limit of the scattering is better behaved and, as a result, the Fourier-transformed pair distribution function will contain fewer of the Fourier truncation ripples that arise through the limited $Q$-range of the data.

\subsection{Novel Material Synthesis}

The recovery of cubic diamond from shocked graphite was first reported 1961, sparking great interest in shock-induced transitions in carbon \cite{DeCarli11961}. The possibilities of using shock compression to generate novel polymorphs was given further impetus in 1967 by the discovery of a shock-induced hexagonal form of diamond \cite{Hanneman1967}, which had also been synthesized using static techniques \cite{Bundy1967}, and later named lonsdaleite \cite{Frondel1967}, in fragments recovered from Barringer Crater in Arizona. Since these pioneering studies, shock compression has become a standard technique for synthesising 
micro polycrystalline diamond powders (see for example \cite{Fujihara1992}), as well as being used to synthesis other industrially useful polymorphs \cite{KOMATSU1999}.

The successful generation and subsequent recovery of shock-generated high-pressure forms clearly requires the relevant polymorphs to exist at extreme conditions, and then be stable, or perhaps more likely metastable, at ambient conditions. While early studies utilised a ``shock and look" approach to synthesis, the advent of electronic structure calculations, coupled to structure searching, now enables novel structural forms to be ``found'' on the computer, and their (meta)stability at ambient conditions established prior to experimental study \cite{Oganov2010}.

\begin{figure}[!htb]
\begin{center}
\includegraphics[width=0.95\columnwidth]{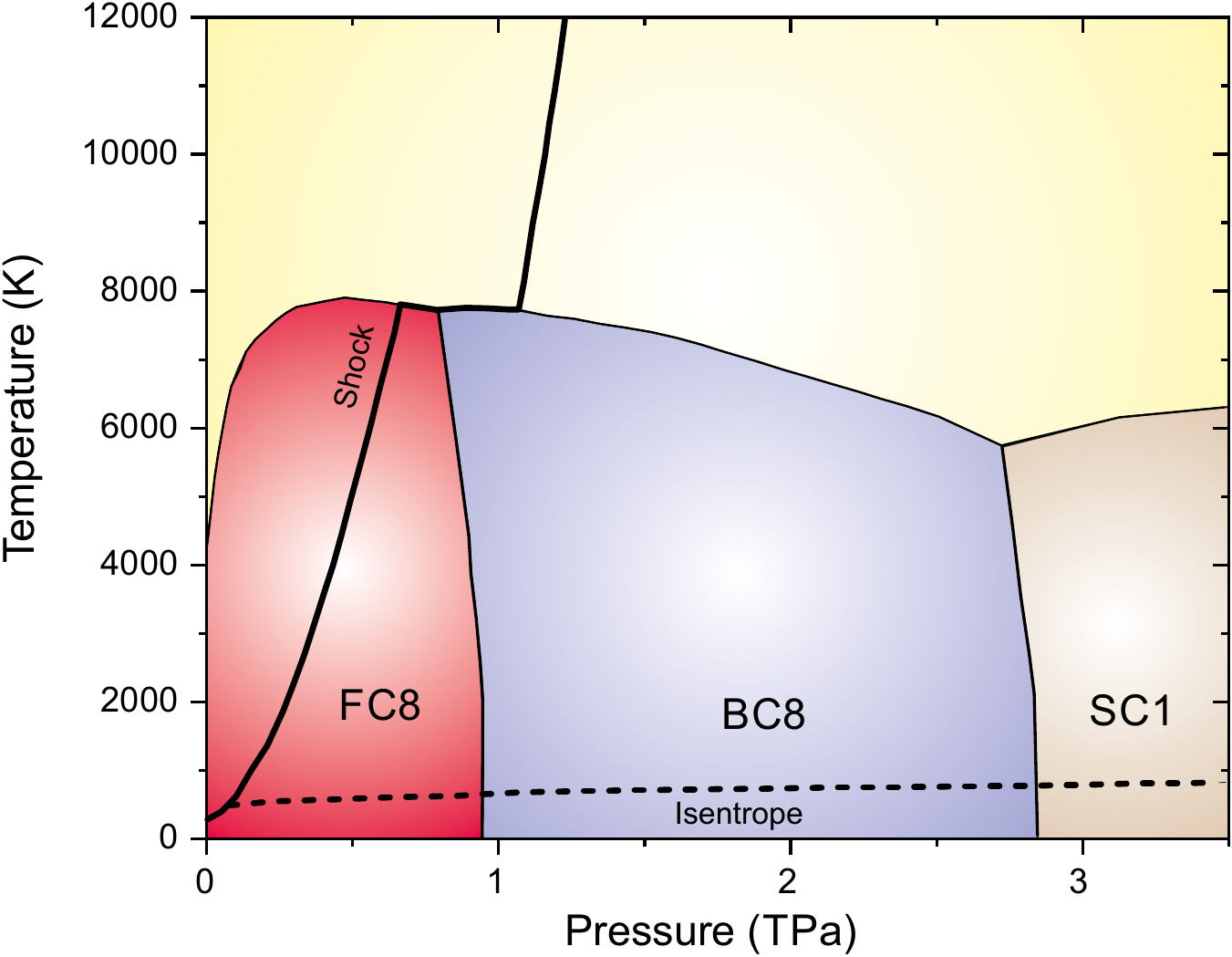}
\caption{The phase diagram of carbon to 3.5 TPa and 12,000 K, showing the DFT predicted phase boundaries between the diamond (FC8), cubic-BC8 and cubic-SC1 phases, and the Hugoniot and isentrope. Adapted from Figure 1 in Ref, \cite{Lazicki2021}.}

\label{F:Carbon}       
\end{center}
\end{figure}

A key example is the so-called BC8 phase of carbon, a metastable structural form long known to be created on pressure decrease from the high-pressure phases of Si and Ge \cite{Wentorf1963,Nelmes1993}. The same BC8 structure is calculated to be the stable and metallic form of diamond above 1.2 TPa \cite{Biswas1984,Biswas1987}, and so its equation of state and physical properties are of great importance for devising models of Neptune, Uranus, and white dwarf stars, as well as of carbon-rich exoplanets \cite{Madhusudhan2012,Mashian2016}. BC8-C is also calculated to a superhard material, and to be metastable at ambient conditions \cite{Mailhiot1991}. Its synthesis might thus provide a new important industrial abrasive.  A schematic of the carbon phase diagram is shown in Fig. \ref{F:Carbon}.

Experimental searches for BC8-C have so far been unsuccessful to pressures as high as 2 TPa, well in excess of the calculated equilibrium phase transition pressures of 1 TPa \cite{Lazicki2021}. At present, only ramp compression experiments at the NIF are capable of generating the
P-T conditions necessary to synthesizing BC8-C on compression. And if such experiments are successful in making the high-density phase, then the sample volume recovered will be measured in $\mu$m$^3$

However, the recent development of high repetition rate lasers now offers the exciting opportunity of the mass production of shock-synthesised materials at lower pressures. For example, silicon possesses a number of dense polymorphs that can be recovered as metastable phases at ambient pressure after initial compression to $>$15 GPa \cite{Haberl2016}. The dense R8 polymorph of Si, a rhombohedral distortion of BC8-Si, is a narrow bandgap semiconductor which is predicted to have a greater light absorption overlap with the solar spectrum than diamond-structure Si \cite{Malone2008}, making it well matched for solar power conversion. As said, the DiPOLE laser installed on the HED beamline at the European XFEL has a repetition rate of 10 Hz, and with suitable targets will enable high-density Si polymorphs to be recovered in cm$^3$
volumes. Indeed, in the case of BC8-Si, more material could be synthesized in 30 seconds using DiPOLE than has been made worldwide since its discovery in 1963 \cite{Wentorf1963}.

Our ability to predict what phases will be observed in such experiments has been greatly improved by the wide spread use of structure prediction methods to determine the lowest enthalpy structure at a particular pressure \cite{Oganov2010,Oganov2019}. Such predictions have been made to multi-TPa pressures in some materials, and even to a petapascal in carbon \cite{Martinez-Canales2012}, well in excess of those currently accessible via experiment. However, such calculations are typically conducted at 0 K.

Fortunately, many structure searching studies also now publish lists of candidate structure which, while not having the lowest enthalpy (at 0 K) at any pressure, are competitive in enthalpy - see for example \cite{Pickard2010,He2018}. Such structures then offer finite-temperature alternatives 
to those identified at zero-temperature. By precalculating the diffraction patterns from such structures, and comparing them to those observed in experiments, it will be possible to interpret the high rep-rate diffraction data in real time.

\subsection{Small Angle Scattering and Line Profile Analysis}
We note that the diffraction that we have been discussing up to now has been wide angle scattering (WAXS), that is to say Bragg diffraction, allowing the discernment of crystal structure.  In addition to such scattering, however, information obtained with much lower momentum transfer could also be of great use in dynamic compression studies, as small angle scattering (SAXS) provides information on irregularities within the crystal that exist on length scales far greater than the lattice spacing.  To date very little work has been performed with FELs using SAXS to study shock or ramp compressed samples, however, that which has been undertaken has been highly successful.  Coakley and co-workers used SAXS (in combination with WAXS) to monitor void evolution and growth during the rarefaction and spall phase that ensued when a laser-shocked copper target unloaded~\cite{Coakley2020}.

Our perspective is that the SAXS technique could also find applications in dynamic compression experiments.  For example, one immediate lesson from the work of Coakley {\it et al.} is that the targets they were shock compressing, having been manufactured by vapor deposition, were not initially fully dense - that is to say voids were present.  This could be deduced from the SAXS signal of the unshocked target.  Indeed, their data shows these voids closing up as the target was initially compressed.  This void closure will of itself lead to another heating mechanism, beyond standard plastic work.

A further question of interest is to what degree can we learn about the dislocation densities present under shock conditions from an analysis of the line profiles in WAXS? The strained regions of the crystal that surround dislocations give rise to distinctive changes in the line profiles of the diffracted x-rays, and there is a long history of using line-profile analysis in static experiments to deduce dislocation density.  Furthermore, in the particular case of stacking faults in face-centered metals, the faults give rise to angular shifts in lines with certain values of $\{hkl\}$  which can be used to determine stacking fault densities~\cite{Warren1990,Rosolankova2006}.  This comes about as a stacking fault looks like a thin sheet of hcp material embedded in the fcc lattice, and as such each fault introduces a phase difference between the x-rays and bulk fcc material. Indeed, an analysis of x-ray line shifts and line profiles has allowed stacking fault densities to be measured in laser-shocked gold at pressures up to 150 GPa using 100 psec pulses of x-rays from a synchrotron source~\cite{Sharma2020a}.

\begin{figure}[!htb]
\begin{center}
\includegraphics[width=0.95\columnwidth]{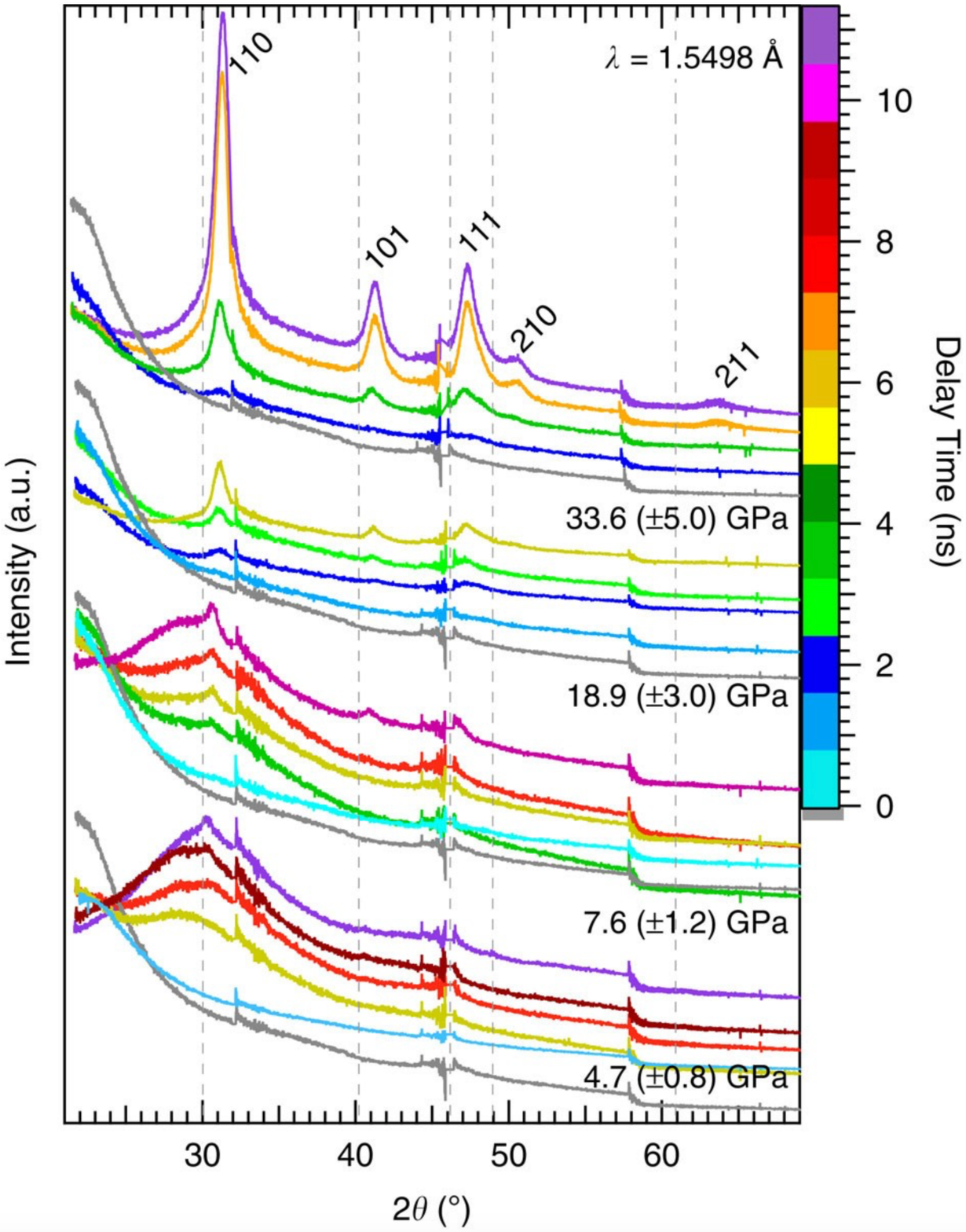}
\caption{Diffraction data showing the emergence of the stishovite crystalline phase from shock compressed amorphous silica for various shock pressures as a function of time.  The stishovite peaks are labelled at the top; ambient condition  positions (grey dashed lines). The traces are clustered according to applied pressure where each colour indicates a different delay time (grey is x-ray only - i.e. the unshocked amorphous phase). The data is offfset along the y-axis for viewing clarity. Discontinuities in the traces are seen at scattering angles (2$\theta$) of 32.5$^{\circ}$, 46.0$^{\circ}$ and 58.0$^{\circ}$ are due to spacing between the application-specific integrated circuits of the detectors.  Diagram taken from the paper of Gleason {\it et al.}~\cite{Gleason2015}}
\label{fig:gleason1}       
\end{center}
\end{figure}

\begin{figure}[!htb]
\begin{center}
\includegraphics[width=0.95\columnwidth]{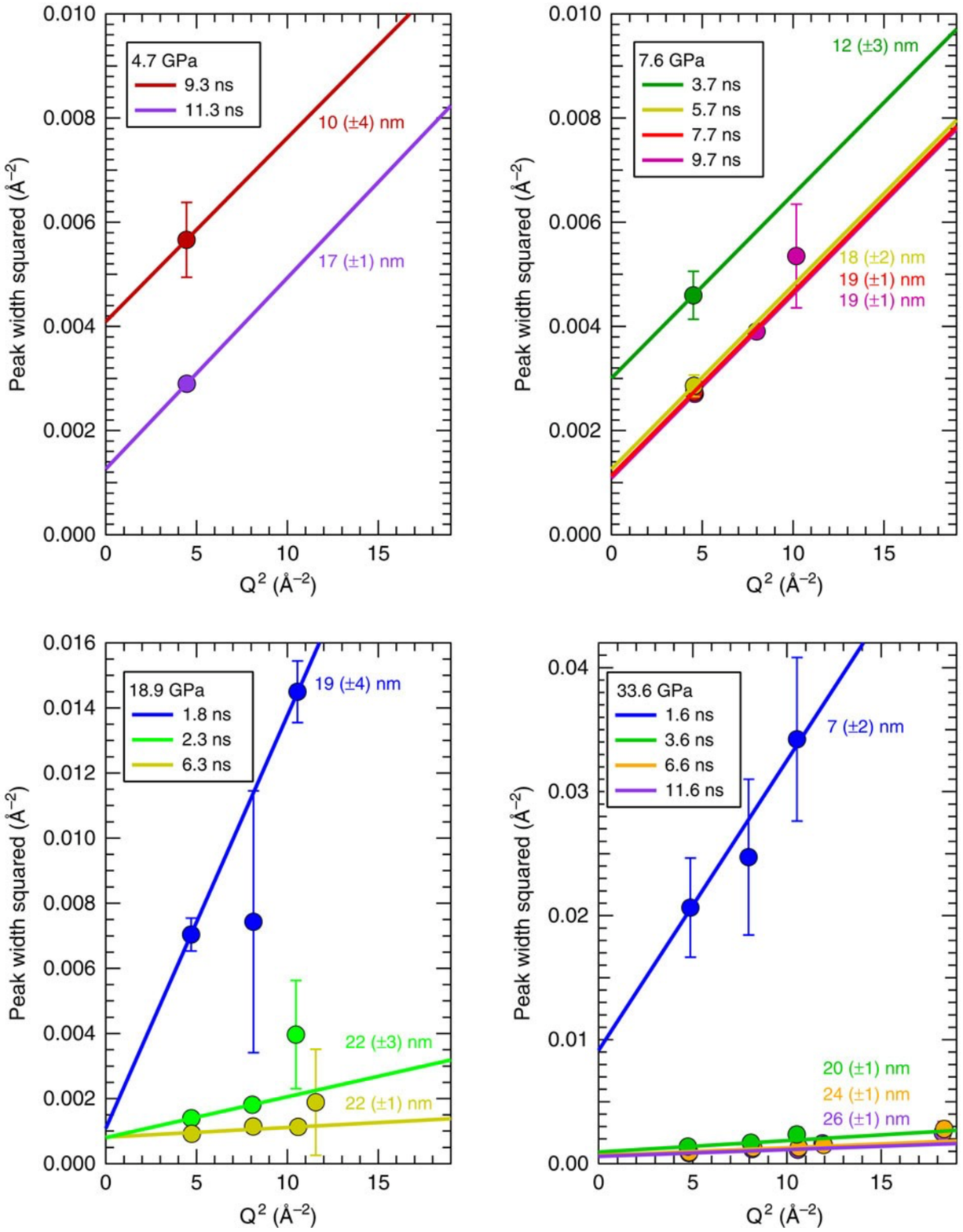}
\caption{The results of an analysis of the peak widths in Fig. \ref{fig:gleason1} showing the peak width as various pressures and times as a function of the momentum transfer, $Q=4 \pi \sin \theta / \lambda$. The solid lines indicate expected peak widths for various grain sizes of the stishovite phase.  Diagram taken from the paper of Gleason {\it et al.}~\cite{Gleason2015}}
\label{fig:gleason2}       
\end{center}
\end{figure}

We envisage that a combination of knowledge of defect density gleaned from line profiles, alongside information on the slip systems which are active, and overall degree of plastic strain, will provide a rich array of information that can inform higher-level plasticity models and our understanding of the dynamic deformation processes.  In addition to providing information on defect densities, the line-profile of the diffracted x-rays can also provide information on the size of the diffracting crystallites, and hence can give insight into the mechanisms and rates of phase transformations.  For example, Gleason {\it et al.} have already used this technique at LCLS to monitor the nanosecond-timescale nucleation and growth of nanocrystalline stishovite grains in laser-shocked amorphous fused silica~\cite{Gleason2015}.  We show the results of those experiments in Figs. \ref{fig:gleason1} and \ref{fig:gleason2}.  Fig. \ref{fig:gleason1} shows the diffraction from the amorphous silica sample, and the evolution on a nanosecond timescale of the distinct diffraction peaks from the crystalline stishovite phase.  There is clearly a narrowing of these peaks as a function of time, and an analysis of their width enabled Gleason {\it et al.} to deduce the growth of the nanometre-scale grains as a function of time, as shown in Fig. \ref{fig:gleason2}.

\subsection{Convergent Geometries}
\label{SS:Convergent}
The combination of dynamical compression techniques with ultrafast x-ray diffraction has, to date, been undertaken almost exclusively in planar geometries, with the diameter of the laser pulse being far greater than the thickness of the material of study, ensuring that the material is compressed under conditions of uniaxial strain.  We have discussed above how it is the relief of the large shear stresses induced by this approach that leads to significant plastic work heating of the target. However, planar compression is inherently inefficient at producing high pressures, and it is well known that pressure amplification occurs in convergent geometries - indeed, this is the basis of the underlying physics behind the design of experiments aimed at achieving laser-driven inertial confinement fusion (ICF)~\cite{NUCKOLLS1972,Lindl1995}, where keeping the fuel on a low adiabat (i.e. closer to an isentrope) is also of importance. The question thus arises as to why such a planar geometry is exclusively used? Our view would be that perhaps the primary reason is that such a geometry allows the use of the VISAR diagnostic (see section \ref{S:Visar}) for pressure measurements, and if such a direct measure of pressure were to be sacrificed for the achievement of higher pressures, there may be some merit in considering  non-planar geometries for future studies.  Whilst such a suggestion might be considered somewhat speculative, we note that the study of high energy density matter up to 50 TPa in the plasma state has been possible in such a geometry, albeit under shock compression~\cite{Doppner2018,Kritcher2020}.  If such an approach were to be used, one might be able to make some reasonable estimates of the pressure if pressure standards could be developed in this regime - for example if the lattice parameters of certain materials, known not to change phase at high pressures, could be tabulated.  In this regard we note recent work that has established such a standard for gold and platinum up to TPa pressures~\cite{Fratanduono2021}.  It is also the case that the vast majority of simulation studies at the atomistic level - that is to say MD simulations - have also almost exclusively concentrated on planar loading, though a small number of studies of other geometries are starting to be made~\cite{Tan2021}, and once more we consider this an area ripe for further exploration.

\subsection{Complementary Techniques: RIXS}
\label{SS:RIXS}

We have discussed above the role that femtosecond x-ray diffraction can play in identifying the crystal structure of novel materials that can be produced under transient high-pressure compression. As noted above, many of these materials, under such conditions, are thought to have interesting electronic properties - e.g. transforming to  electrides, topological insulators or superconductors. Furthermore, it is well known that pressure can transform metals to insulators, and vice-versa~\cite{Mott1982,Ma2009}. The distribution of the electrons in real space (which is essentially the information gleaned by the elastic x-ray scattering process of diffraction) does not of itself provide information on these electrical properties - for that we will need to be able to probe the actual electronic structure of the material whilst it is under high pressure.  In the past electrical or optical properties of matter under transient compression has been gleaned from surface reflectivity measurements~\cite{Millot2018} but the information gathered by this means is extremely limited.

\begin{figure}[!htb]
\begin{center}
\includegraphics[width=0.45\columnwidth]{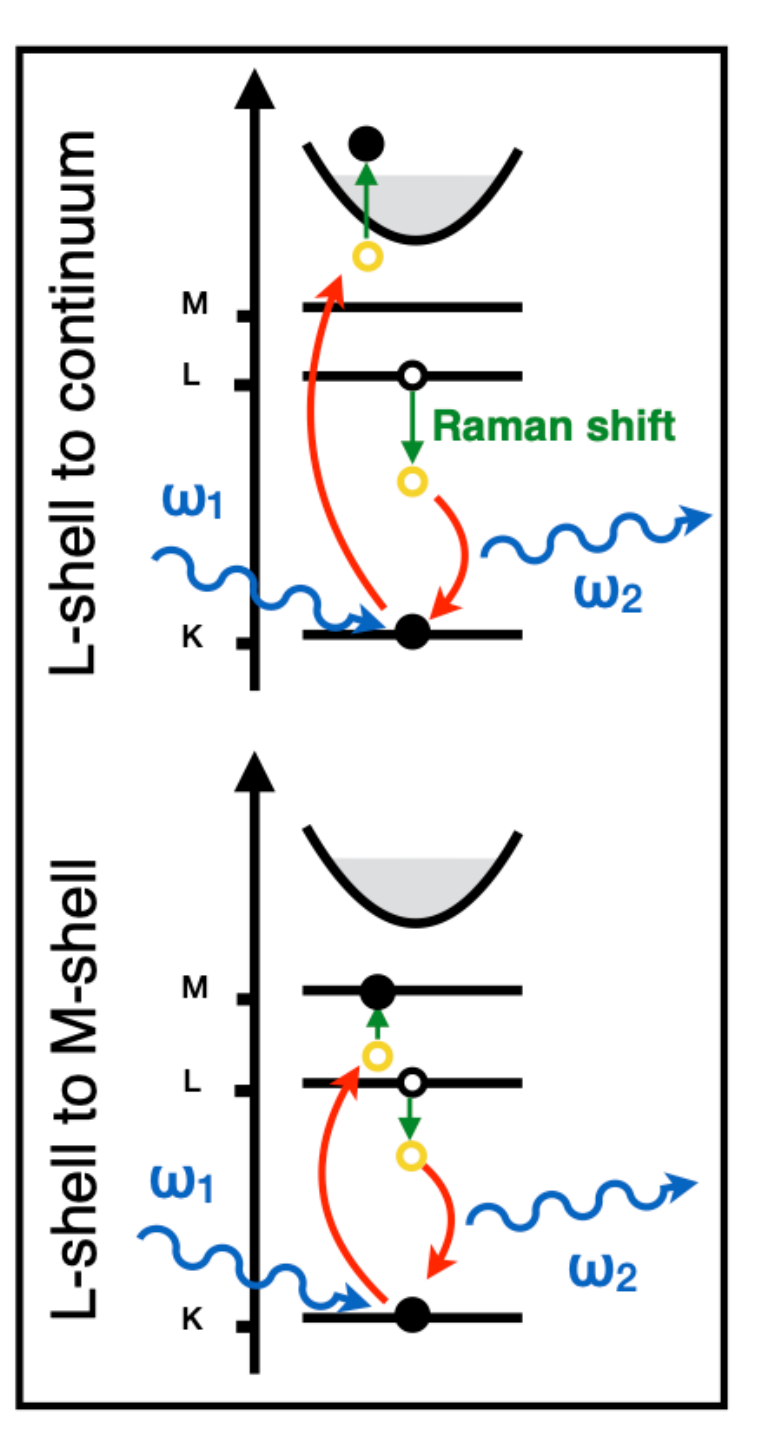}
\caption{A schematic diagram of resonant inelastic x-ray scattering (RIXS). Empty states in the continuum (upper plot) or in a bound level (lower plot) can be probed by observing the intensity of outgoing Raman-shifted x-ray photons.}
\label{fig:rixs}       
\end{center}
\end{figure} 

One potential method of gathering far more information about the electronic structure, which has been proven in principle on the pertinent timescales is RIXS \cite{humphries2020}: Resonant Inelastic X-ray Scattering. Whilst x-ray diffraction relies on elastic scattering of the incident photons by the electrons in the system, be they bound or free (and thus is effectively Thomson scattering), RIXS is an inelastic scattering technique where the outgoing photon has lower energy than the incident FEL photon.  The process is shown schematically in Fig. \ref{fig:rixs}.

We can model the interaction of x-rays with the electrons in matter by noting that the vector potential of the electromagnetic field of the x-rays gives rise to an interaction Hamiltonian containing the terms
\begin{equation}
    H_{I} = -\frac{\alpha}{2}(\vec{p}\cdot \vec{A} + \vec{A}\cdot \vec{p}) + \frac{\alpha^2}{2} \vec{A} \cdot \vec{A}  \quad ,
\end{equation}
where $\alpha$ is the fine structure constant.  When applied as a perturbation in first order, the $\vec A \cdot \vec A$ term gives rise to Thomson scattering, and the $\vec{A} \cdot \vec{p}$ term  photon absorption or emission (these being separate processes).  RIXS, being a single process that gives rise to a photon-in and a photon-out, is a term that is second order in $\alpha$ (like Thomson scattering), but in contrast arises from the second order perturbation of the $\vec A \cdot \vec p$ term, and thus depends on the electronic structure of the material.  

As an example, and as can be seen from Fig. \ref{fig:rixs}, if the x-rays are tuned below the K-edge of a material to an intermediate state, then an electron can be excited into the valence band, and simultaneously, to conserve energy, a photon is then emitted with an energy lower than the center of the K-$\alpha$ energy.  If the lifetime of this process is dominated by the K-shell core-hole lifetime, and considering the process whereby an electron is excited into the valence band, 
for FEL photons of frequency $\Omega$, photons at lower energies $\omega$ are scattered with a cross section that is proportional to a Lorentzian profile dictated by the natural lifetime of the K$_{\alpha}$, and the empty density of states of the valence bands (multiplied by an appropriate dipole matrix element).  As such,  the empty density of states (DOS) of the valence band can be probed by such scattering in a single few-femtosecond shot over a wide range of energy.  Indeed, the first experiments demonstrated that the DOS could be probed over several hundred eV in this way.  These proof of principle experiments used an FEL pulse with a wide spectral bandwidth ($\sim$ 30 eV at a photon energy of $\sim$ 8 keV), dictated by the SASE nature of the beam, and thus the spectral resolution would have been insufficient to see features such as the opening or closing of bandgaps.  However, similar numbers of FEL photons can now be produced in seeded beams with bandwidths closer to 0.5 eV or below~\cite{Amann2012}, and thus exploring such electronic properties in this way does seem feasible.  

It is also worth noting that such an increase in spectral resolution could also potentially give access to a measurement of the temperature of the electrons in metallic systems (assuming it is low compared with the Fermi temperature), in that population of bound states in the band will be determined by the temperature-dependent Fermi function.  With improvements in the bandwidth of the FEL pulses RIXS may thus provide a complementary means of measuring temperature within a system compared with those already discussed in subsection \ref{SS:Temperature}.  In certain cases temperature might also be inferred from the depopulation of bound states via ionisation, as depicted in the lower plot in Fig. \ref{fig:rixs}.  Indeed, it was this method that was used to determine temperatures in the intial FEL RIXS experiments~\cite{humphries2020}. 

Care will need to taken that the x-ray beam itself does not significantly perturb the electron distribution function of the system, which will place an upper limit on the intensity of the FEL pulse on the target of interest.  In the preliminary experiment described above, the FEL beam was deliberately focused to a small spot (down to a diameter of 2 $\mu$m) in order to heat the system to temperatures in the few eV regime.  Given that for many laser compression experiments diameters of more than 100 $\mu$m would be more appropriate, and absorption is by definition low as one remains below an absorption edge, it should be possible to keep the heating effect of the scattering beam such that it does not overly perturb the system.

Ideally one would wish to employ RIXS at the same time as obtaining structural information via x-ray diffraction.  This may not always be possible, or conflicts may arise, because diffraction will require high energy x-rays in order to probe a sufficiently large volume of reciprocal space (leading to high spatial resolution), whereas RIXS relies on tuning the x-rays to a particular energy based on the atomic number of the element of interest.  It may thus not be possible to employ both techniques with a single color FEL pulse, but may require simultaneous probing of the sample with two different FEL pulses of two differing photon energies, if and when such technology becomes available.

\section{Technical Developments}
\label{S:Tech_Dev}

The evolution of our ability to study materials that have been dynamically compressed to high pressures via femtosecond x-ray diffraction has been reliant on the astonishing development of x-ray FELs, and upon the optical drivers required for the compression process.  We thus conclude our perspectives by noting that future advances within this field will themselves be influenced largely by further advancements in the underpinning FEL and driver technologies.

Our first comment is to note the huge disparity to date in the repetition rate between the x-ray sources and the optical lasers that provide the ablation pressure to compress the sample.  When it first came online in operating mode, LCLS could provide x-ray pulses at 120 Hz.  In contrast, the several 10s of Joules nanosecond-pulselength laser that operated at the Materials in Extreme Conditions (MEC) end-station, allowing dynamic compression, is a 2-arm system, where each arm of the laser can only produce a shot once every 3 minutes.  We note that FELs based on superconducting accelerator technologies, such as the European XFEL~\cite{Decking2020}, have even higher x-ray repetition rates, bringing the disparity between the rate at which the optical laser can created samples under compression, and that at which the x-ray FEL probe such samples, into even starker relief.  It is in this context recent development in driver technologies should be viewed, whereby the use of diodes to pump the gain media of the drive lasers (as opposed to flashlamps), combined with active cooling of the amplifying media, has led to the development of 100 J class nanosecond optical lasers that have repetition rates of 10 Hz.  Specifically, the DIPOLE laser~\cite{Mason2017,mason2018} that has recently been sited at the European XFEL has been shown to exceed the parameters described above.  Furthermore, it has been developed such that the temporal profile of the laser pulse can be sculpted in time to a user-defined shape in order to allow ramped pressure pulses to be applied to the target in order to keep it closer to an isentrope, as described in section \ref{SS:Shock-Ramp-JSW}.  The system as a whole also incorporates adaptive optics with the aim of optimising the spatial uniformity of the beam at focus on the target.  At the time of writing this system has yet to be fully commissioned, but we anticipate it being able to vastly increase the number of experiments that can be performed, and the amount of data collected.  The capacity to vary the temporal shape of the driving pulse in real time, and thus the rate of loading, should influence the final temperatures reached in the target.  In addition, repetition rates of this order will be necessary if techniques that are photon-hungry from the x-ray perspective are also going to be exploited - e.g. the inelastic scattering from phonons discussed in section \ref{SS:Temperature}.  The full exploitation of these higher repetition rates will itself be reliant on novel methods of mass-producing the targets to be compressed, as discussed in section \ref{SS:Targets}.

Whilst drivers with increased repetition rate comprise one obvious direction for the field, we also consider that drivers with increased overall energy will have significant future impact.  The nanosecond lasers that have been placed alongside FELs to date have been in the 10 to 100 J class.  For the experiments that have been conducted thus far - i.e. those in planar geometry - the requirements to have laser spot sizes far greater than the sample thickness, along with a laser pulse length itself determined largely by the target thickness and sound speed - leads to an overall limit on the peak laser intensity that can be achieved on the target to be compressed, and thus a limit on the pressures achievable. Whilst it will be a sample-dependent figure, realistically we would estimate that the peak pressures achievable, at least for ramp compression combined with x-ray diffraction, will be of order 0.5 TPa for a 100 J class optical laser.  As shown in section \ref{S:Ablation}, simple ablation physics models show that the applied pressure scales approximately as the intensity to the two-thirds power, i.e. less than linearly, implying that a kJ class system will be required to enable multi-TPa physics to be explored, unless splitting of the beams to enable convergent geometry compressions, as described in section \ref{SS:Convergent} is implemented.

In the context of the suggestion above that increased laser energies (and intensities) will lead to high-pressures, an interesting question that has yet to be addressed in any great detail is what the ultimate limitations might be of this approach.  The laser-ablation technique by its very nature produces a high temperature plasma at the target surface, and the temperatures reached can easily attain a few hundred eV to a keV for intensities associated with multi-TPa pressures (note 1 eV is about 11,000 K).  This plasma thus acts as a source of copious x-rays that can preheat the underlying material, which would render unusable the whole approach.  Furthermore, it is well known from studies of relevance to ICF that laser-plasma instabilities can lead to the production of electrons with energies far greater than those in the mean thermal background, and these `hot electrons' can also be a source of preheat~\cite{Kidder1981}.  It is for these reasons that in experiments on the NIF discussed in section \ref{SS:Some-History-JSW} the targets have often incorporated a thin layer of gold between the ablation surface and the material of interest, to act as an x-ray absorber, and we consider such shielding will be required in most targets as higher pressures are sought, and knowledge gleaned from the ICF community, where the DT fuel needs to be kept cool during laser ablation, will prove extremely useful in future target designs.

In addition to the advances in drive laser technology discussed above, advancements in the performance and operational modes of the FELs themselves will no doubt advance the field, and we have already commented on the impact of several of these in previous parts of this manuscript. Moving from SASE operation to self-seeded operation can reduce the bandwidth of the FEL from about several 10s of eV with irreproducible multimode
structure to several eV FWHM and a nearly Fourier-transform limited bandwidth~\cite{Amann2012,Emma2017,Inoue2019}. This advance will allow much improved high resolution spectroscopy (HRIXS) and higher-resolution diffraction.  In section \ref{SS:Temperature} we emphasized that extremely high spectral resolution is required to discern inelastic scattering from phonons, and thus increases in FEL spectral brightness by seeding methods - providing similar numbers of phonons in a greatly reduced spectral bandwidth - will be extremely beneficial.  Such advances provide similar benefits for the RIXS measurements discussed in section \ref{SS:RIXS}.

In section \ref{SS:liquids} we noted that the operation of FELs at higher photon energies will significantly improve our understanding of high pressure liquids by enabling higher-$Q$ data to be collected above the current limiting value of 10-12\AA$^{-1}$. A lower $Q$ limit of 15 \AA$^{-1}$ would be both achievable and provide useful structural data. With 20 keV x-rays this requires data to 2$\theta\sim$95$^{\circ}$ while with 25 keV x-rays it would required data to 2$\theta\sim$73$^{\circ}$. It should be noted that the next generation of diffraction limited synchrotrons, such as ESRF-EBS, APS-U, and PETRA-IV, will produce extremely intense (10$^8-10^9$ ph/pulse), essentially monochomatic ($\Delta E/E$ $\sim$5$\times 10^{-3}$) x-ray pulses at energies up to 60 keV. While the peak brightness from such storage rings will still be below that obtainable at a FEL, the difference is reduced at 60 keV.

Finally, the new ability at LCLS and elsewhere to operate in a double or multi-bunch mode where x-ray pulses are separated by intervals on the 10s or 100s of ps timescales \cite{Decker2019, Sun2018}, and the future optimization of area detectors capable of single-bunch isolation\cite{Hart2019}, will revolutionize the study of time-resolved phase transformation and many other pump-probe type experiments including isochoric heating with subsequent compression
and probing by the following pulses.

\section{Summary}
\label{S:Summary}
In summary, our ability to study the structure and deformation pathways of laser-shock-compressed matter, and, via ramp compression, solids created via laser-ablation at pressures exceeding those that can be achieved by any diamond anvil cell has advanced enormously over the past few years.  These advances have been made possible by the attendant developments in technology, in that exquisitely temporally-tailored laser pulses allow us to compress materials on the timescales of several nanoseconds, before the wave has steepened into a shock.  Structural diagnosis of materials on ultra-fast timescales has been revolutionised by the development of x-ray FELs.  These two areas are starting to converge with the siting of new high repetition-rate optical drive lasers alongside FEL facilities.

Whilst much has been learnt over the past decade since the LCLS produced its first x-rays, in many ways the field is still in its infancy, and many questions still remain.  Within this article we have attempted to outline our perspectives on those issues we deem to be the most pertinent. From the early FEL experiments that have been performed much is starting to be understood, at least for a few materials, on what the underlying deformation processes are at the lattice level that lead to the relief of high shear stresses induced by the uniaxial compression process.  That said, our overall understanding of plastic deformation as a function of both strain-rate and strain is clearly far from complete.  As it is the amount of plastic work to which the sample is subject that dominates the temperature rise in the target, these gaps in our knowledge clearly need to be addressed.  However, not only are there uncertainties in how much heating may be taking place in these types of experiments, but reliable wide-ranging and accurate measurements of temperature of dynamically compressed samples are lacking.

In addition to our inability to know accurately the temperature, it is also clear that rapid compression, beit by shock or by ramp,  leads to the observation that phase boundaries for polymorphic transitions often do not match those seen in static cases, and indeed certain phases may appear absent, or new unexpected ones may emerge.  The degree to which these observations can be accounted for by residual shear stresses (or the non-hydrostatic pathway taken to a largely hydrostatic state) or kinetic effects, or both, is a question that has not, in our view, received as much attention as it might eventually merit.  However, if we wish to apply findings about the structure of dynamically-created high pressure matter to environments where high pressures exist on longer timescales, for example within the interior of planets, then a more nuanced understanding of the states reached on these short timescales is surely in order.

Finally, we note that the increased understanding of the response of matter at the lattice level to rapidly applied pressure that has been gleaned over the last decade or so has been largely driven by the spectacular improvements and developments in  both optical lasers and x-ray FELs, along with concomitant improvements in simulation and diagnostic techniques: the furtherance of this burgeoning field will no doubt be shaped by the future evolution of these underpinning technologies.

\section{Acknowledgments}
MIM would like to thank Prof David Keen of Rutherford Appleton Laboratory and Prof Graeme Ackland of The University of Edinburgh, for useful discussions about structural studies of liquids, and DFT calculations, respectively. JSW is grateful to Dr Patrick Heighway from The University of Oxford for useful discussions about the details of plasticity at ultra-high strain rates, and for support from AWE through the Oxford Centre for High Energy Density Science (OxCHEDS).  He is further grateful to EPSRC for support under grant No.\ EP/S025065/1.  JHE would like to thank Dr Peter Celliers for useful discussions about VISAR analysis and to acknowledge that his work was performed under the auspices of the U.S. Department of Energy by Lawrence Livermore National Laboratory under Contract DE-AC52-07NA27344 and was supported by the LLNL-LDRD Program under Project No. 21-ERD-032.

\bibliography{JSW-bib-file}

\end{document}